\documentclass[11pt]{article}
\usepackage{epsfig}

\normalbaselines
\newcommand{\be}{\begin{eqnarray}}
\newcommand{\ee}{\end{eqnarray}}

\def\fr{\frac{1}{2}}

\def\mref#1{(\ref{#1})}
\def\zero{\setcounter{equation}{0}}

\def\bd{\begin{displaymath}}
\def\ed{\end{displaymath}}

\def\ba#1{\begin{array}{#1}}
\def\ea{\end{array}}
\def\nn{\nonumber}
\newfont{\Bbb}{msbm10 scaled 1200}
\renewcommand{\theequation}{\thesection.\arabic{equation}}

\begin{document}

\pagestyle{empty}

\begin{center}

{\LARGE\bf         Topological Expansion and Exponential Asymptotics
in 1D Quantum Mechanics }

\vskip 24pt

{\large\bf  Stefan Giller \footnote{Supported by the KBN 2 PO3B 134 16}}

\vskip 18pt

Theoretical Physics Department II, University of {\L}\'od\'z,\\
Pomorska 149/153, 90-236 {\L}\'od\'z, Poland\\ 
e-mail: sgiiler@krysia.uni.lodz.pl
\end{center}
\vspace{18pt} 
\thispagestyle{empty}

\begin{abstract}
Borel summable semiclassical expansions in 1D quantum
mechanics are considered.  These are the Borel summable
expansions of fundamental solutions and of quantities
constructed with their help. An expansion, called topological,
is constructed for the corresponding Borel functions. Its main
property is to order the singularity structure of the Borel
plane in a hierarchical way by an increasing complexity of this
structure starting from the analytic one. This allows us to
study the Borel plane singularity structure in a systematic way.
Examples of such structures are considered for linear, harmonic
and anharmonic potentials. Together with the best approximation
provided by the semiclassical series the exponentially small
contribution completing the approximation are considered. A
natural method of constructing such an exponential asymptotics
relied on the Borel plane singularity structures provided by the
topological expansion is developed. The method is used to form
the semiclassical series including exponential contributions for
the energy levels of the anharmonic oscillator.

\end{abstract}

\vskip 6pt
\begin{tabular}{l}
{\small PACS number(s): 03.65.-W , 03.65.Sq , 02.30.Lt , 02.30.Mv} \\[6pt]
{\small \parbox{14cm}{Key Words: Quantum mechanics, fundamental solutions,
  semiclassical expansions, Borel summability, exponential asymptotics.}}
\end{tabular}

\newpage

\pagestyle{plain}

\setcounter{page}{1}
        
\section{Introduction\label{s1}}
\zero

\hskip+2em In this paper we continue our investigations to
represent basic quantities of the quantum mechanics in the form
of Balian - Bloch representation i.e. in the form of the
Laplace-Borel transforms in which the cojugate variables are an
action and the Planck constant \cite{1}. The key results, which the
paper is relied on, has been published earlier \cite{2}. The present
work develops these key ideas and using the explicit form of the
fundamental solutions \cite{11,12} to the 1D Schr\"odinger equation
expresses the Balian-Bloch representation in the form of what we
call a topological expansion. We describe also the way of using
the representation to construct extended JWKB approximations in
the form of so called exponential asymptotics (sometimes called
also the hyperasymptotics, see \cite{23,24,25,26,27} and the references cited
there) and we consider some particular applications of the
Balian-Bloch method as well.  

For simplicity the potentials
considered in this paper are assumed to be polynomial but its
main results are valid for more general meromorphic potentials as well.  

Being Borel summable the fundamental solutions define
their corresponding Borel functions with the help of which they
can be represented in the form of the Borel transformation from
the Borel plane of the action variable to the complex plane of
the $\hbar^{-1}$-variable. For the polynomial potentials these Borel
functions are in fact all the same despite the fact that they
are defined by different fundamental solutions \cite{3}. This means
of course that the fundamental solutions themselves are in close
relations to each other being in fact a mutual analytical
continuation of each other in the $\hbar^{-1}$ plane \cite{2,3}.

Therefore, to get any of the fundamental solutions it is necessary only to
choose properly an integration path in the Borel plane. However,
to do it a detailed knowledge of singularity distribution of the
Borel function in the Borel plane is necessary. It is the aim of
this paper to provide us with an effective tool for studying
these singularities. Namely, we develope an expansion for the
Borel function called toplogical in which an expansion parameter
is the complexity of the Borel plane corresponding to successive
terms of the expansion.

  With the help of the fundamental
solutions we can solve most of the 1D quantum mechanical
problems so that the corresponding quantities involved in the
problems considered depend on different pieces of the
fundamental solutions used. These quantities themselves can have
then semiclassical representations which can be Borel summable
and can serve as a source of their semiclassical approximations
as well. It is clear that the corresponding Borel plane
singularities of these quantities are defined then by the pieces
of the fundamental solutions constructing them. Therefore the
topological expansion method can be applied also to determine
the approximate singularity structure for these quantities as well.  

The semiclassical expansions used as a source of
approximations are considered as insufficient providing us with
unavoidable nonvanishing errors. It is well known that the
reasons for these errors are immanent due to the divergence of
the semiclassical series so that the latter as asymptotic
neglect the exponentially small contributions. Nevertheless,
since the series are Borel summable they have to contain the
full information about such exponential contributions.  A common
goal of many approaches was just to recover these contributions
leading to a formulation of so called resurgent theory \cite{24,25,26,27}.

Let us note, however, that the
exponentially small contributions is of its own importance
since in many cases of quantities considered these
contributions are $dominant$. Among the latter cases the
most well known one is the difference between the energy levels
of different parities in the symmetric double well \cite{30}. But
these are also the cases of transition probabilities in the
tunnelling phenomena \cite{30} or their adiabatic limits in the time
dependent problem of transitions between two (or more) energy
levels (see \cite{32,33} and references cited there) or the
exponential decaying of resonances in the week electric field
(see \cite{34,35} and references cited
there).

In an approach of our paper we make full use of the Borel
summability of the quantities considered as well as of the
corresponding topological expansions to construct the relevant
exponential asymptotics.  

However, as a necessary step of our
fomulation is the knowledge of the Borel plane singularity
structure of any considered quantity. It is just the topological
expansion which allows us to built this knowledge step by step.

The toplogical expansion is constructed directly from the Fr\"oman
and Fr\"oman representation of the fundamental solutions which
themselves are given in the forms of functional series \cite{2,3,4}.
Therefore, we shall start in the next section with the detailed
presentation of the series.  

In Section 3 the topological series
representation for the Borel functions is introduced and its
convergence is proved. This representation provides us with an
algorithm for approximate calculations of Borel functions being
alternative to the ones relied on the Pad{\'e} approximants \cite{5,8,9,14},
continued fractions \cite{10} or conformal transformations \cite{14}.  

In Section 4 singularity structures of the topological series
expansion are analyzed and their hierarchic form (which gives
rise to the name of the series) is established. We consider
there as the simplest examples the 'first sheet' singularity
structures of the linear and harmonic potentials. In particular
we describe completely the singularity structure of the Borel
plane of the harmonic oscillator Joos function found first by
Voros by a different method \cite{19}.  

The results obtained in
Sections 3 and 4 are applied in the next section where the
solution of the so called connection problem within the
framework of the Balian-Bloch representation is discussed.  

In Section 6 we discuss the problem of the exponential asymptotics
\cite{23,24,25,26,27}. We show that this problem has a natural solution in the
framework of the Balian-Bloch representation and gets a natural
support from the topological expansion method.

  In Section 7 the
energy levels of the single-well anharmonic potential are
considered in order to show how to use the topological expansion
to construct their extended exponential asymptotics.  

Finally, in Section 8 we summarise our results.

\section{Fundamental solutions \label{s2}}
\zero

Let us remind basic notions of our
considerations (see \cite{2}, for details).  

The fundamental
solutions satisfy the Schr\"odinger equation: 
\be\label{2.1}
& \Psi''(x,\lambda,E)-\lambda^2q(x,E)\Psi(x,\lambda,E)=0 &
\ee
where: $q(x,E)=V(x)-E, \; \lambda=\sqrt{2m}\hbar^{-1}$. Both $\lambda$
 and $E$ can take on complex values.
$V(x)$ is assumed to be a polynomial of any degree $n \geq 1$. A Stokes
line pattern (see \cite{11,12} for necessary definitions) relevant
for our considerations is shown in Fig. 1. (A total number p of
sectors equals to $n+2$ in this case.) 

The following fundamental
solution $\Psi_1^{\sigma}(x,\lambda,E)$ to \mref{2.1} can be attached 
to the sector $S_1$:

\be
\Psi_1^{\sigma}(x,\lambda,E) = q^{-\frac{1}{4}}(x,E) 
e^{\sigma \lambda \int_{x_o}^{x} q^{\fr}(y,E)dy}
\chi_1^{\sigma}(x,\lambda,E) \nn \\
\label{2.2}
\Re \left[ \sigma \int_{x_o}^{x} q^{\fr}(y,E)dy \right] < 0, \;\;\;
x\in S_1 ,\;\;\;\; \sigma=\pm 1 \\
q(x_0,E) = 0 \nn
\ee

with the "amplitude factor" $\chi_1^{\sigma}(x,\lambda,E)$ given 
by the following functional series:

\be\label{2.3}
\chi_1^{\sigma}(x,\lambda,E)
=1+\sum\limits_{n\geq 1}\left(\frac{\sigma_k}{2\lambda}\right)^n
\int\limits_{\gamma_1^{\sigma}(x)}dy_1 \ldots 
\int\limits_{\gamma_1^{\sigma}(y_{n-1})}dy_n
\omega(y_1) \ldots \omega(y_n)
\cdot \\
\cdot \left[1-e^{2\lambda\xi(y_1,x)}\right]
\cdot \left[1-e^{2\lambda\xi(y_2,y_1)}\right] \cdot \ldots \cdot
\left[1-e^{2\lambda\xi(y_n,y_{n-1})}\right] \nn
\ee
where:

\be\label{2.4}
 \omega(y)= \frac{1}{4} \left[
\frac{\tilde{q}''(y)}{\tilde{q}^\frac{3}{2}(y)}-\frac{5}{4}
\frac{\tilde{q}'^2(y)}{\tilde{q}^\frac{5}{2}(y)}\right] = 
- q^{-\frac{1}{4}}(y) \left(q^{-\frac{1}{4}}(y) \right)'' \\
 \xi(x_0,x) = -\sigma \int_{x_0}^{x} q^{\fr}(y,E)dy \nn
\ee
and where an obvious dependence of $\omega,\; q, \; \xi$, etc. on $E$ has
been dropped. We shall also put $\sigma = -1$ in (\ref{2.2})-(\ref{2.4}) 
assuming that in \mref{2.2} the corresponding inequality is satisfied in this
case.  

\begin{tabular}{c}
\psfig{figure=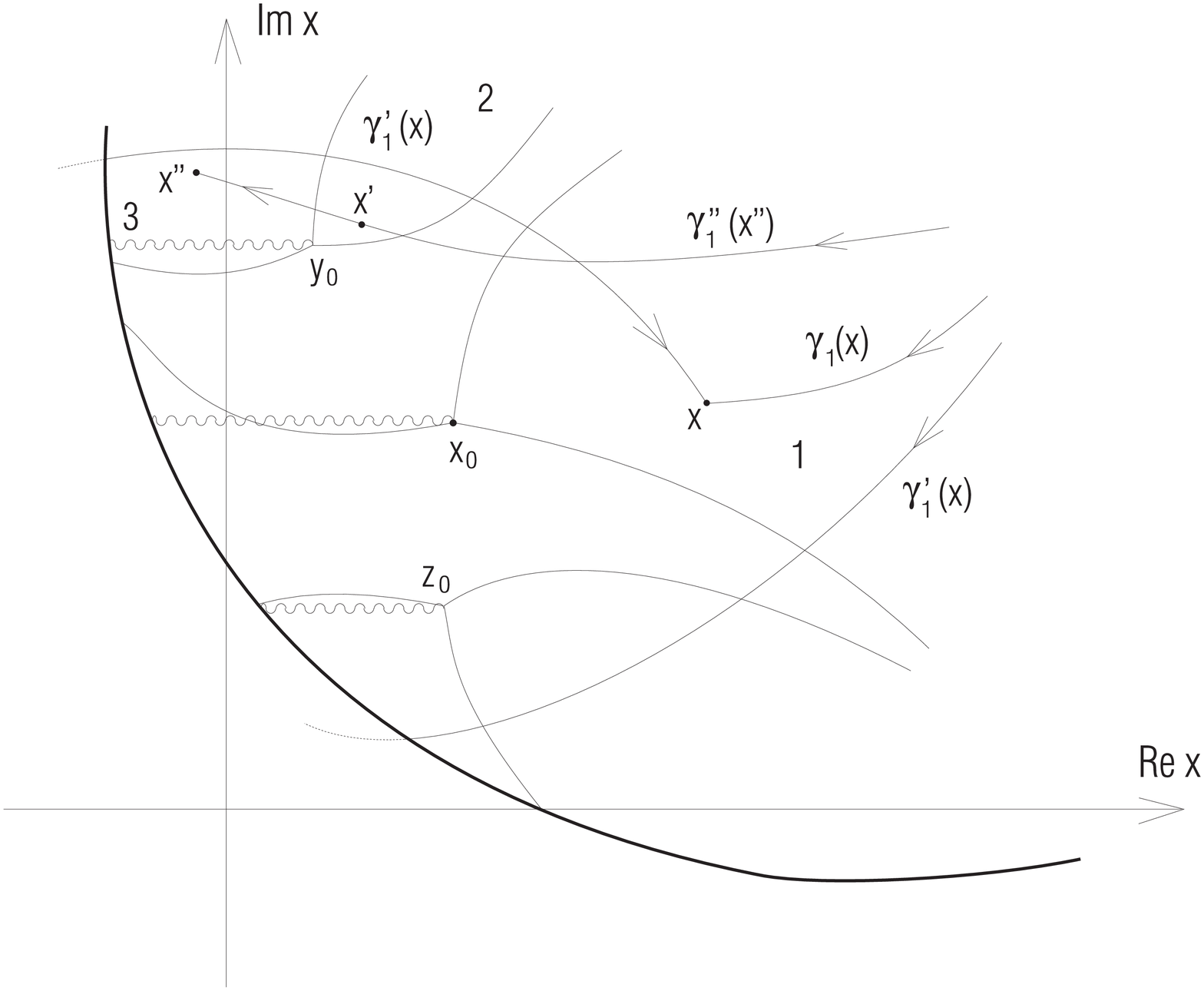, width=10cm} \\
\parbox{14cm}{Fig. 1. $\;\;\;$ The Stokes graph for a general 
polynomial potential}
\end{tabular}
\vskip 12pt

In the sector $S_1$ the following semiclassical expansion
for $\chi_1(x,\lambda)$ takes place: 
\be
 \chi_1 (x,\lambda) \sim  \chi_1^{as} (x,\lambda)= 
1 + \sum_{n\geq1}\frac{\kappa_{1,n} (x)}{2\lambda)^n} \nn\\
 \kappa_{1,n} (x) =
\int_{\infty_k}^{x}d\xi_{n}q^{-\frac{1}{4}}(\xi_{k}) 
\left( q^{-\frac{1}{4}}(\xi_{n}) 
\int_{\infty_k}^{\xi_{n}}d\xi_{n-1}q^{-\frac{1}{4}}(\xi_{n-1}) \label{2.5}
\right. \\
 \left.  
\cdot \left(\ldots q^{-\frac{1}{4}}(\xi_{2}) 
\int_{\infty_k}^{\xi_{2}}d\xi_{1}q^{-\frac{1}{4}}(\xi_{1}) 
\left( q^{-\frac{1}{4}}(\xi_{1}) \right)^{\prime{\prime}} 
 \ldots \right)^{\prime{\prime}} \right)^{\prime{\prime}}  ,\;\;\;\;
k = 1,2,\ldots \nn
\ee

As it has been shown in \cite{2} if $x$ stays
in $S_1$ of Fig. 1 then we can define for $\Re s < 0$ the following
Laplace transformation of the amplitude factor $\chi_1 (x,\lambda)$: 
\be\label{2.6}
\tilde{\chi}_1 (x,s) =
\frac{1}{2\pi i} \int_C e^{-2\lambda s}
\frac{\chi_1 (x,\lambda)}{\lambda}d\lambda
\ee
with
the integration contour $C$ shown in Fig. 2. (The factor $2$ in the
exponential function in (\ref{2.5}) is introduced for convenience). By
the form (\ref{2.6}) $\tilde{\chi}_1 (x,s)$ is defined holomorphically in the
half-plane $\Re s < \Re \xi(x_0,x)$ and since 
$\Re \xi(x_0,x)$ is positive $\tilde{\chi}_1 (x,s)$
appears to be, in fact, the Borel transform of $\chi_1 (x,\lambda)$. The
contour C in (\ref{2.6}) can be chosen as a circle with its radius $\lambda$
to be large enough to substitute $\chi_1 (x,\lambda)$ by its semiclassical
series (\ref{2.5}). Then for $|s| < |\xi(x_0,x)|$ the LHS of 
(\ref{2.6}) can be integrated to give the following Borel series: 
\be\label{2.7}
\tilde{\chi}_1 (x,s) = 1 +
\sum_{n\geq 1}\kappa_{1,n}(x)\frac{(-s)^n}{n!}
\ee
convergent in the
circle $|s| < |\Re \xi(x_0,x)|$. The point $s_0(x) = \xi(x_0,x)$ is a
singularity for $\tilde{\chi}_1 (x,s)$ closest to the origin.  

\begin{tabular}{c}
\psfig{figure=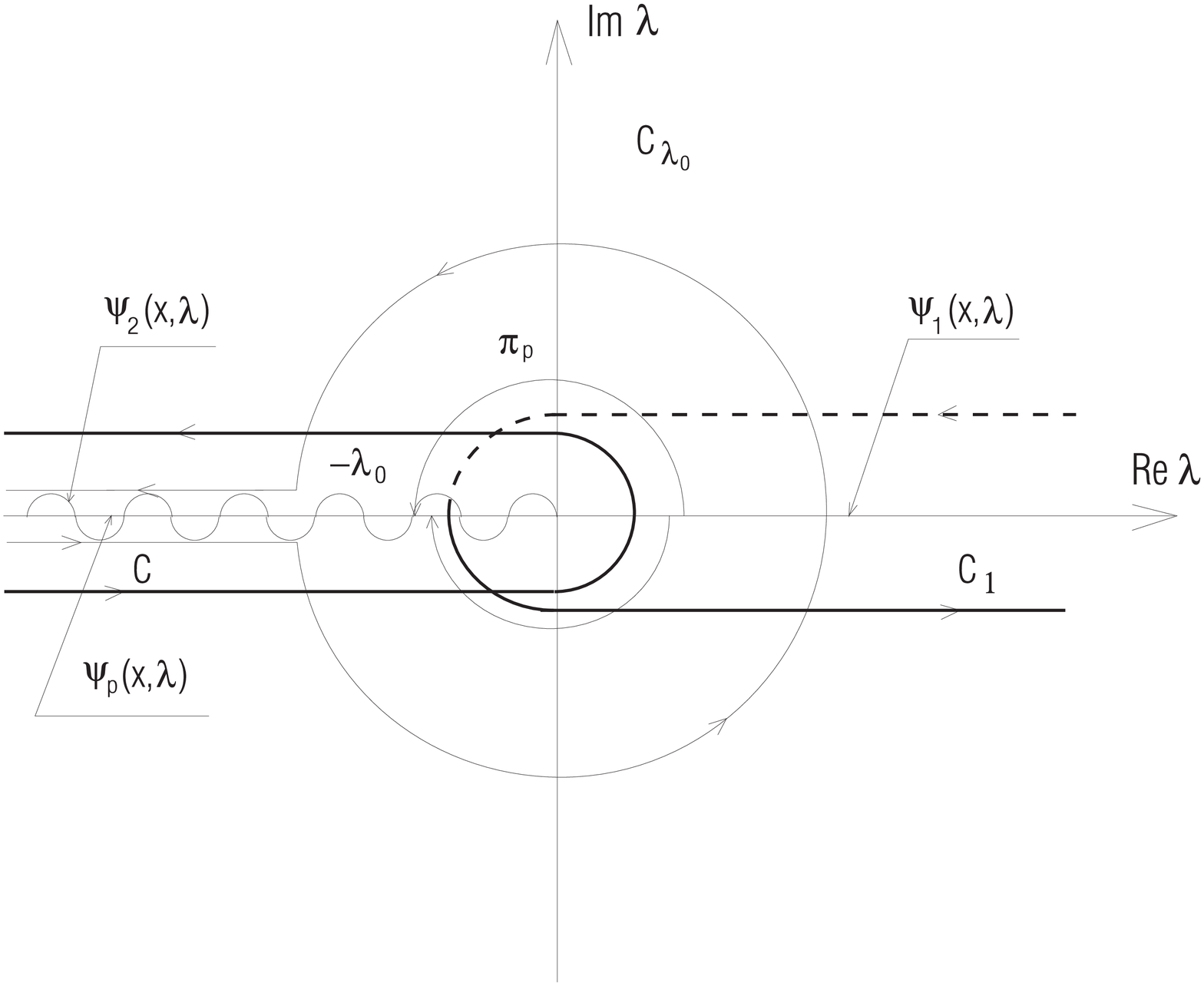,width=10cm} \\
\parbox{15cm}{Fig. 2. $\;\;$ 
The cut $\lambda$-plane corresponding to the global
solution $\Psi(x,\lambda)$}
\end{tabular}
\vskip 12pt

The transformations (\ref{2.5}) can be inverted to give:

\be\label{2.8}
\chi_1 (x,\lambda)=
2\lambda \int_{\tilde{C}} e^{2\lambda s}
\tilde{\chi}_1 (x,s) ds
\ee
 where the contour
$\tilde{C}$ starts at the infinity $\Re(\lambda s) = -\infty$ and ends 
at $s = 0$. Since $\tilde{C}$
can be freely deformed in the half plane $\Re s \leq 0$ the formula
(2.8) define $\chi_1 (x,\lambda)$ in the whole sheet shown in Fig. 2
excluding the points of the negative real half-axis.  

 The following is worth to note.  

The formula (2.8) is certainly
valid for $\Re \lambda > 0$ when the contour $\tilde{C}$ stays 
in the half plane $\Re s< 0$. It can be continued, however, to other domains 
of the Riemann $\lambda$-surface corresponding to $\chi_1 (x,\lambda)$ 
if accompanied with suitable changes of the variable $x$. 
Thus, for example, when
continuing $x$ to the sector $S_2$ and deforming the contour $C$ in
Fig. 2 into $C_1$ the formula (\ref{2.5}) will then 
define $\tilde{\chi}_1 (x,s)$  in
the half plane $\Re s > 0$. On the other hand the inverse formula
(2.8) defines then $\chi_1 (x,\lambda)$ in the half plane $\Re \lambda < 0$ 
with the contour $C$ in the formula deformed (anticlockwise) from its
position in the left half-plane to its new position in the right
half of the $s$--plane. The function $\chi_1 (x,\lambda)$ fulfils then for 
$\lambda > 0$ the condition: $\chi_1 (x,-\lambda) \equiv \chi_2 (x,\lambda)$. 
Possible singularities of
$\tilde{\chi}_1 (x,s)$ existing in the corresponding half-plane 
$\Re s > 0$ when $x$
stays in the sector $S_1$ move to the half-planes $\Re s < 0$ when $x$
moves to the sector $S_2$ (see also Section 5 for a relevant
discussion).

\section{Topological expansion of Borel function $\tilde{\chi}_1 (x,s)$ 
\label{s3}}
\zero 

\hskip+2em As it follows from the
definition of $\tilde{\chi}_1 (x,s)$ if we want to learn something about it we
have to analyze $\chi_1 (x,\lambda)$ as given by (\ref{2.3}). We shall show 
below that if $x$ stays in $S_1$ (see Fig. 2) then it is possible to
represent each term of the series in (\ref{2.3}) in the form of the
Laplace transformation i.e. we shall show that $\tilde{\chi}_1 (x,s)$ can also
be represented by some convergent functional series. The series
however can still be continued to almost the whole $x$-plane when
the latter is deprived of some vicinities of their turning points.  

To this end let us consider the $n^{th}$ term of the series
in (\ref{2.3}) and particularly its $n$-fold integral. Introducing to it
$\xi = \xi (x) = \xi(x_0,x)$ as a new integration variable we get: 

\be\label{3.1}
 Y_n(\xi,\lambda)=
\int\limits_{\tilde{\gamma}_1(\xi)} d\xi_1 \ldots 
\int\limits_{\tilde{\gamma}_1(\xi_{n-1})} d\xi_n
\tilde{\omega}(\xi_1) \ldots \tilde{\omega}(\xi_n)
 \left(1-e^{2\lambda(\xi - \xi_1)}\right)
  \ldots 
\left(1-e^{2\lambda\xi(\xi_{n-1} - \xi_{n})}\right)
\ee
where $\tilde{\omega}(\xi(x)) \equiv \omega(x) q^{-\fr}(x)$ and the path 
$\tilde{\gamma}_1 (\xi)$ starts from $\Re \xi = + \infty$
and ends at the point $\xi$ of the $\xi$-plane.

Let us point to some
basic properties of the transformation $\xi= \xi(x)$. First it maps
the two sheeted Riemann surface which the $x$-plane actually is
into another (in general infinitely sheeted) Riemann surface on
which each sector of Fig. 2 is represented by the right (left)
half planes. In particular sector $1$ in Fig . 2 is mapped into the
right half of this cut $\xi$-plane whilst sectors $2,\; 3$ and $p$ into
the left ones (see Fig. 3 where the sectors $2$ and $p$ lie below
the sheet shown). Zeros of $q(x)$ which are singular points for
$\omega(x) q^{-\fr}(x)$ are also suitably transformed into the
corresponding root branch points (of the third degree) of 
$\tilde{\omega}(\xi(x))$
on the $\xi$-Riemann surface (see Fig. 3 and Appendix 2). On this
surface $\tilde{\omega}(\xi(x))$ becomes additionally infinitely 
periodic with its
complex periods acting however between $different$ sheets of the
surface. As a result of this an image of each root of $q(x)$
proliferates infinitely on the $\xi$-Riemann surface with every such
a copy giving rise to still new branch point and sheet. The only
exception of the latter rule is the linear potential case the
$\xi$-Riemann surface of which is three sheeted with a single root
branch point of the third degree.  

\begin{tabular}{c}
\psfig{figure=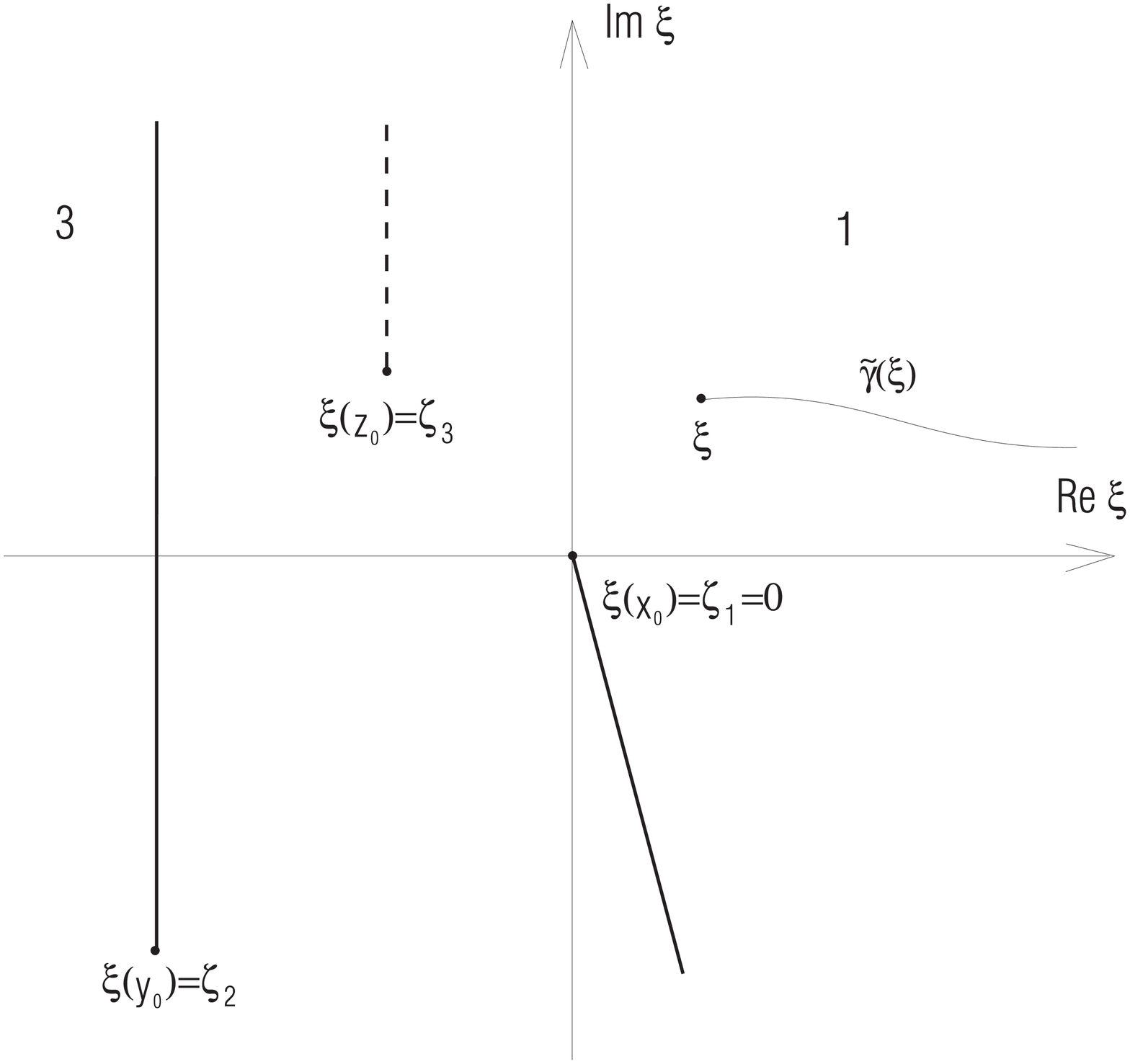,width=10cm} \\
\parbox{15cm}{Fig. 3. $\;\;\;$ The $\xi$-plane singularities corresponding to
$\tilde{\Phi}_1^{(0)}(\xi,s)$ (case $q=0$)}
\end{tabular}
\vskip 18pt

Opening the brackets in (\ref{3.1}) we get: 

\be\label{3.2}
Y_n(\xi,\lambda)
= \sum_{\begin{array}{c}
{\small 0\leq r_1< \ldots \leq n } \\
{\small 0 \leq 2q \leq n}
\end{array}}
Y_{n;r_1 \ldots r_{2q}}(\xi,\lambda)
(-1)^{r_1 - r_2 +r_3 - \ldots + r_{2q-1} - r_{2q}}
\ee
where 

\be\label{3.3}
Y_{n;}(\xi)
=
\int\limits_{\tilde{\gamma}_1(\xi)}d\xi_1 \ldots 
\int\limits_{\tilde{\gamma}_1(\xi_{n-1})}d\xi_n
\tilde{\omega}(\xi_1) \ldots \tilde{\omega}(\xi_n) = 
\frac{\Omega^n(\xi)}{n!}
\ee
with 

\be\label{3.4}
\Omega(\xi) = \int\limits_{\tilde{\gamma}_1(\xi)}d\eta 
\tilde{\omega}(\eta)
\ee
and 

\be\label{3.5}
 Y_{n;r_1 \ldots r_{2q}}(\xi,\lambda) =
\int\limits_{\tilde{\gamma}_1(\xi)}d\xi_1 \ldots
\int\limits_{\tilde{\gamma}_1(\xi_{n-1})}d\xi_n
\tilde{\omega}(\xi_1) \ldots \tilde{\omega}(\xi_n) 
e^{2\lambda (\xi_{r_1} - \xi_{r_2} + \xi_{r_3} - 
\ldots + \xi_{r_{2q-1}} - \xi_{r_{2q}})} \\
 \xi_0 \equiv \xi , \;\;\;\;\;\;\;\;\;\;\;\;\;\;\;\;\; q=1,2,3,\ldots
\;\;\;\;\;\;\;\;\;\;\;\;\;\;\;\;\;\;\;\; \nn
\ee

Note, that all the integrals in (\ref{3.5}) are
absolutely convergent. Therefore, it should be now obvious that
to each integral in (\ref{3.5}) the following Laplace transformation
form can be given: 
\be\label{3.6}
Y_{n;r_1 \ldots r_{2q}}(\xi,\lambda)
=
\int\limits_{\tilde{C}}ds 
e^{2\lambda s}
\tilde{Y}_{n;r_1 \ldots r_{2q}}(\xi,s)
\ee
where the integration contour $\tilde{C}$ starts at $\Re s = -\infty$ 
and ends at $s=0$ and the Laplace transform $Y_{n;r_1...r_{2q}}(\xi,s)$ 
is to be determined. We do it in Appendix 1. An important
observation done there is that it is possible to rearrange the
order of terms in the series (\ref{2.3}) in such a way to sum it in an
accordance with the increasing $q$ rather than $n$ - the number of
the integrations in (\ref{3.5}).  (All these are still possible since
the series (\ref{2.3}) is absolutely convergent). As a result of such
reordering $\chi_1(\xi,\lambda)$ can be represented as the following sum:
\be\label{3.7}
\chi_1(\xi,\lambda) = 2 \lambda
\sum_{q\geq 0} \chi_1^{(q)} (\xi,\lambda) \\
\xi \in S_1 , \;\;\;\;\;\;\; |\arg \lambda| < \pi \nn
\ee
where
\be\label{3.8}
\chi_1^{q} (\xi,\lambda) =
\int\limits_{\tilde{C}}ds
e^{2\lambda s} \tilde{\Phi}_1^{(q)} (\xi,s)
\ee
with $\tilde{\Phi}^{(q)} (\xi,s)$, $q \geq 0$ given by formulae (\ref{A1.12}) 
of Appendix 1 and with the contour $\tilde{C}$ shown in Fig. 5. Of course, 
since the series (\ref{3.7}) is absolutely and uniformly convergent we have 
also:
\be\label{3.9}
\chi_1 (\xi,\lambda) =
2\lambda \int\limits_{\tilde{C}}ds
e^{2\lambda s} \tilde{\Phi}_{1}(\xi,s)
\ee
with $\tilde{\Phi}_{1}(\xi,s)$ given by (\ref{A1.11}) so that the corresponding
Laplace transform $\chi_1(\xi,s)$ defined by (\ref{2.6}) can be identified
as: 
\be\label{3.10}
\chi_1(\xi,s) \equiv
\tilde{\Phi}_{1}(\xi,s)
\ee

The expansions (\ref{3.7}) and (\ref{A1.10}) shall be called further
topological expansions for the following two reasons: 
\begin{enumerate}
\item the higher term of the series in (\ref{A1.12}), 
the more complicated is its Riemann surface;
\item the Riemann surface $R_q$ corresponding to
the term $\tilde{\Phi}_1^{(q)} (\xi,s)$ in (\ref{A1.11}) can be reduced 
to some $R_{q'}$ with $q'<q$ when deprived of some singular points of 
$\tilde{\Phi}_1^{(q)} (\xi,s)$ i.e. a set $S_q$  of all singularities 
of $\tilde{\Phi}_1^{(q)} (\xi,s)$ includes a set $S_{q'}$ corresponding to 
$\tilde{\Phi}_1^{(q)} (\xi,s)$ (see the next section).
\end{enumerate}

\vskip 12pt
{\it 3.2. Analytic properties of $\tilde{\chi}_1(\xi,s)$}
\vskip 12pt

The analytic properties of $\tilde{\chi}_1(\xi,s)$ have been established in 
Section 3 of Appendix 1. As it
follows from Appendix A1.3 the Laplace transform $\tilde{\chi}_1(\xi,s)$ is
holomorphic in some vicinity of the point $s=0$ for $\xi \in {\bf R}(d")$ i.e.
it is the Borel function (\ref{2.7}) corresponding to $\chi_1(\xi,\lambda)$. 
For $\Re \xi > 0$, however, $\tilde{\chi}_1(\xi,s)$ is holomorphic 
in the half plane $\Re s<0$.
Therefore, the asymptotic series constructed for $\chi_1(\xi,\lambda)$ when
$\lambda \to \infty$ is Borel summable to the function itself - a result 
which is in a full accordance with the corresponding one obtained in \cite{2}
and mentioned in Section 2.

\section{Singularity structure of $\tilde{\chi}_1(\xi,s)$ \label{s4}}  
\zero

\hskip+2em Because of (\ref{3.1}0) this is the singularity 
structure of $\tilde{\Phi}_1 (\xi,s)$ and the latter structure is determined 
by the corresponding singularity structures of $\tilde{\Phi}_1^{(q)} (\xi,s)$
due to (\ref{A1.10}). These structures can be investigated
by the analytic continuation procedure of the formulae (\ref{A1.11}) -
(\ref{A1.12}) with respect to $s$ and $\xi$ and are, on its own, determined
completely by the corresponding singularity structures of
$\tilde{\omega}(\xi)$ and
the integrations present in (\ref{A1.11}) and (\ref{A1.12}) (see Appendix 1).
These integrations can give rise to singularities due to the
following two mechanisms \cite{13}: 

\begin{enumerate}
\item moving singularity of the
integrand approaches a fixed limit of the integration or,
inversely, a moving limit of an integration approaches a fixed
singularity of the integrand (so called end point (EP-)
singularities).  
\item moving singularity of the integrand
approaches some another singularity pinching unavoidably in that
way the integration contour (so called pinch (P-) singularities).  
\end{enumerate}

In the convolution integrals of the formula
(\ref{A1.11}) only the functions  $\tilde{\omega}(\xi)$ and $\Omega(\xi)$ 
can give rise to both the (EP- and P-) 
singularity mechanisms since a dependence of the integrals on 
the remaining partners of the convolutions are holomorphic.  

From the defining formulae (\ref{A1.12}) and from the $\xi$-Riemann surface 
structure on which $\tilde{\omega}(\xi)$ and $\Omega(\xi)$ are defined
(this structure was sketched in the previous section) it follows
also that even for the simplest cases of first few $\tilde{\Phi}_1^{(q)} 
(\xi,s)$'s
their global $(\xi,s)$-Riemann surface structures are too
complicated to be fully handled and only some crude descriptions
of them are possible limited to a few first sheets and a few
singularities.  

However, in making the corresponding analysis by limiting ourselves to 
first few $q$'s we are free in deforming the integration 
contours in (\ref{A1.12}) i.e. the limitation of  $\tilde{\gamma}_1(\xi)$ to
the canonical choices is no longer valid. This observation is
very important and prooves that the Borel function $\tilde{\chi}_1(\xi,s)$
constructing initially for the fundamental solution of the
sector $S_1$ is universal i.e. each Borel summable solution to the
Schr\"odinger equation \mref{2.1} can be obtained by the Borel
transformation of $\tilde{\chi}_1(\xi,s)$ with a properly chosen integration
path in the Borel plane.  A discussion of the latter property of
the Borel summable solutions and some of its consequences is
postponed however to another paper \cite{3}.  

Having in mind the
incredible (in general) complexity of the $(\xi,s)$-Riemann surface structure 
of $\tilde{\Phi}_1 (\xi,s)$ we shall describe first a general procedure 
of getting this structure for first few $\tilde{\Phi}_1^{(q)} (\xi,s)$'s 
taking into account also a few singularities of 
$\tilde{\omega}(\xi)$ and $\Omega(\xi)$
and next we try to give as full as possible a description of such
structures for the linear and harmonic potentials.

\vskip 12pt
{\large\bf q = 0 }
\vskip 12pt

It is seen from (\ref{A1.12}) that $\tilde{\Phi}_1^{(0)} (\xi,s)$ is an entire
function of $s$ for any $\xi$ not coinciding with singularities of
$\tilde{\omega}(\xi)$. Its singularities in the $\xi$-variable coincide 
therefore with those of $\Omega(\xi)$ and consequently with those of 
$\tilde{\omega}(\xi)$ as the EP-
singularities shown in Fig. 3.

\vskip 12pt
{\large\bf q = 1 }
\vskip 12pt

 This is the Riemann surface structure of $\tilde{\Phi}_1^{(q)} (\xi,s)$ 
as defined by (\ref{A1.12}) for q=1.  

\be\label{4.1}
\tilde{\Phi}_1^{(1)} (\xi,s) = \int_{\tilde{C}(s)} d\eta
\tilde{\omega}(\xi - \eta)(2s - 2\eta)
\frac{I_1 \left( [8(s - \eta)\Omega(\xi - \eta) -4(s - \eta)\Omega(\xi)]^{\fr}
\right)}{[8(s - \eta)\Omega(\xi - \eta) -4(s - \eta)\Omega(\xi)]^{\fr}}
\ee

It follows from (\ref{4.1}) that
singularities of the subintegral function are essential
singularities coniciding with the branch points of $\Omega(\xi)$ and
$\Omega(\xi-\eta)$. Because of the single $\eta$-integration in (\ref{4.1}) 
only the EP-mechanism can generate singularities in the '$s$-plane' since
all the $\eta$-singularities are the moving ones (depending linearly
on $\xi$) so that the positions of all the (essential) singularities
of $\tilde{\Phi}_1^{(1)} (\xi,s)$ coincide again with those of
$\Omega(\xi-s)$ and $\Omega(\xi)$.
Therefore, these positions on the $\xi,s$-Riemann surface are the
following: 

\be\label{4.2}
\xi = \zeta_k, & \xi -s = \zeta_k, & k=1,2,... ,\;\; etc.
\ee

The nature of all these singularities is not altered
by the integrations i.e. all they are branch points. Therefore,
the resulting pattern of cuts on the corresponding Riemann
surface which follows from Fig. 3 is sketched in the figuers 4-5.

\vskip 12pt
{\large\bf q = 2}
\vskip 12pt
 
This is the Riemann surface structure of $\tilde{\Phi}_1^{(2)} (\xi,s)$
as defined by (\ref{A1.11}) for q = 2. From (\ref{A1.14}) we have: 

\be\label{4.3}
 \tilde{\Phi}_1^{(2)} (\xi,s) = \int\limits_{\tilde{C}(s)} d\eta
\int\limits_{\tilde{\gamma}(\xi)} d\xi_1
\tilde{\omega}(\xi_1 - \eta)\tilde{\omega}(\xi_1)(2s - 2\eta)^2
\frac{I_2 \large( z^{\fr}\large)}{z} \\
 z=
4(s - \eta)\left(\Omega(\xi)-2\Omega(\xi_1)+2\Omega(\xi_1 - \eta)\right)\nn
\ee

Note that
the $\xi$-integration in (\ref{4.3}) runs across a sheet of the $\xi$-Riemann
surface shown in Fig. 4 (where the $s$ variable is to be
substituted by the $\eta$ one). However, contrary to the close
correspondence between the distributions of sectors and turning
points on the Stokes graph of Fig. 1 and of sheets and the
corresponding cuts on Fig. 3 such a correspondence is lost in
the case of Fig. 4 i.e. we are left only with some properly
arranged system of branch points and cuts.

  Since the half of the cuts in Fig. 4 are moving then except of the
EP-singularities the P-singularities are also generated by both
the $\xi$- and $\eta$-integrations in (\ref{4.3}).  

\begin{tabular}{c}
\psfig{figure=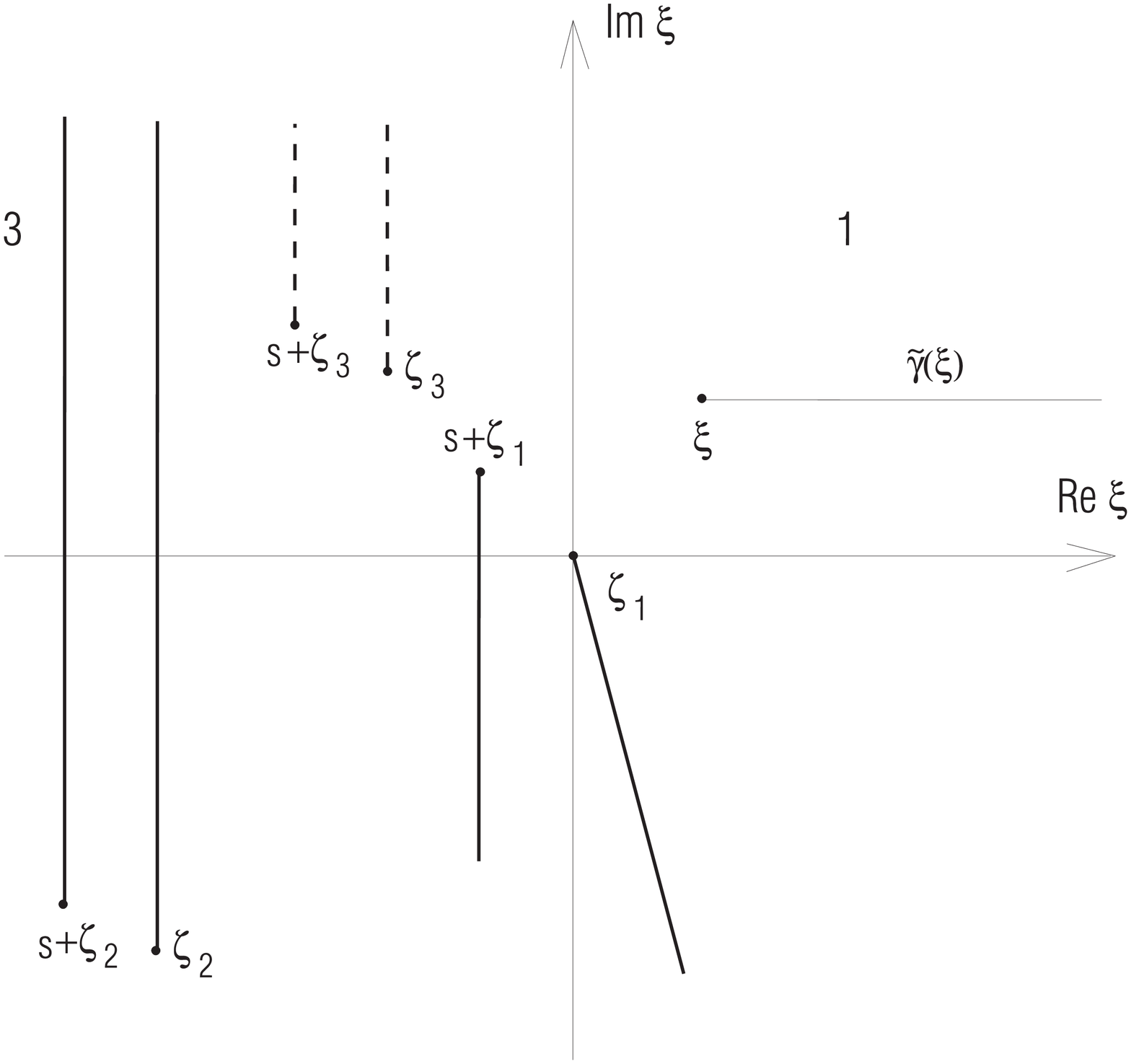,width=10cm} \\
\parbox{15cm}{Fig. 4. $\;\;\;$ The $\xi$-plane singularities corresponding to
$\tilde{\Phi}_1^{(1)}(\xi,s)$ (case $q=1$)}
\end{tabular}
\vskip 18pt

Consider first results of the $\xi$-integration in (\ref{4.3}).  

The EP-singularities which follow
from this integration coincide (with the corresponding
substitution $s$ by $\eta$) with those in the figuers 4 and 5 are given
again by (\ref{4.2}).  

\begin{tabular}{c}
\psfig{figure=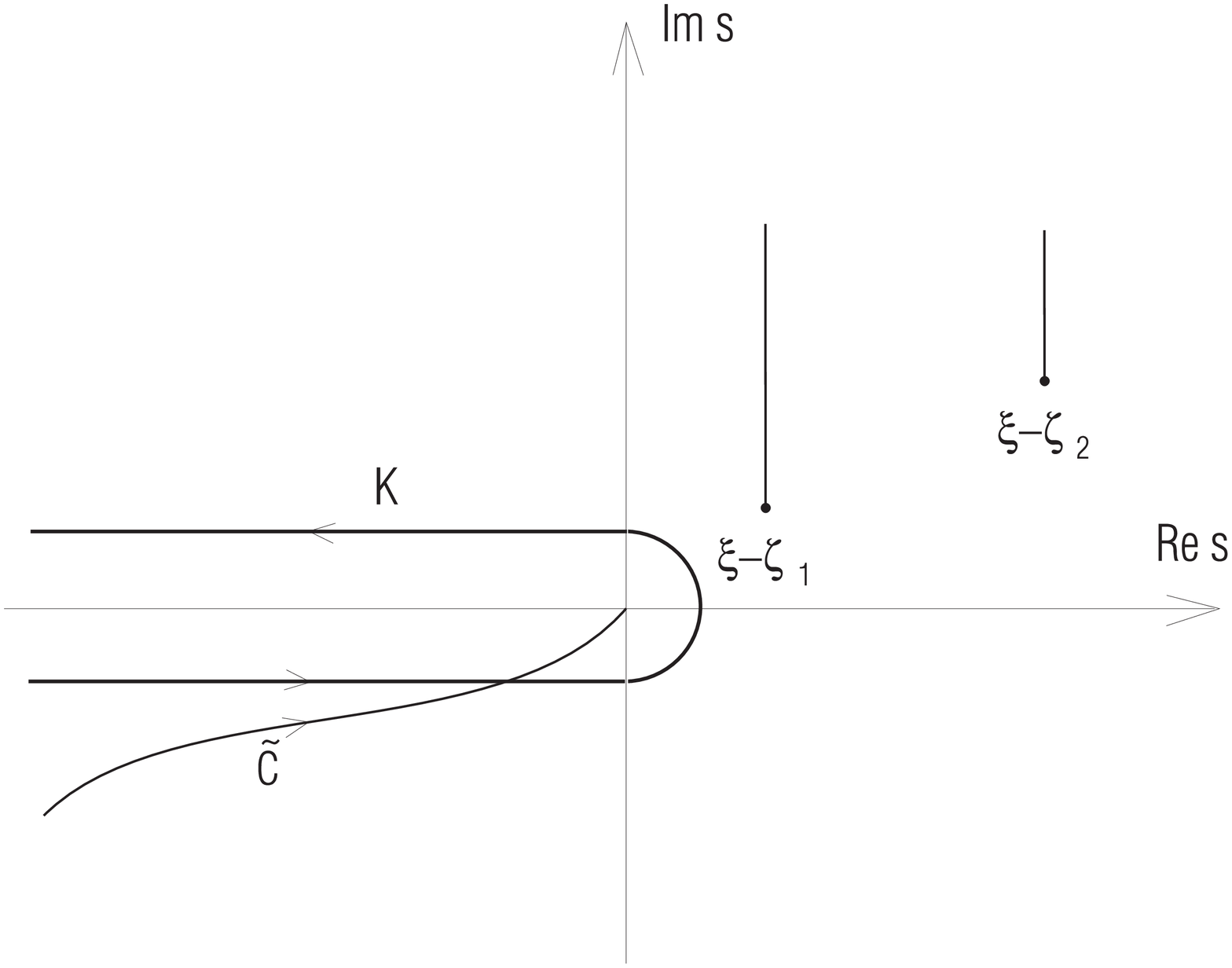,width=10cm} \\
\parbox{15cm}{Fig. 5. $\;\;\;$ The $s$-plane singularities corresponding to
$\tilde{\Phi}_1^{(0)}(\xi,s)$ (case $q=0$)}
\end{tabular}
\vskip 18pt

A generation of P-singularities can be
performed by moving singularities depending on $\eta$ (see Fig. 4).
For example, moving clockwise the singularity $\eta+\zeta_1$ around the
end point $\xi$ of $\tilde{\gamma}_1(\xi)$ and next pinching 
$\tilde{\gamma}_1(\xi)$ against $\zeta_1$ we
generate a singularity of \mref{4.3} at $\eta=0$ in the $\eta$-Riemann 
surface. It is
placed however on another sheet of the surface since to achieve
it we had to go around the branch point singularity $\xi - \zeta_1$, shown
in Fig. 5, in the clockwise direction. 

 To obtain all other $\eta$-plane singularities generated 
by the $\xi$-integration in (\ref{4.3}) we
proceed in the same way as described above. All these
singularities lie on sheats which can be reached by going around
the two branch points (in any direction - clockwise or
anticlockwise) shown in Fig. 5. Therefore all these
singularities are shared by the actual positions of the branch
points cuts of Fig. 5. They can become visible by cutting the
$\eta$-plane in a different way or moving appropriately both the
branch points to the left.  

\begin{tabular}{c}
\psfig{figure=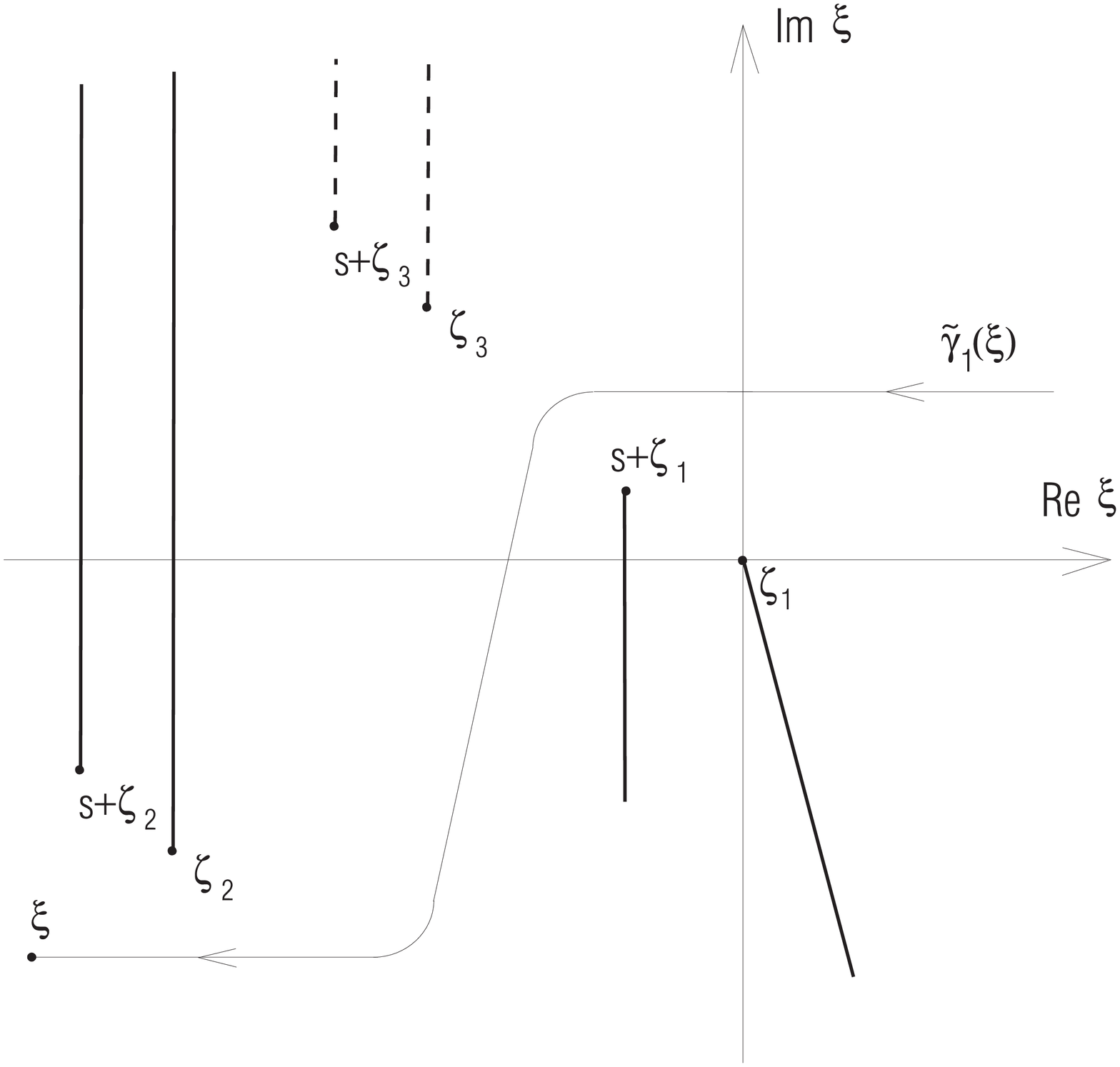,width=10cm} \\
\parbox{15cm}{Fig. 6. $\;\;\;$ The $\xi$-plane singularities corresponding to
subintegral function in \mref{4.3}}
\end{tabular}
\vskip 18pt

Choosing for example the last
possibility and moving $\xi$ toward Sector 3 we shall arrive at the
situation shown in Fig. 6. If $\xi$ and $\eta$ are moved so that
$\Re \xi < \Re \zeta_2 = \Re(\eta+\zeta_1) $
then a further motion of $\eta+\zeta_1$ upwards to the point $\zeta_2$ 
pinches the path $\tilde{\gamma}_1(\xi)$ producing in that way a
singularity at $\eta = \zeta_{21}\equiv \zeta_2 - \zeta_1$. 
It lies to the right from the cut at $\xi - \zeta_1$ in the 
'$\eta$-plane' and is screened therefore by the cut just
mentioned when $\Re \xi > \Re \zeta_2$ (see Fig. 7).

  By the identical analyses applied to each pair $\eta-\zeta_i$, $\zeta_j$ 
of the singularities lying on the sheet in Fig. 4 the singularities 
at $s=\zeta_{ij}$ or at $s=\zeta_{ji}=-\zeta_{ij}$ can be produced being 
screened by cuts at $s=\xi - \zeta_j$ or at $s=\xi - \zeta_i$,
correspondingly. All the singularities produced in this way are
branch points.  

According to (\ref{4.3}) the second, final integration
is performed over the $\eta$-plane providing $\tilde{\Phi}_1^{(2)} (\xi,s)$ 
with all its $\xi$- and $s$-plane singularities. This integration transforms
all the $\eta$- singularities obtained by the first ($\xi$-)integration
into the corresponding $s$-ones (by the EP- mechanism) and
provides us with additional $\xi$-singularities by the pinch mechanism. 
Pinching for example the singularity $\xi - \zeta_2$ against $\zeta_{21}$
we obtain the $\xi$-singularity at $\xi=\zeta_2+ \zeta_{21}$ lying on a sheet
of the $\xi$-Riemann surface originated by the branch point at $\zeta_2$ on 
Fig. 4. This branch point is screened, of course, by the cut at 
$\xi = s+ \zeta_2$ when $\Re s>\Re \zeta_{ij}$ (see Fig. 8). Therefore, 
figures 7-8 show the complete singularity structure of 
$\tilde{\Phi}_1^{(2)} (\xi,s)$ when 
continued in $\xi$ in the way shown in Fig. 6.

\begin{tabular}{c}
\psfig{figure=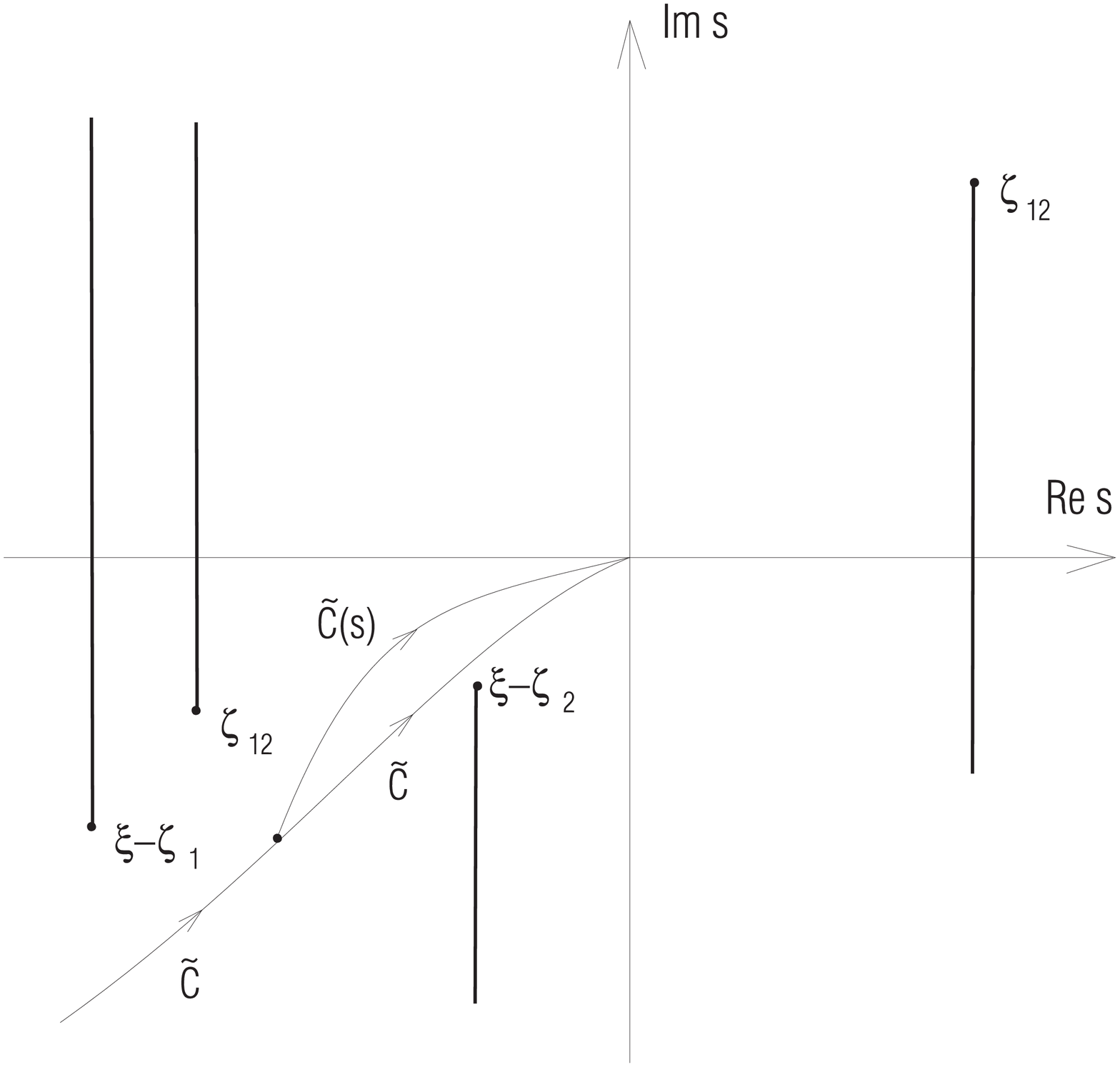,width=8cm} \\
\parbox{15cm}{Fig. 7. $\;\;\;$ The $s$-plane singularities corresponding to
$\tilde{\Phi}_1^{(2)}(\xi,s)$ }
\end{tabular}
\vskip 18pt

{\it 4.1. The analytic structure of the Borel function for the linear
potential}
\vskip 12pt

 We can put for this case $q(x,E)\equiv x$ and $\xi = x^{3/2}$ with the
corresponding Stokes graph shown in Fig. 9. and we shall
consider $\tilde{\Phi}_1 (\xi,s)$ as the Borel function defined by the
fundamental solution $\Psi_1 (x,\lambda)$.
 
\begin{tabular}{c}
\psfig{figure=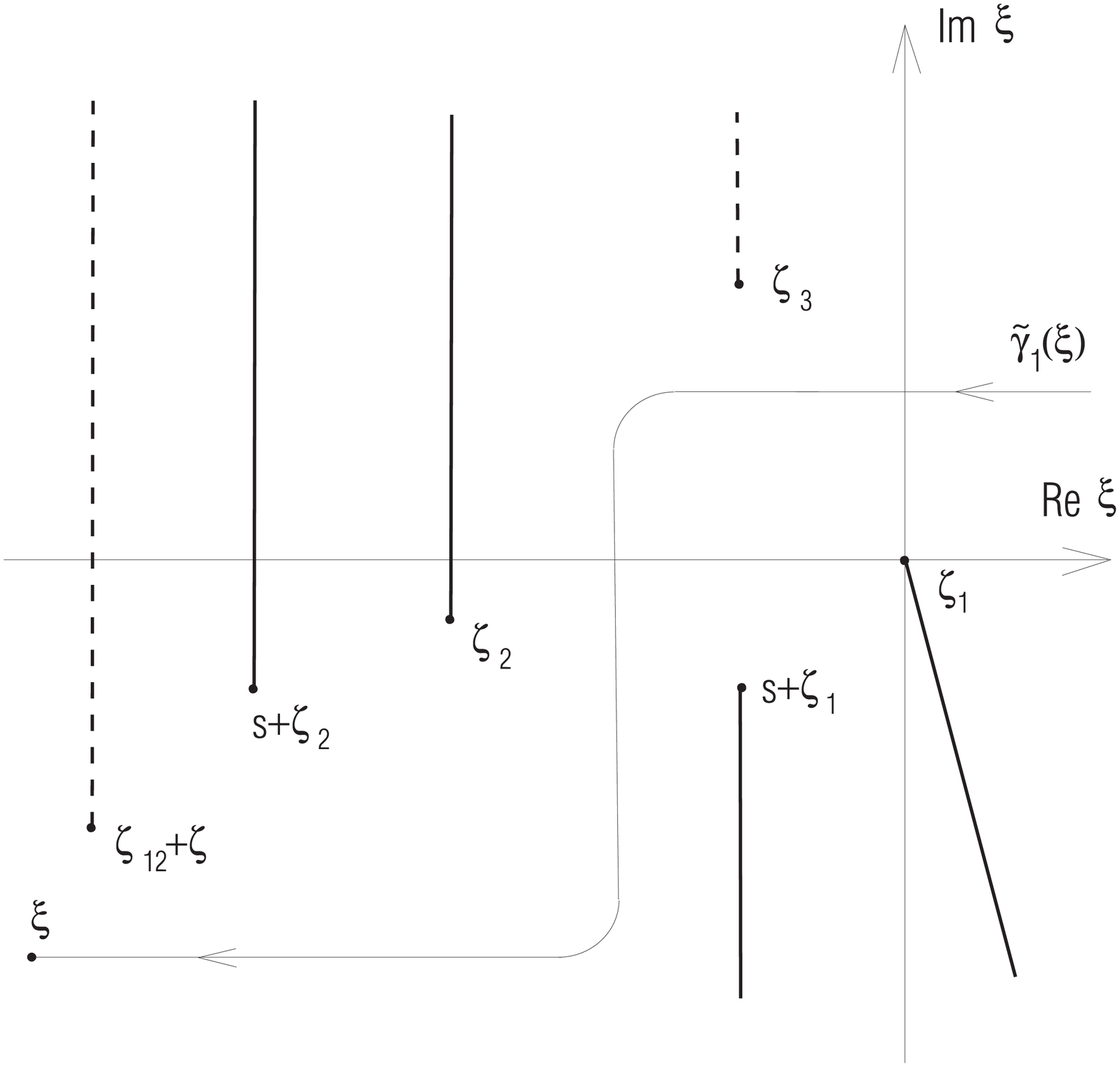,width=8cm} \\
\parbox{15cm}{Fig. 8. $\;\;\;$ The $\xi$-plane singularities corresponding to
$\tilde{\Phi}_1^{(0)}(\xi,s)$}
\end{tabular}
\vskip 18pt

At the first glance the corresponding analysis seems to be simple 
because of the simplicity of the relevant functions 
$\tilde{\omega}(\xi) = - \frac{5}{16}\frac{1}{\xi^2}$ and 
$\Omega(\xi)=\frac{5}{16}\frac{1}{\xi}$ as a result of which
the three sheeted Riemann surface branching at $\xi=0$ (the surface
being the image of the two sheeted $x$-plane by the transformation
$\xi=x^{3/2}$) decouples into three independent sheets. The unity of
the surface is recovered however by the solution $\Psi_1 (x,\lambda)$ which
being holomorphic at $x=0$ branches at this point as $\xi^{2/3}$ when
considered as a function of $\xi$.  
However, since we are interested in the properties of the Borel function
$\tilde{\Phi}_1 (\xi,s)$  determined
rather by $\chi_1(\xi,\lambda)/2\lambda$ it is the latter the 
($\xi,\lambda$)-dependence of which is most important.

\begin{tabular}{c}
\psfig{figure=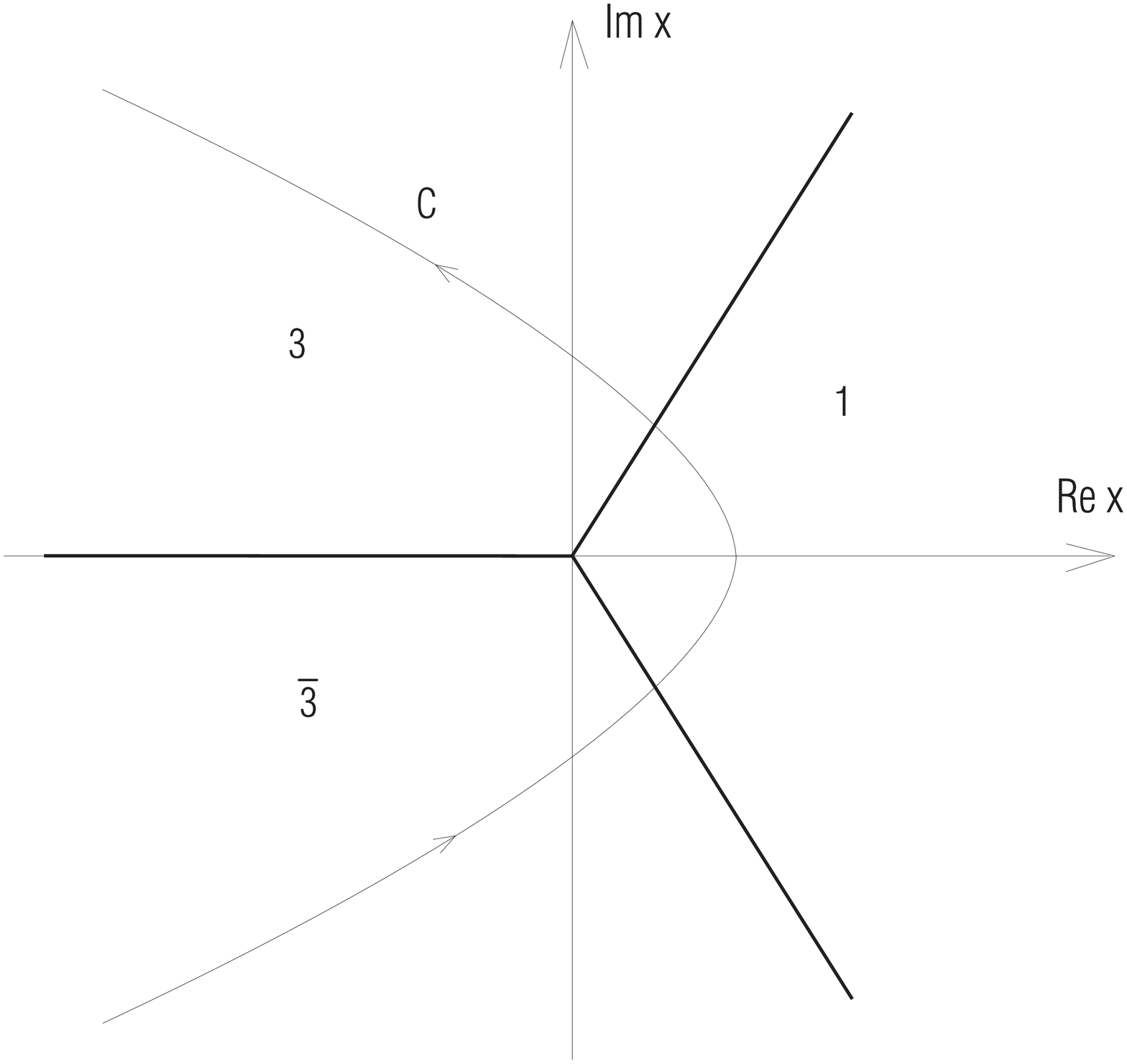,width=8cm} \\
\parbox{15cm}{Fig. 9. $\;\;\;$ The Stokes graph for the linear potential}
\end{tabular}
\vskip 18pt

The latter dependence can be established to some
extent noticing that continuing analitically the solution 
$\Psi(\xi,\lambda)=\xi^{-\frac{1}{6}}e^{-\lambda\xi} \chi_1(\xi,\lambda)$
 in the $\lambda$-plane (whilst $\xi$ is fixed) by rotating $\lambda$
 by the angle $\pm 6\pi$ we come back with the beginning of the
 integration path $\tilde{\gamma}(\xi)$ in  $\chi_1(\xi,\lambda)$ to the
infinity of the first sector. Of course, this path is by the
above continuation deformed from the initial canonical one into
the one surrounding the point $\xi=0$ twice (in the direction
suitable to the sign) to end eventually at the point $\xi$. This is
because continuing $\chi_1(\xi,\lambda)$ in $\lambda$ in 
the above way we have to shift the
infinite end of the path to the neighbour sectors each time when
$\lambda$  changes by $\pm\pi$ (this operation keeps the factor 
$e^{-\lambda \xi}$  always
vanishing in the infinities of the passed sectors) \cite{3}.

However, the above $\lambda$-continuation of $\chi_1(\xi,\lambda)$ is 
equivalent to its 
continuation to the same point $\xi$ along the $deformed$ path
$\tilde{\gamma}(\xi)$ starting from its initial canonical form. Since by this 
latter continuation the argument of $\xi$ changes also by 
$\pm 6\pi$ then the factor $\xi^{-\frac{1}{6}}$ of $\Psi(\xi,\lambda)$
aquires minus by this continuation so does the
factor $\chi_1(\xi,\lambda)$ since by this continuation 
$\Psi(\xi,\lambda)$ can  not change because it branches at 
$\xi=0$ as $\xi^{2/3}$. It follows therefore that $\chi_1(\xi,\lambda)$ 
branches at $\xi=0$ as $\xi^{1/6}$.  

From the latter observation it follows further
directly that the Borel function $\tilde{\Phi}_1 (\xi,s)
(\equiv \tilde{\chi}_1(\xi,s))$ 
 branches at the infinity point
of its $s$-plane also as $s^{1/6}$. This can be seen noticing that to
recover the factor $\chi_1(\xi,\lambda)$ by the Borel 
transformation of $\tilde{\Phi}_1 (\xi,s)$  we have to
change successively the integration path in the transformation
from the negative real halfaxis to the positive one (and vice
versa) according to which the sector the infinite end of the
deformed path $\tilde{\gamma}(\xi)$ is actually in.  These Borel 
transformation paths
are again the deformations of each other obtained by moving the
infinite end of them along the circle of infinite radius i.e.
all the singularities of $\tilde{\Phi}_1(\xi,s)$
 are avoided by these deformations.
Since after six such changes the Borel transformation of 
$\tilde{\Phi}_1(\xi,s)$ has to
change its sign in comparison with its initial value so 
$\tilde{\Phi}_1 (\xi,s)$ itself has to do it.  

Therefore, we conclude that for fixed $\xi$ the
$s$-Riemann surface of $\tilde{\Phi}_1 (\xi,s)$ is built of six sheets.  

The above
situation is however not so simple when the formulae
(\ref{A1.12})-(\ref{A1.14}) defining $\tilde{\Phi}_1 (\xi,s)$ are considered. The Bessel
functions in these formulae convert the simple pole of $\Omega(\xi)$ at
$\xi=0$ into a corresponding root (of the forth order) branch points
accompanied by essential singularities (see Appendix 3.1). Also
the successive $\xi$- and $\eta$-integrations in these formulae have to
generate unavoidably the branch points at $\xi=0$, $s=0$ and $\xi=s$ of
the logarithmitic type. This is of course because the
representation of $\tilde{\Phi}_1 (\xi,s)$ given by (\ref{A1.12})-(\ref{A1.14})
 is singular
providing us with the correct positions of singularities but not
necesserilly with their nature. The above example of the linear
oscillator shows that the proper behaviour of $\tilde{\Phi}_1 (\xi,s)$ close 
to its singularities is obtained only by the full resummation of these
series. Nevertheless, in more complicated cases of potentials an
information the series provide us are certainly very useful.
Also in the case just considered.  

Namely, taking into account
the reccurent relations (\ref{A1.14}) we can establish inductively
that $\tilde{\Phi}_1 (\xi,s)$ being defined on its six sheeted 
($\xi,s$)-Riemann
surface has on its first two sheets singularities shown in the
figures 10a,b (see Appendix 3.1 for details).  The point $s=0$ on
the sheet of Fig. 10b is regular for $\tilde{\Phi}_1 (\xi,s)$, according to
general results of App. 1. According to this analysis the points
$\xi-s=0$ are the four order branch points of $\tilde{\Phi}_1 (\xi,s)$ and
simultaneously its essential singularities but we should have in
mind that the last two properties can be incorrect.

  The same
property concerns the points $\xi=0$ and $s=0$ the latter being on the
second and further sheets of Fig. 10b. All these points arrange
themselves to build in the considered approximation of 
$\tilde{\Phi}_1 (\xi,s)$
infinitely sheeted Riemann surface. However, even for this
simple case the topology of the surface except its first two
sheets is too complicated to be fully described.

\begin{eqnarray*}
\begin{tabular}{cc}
\psfig{figure=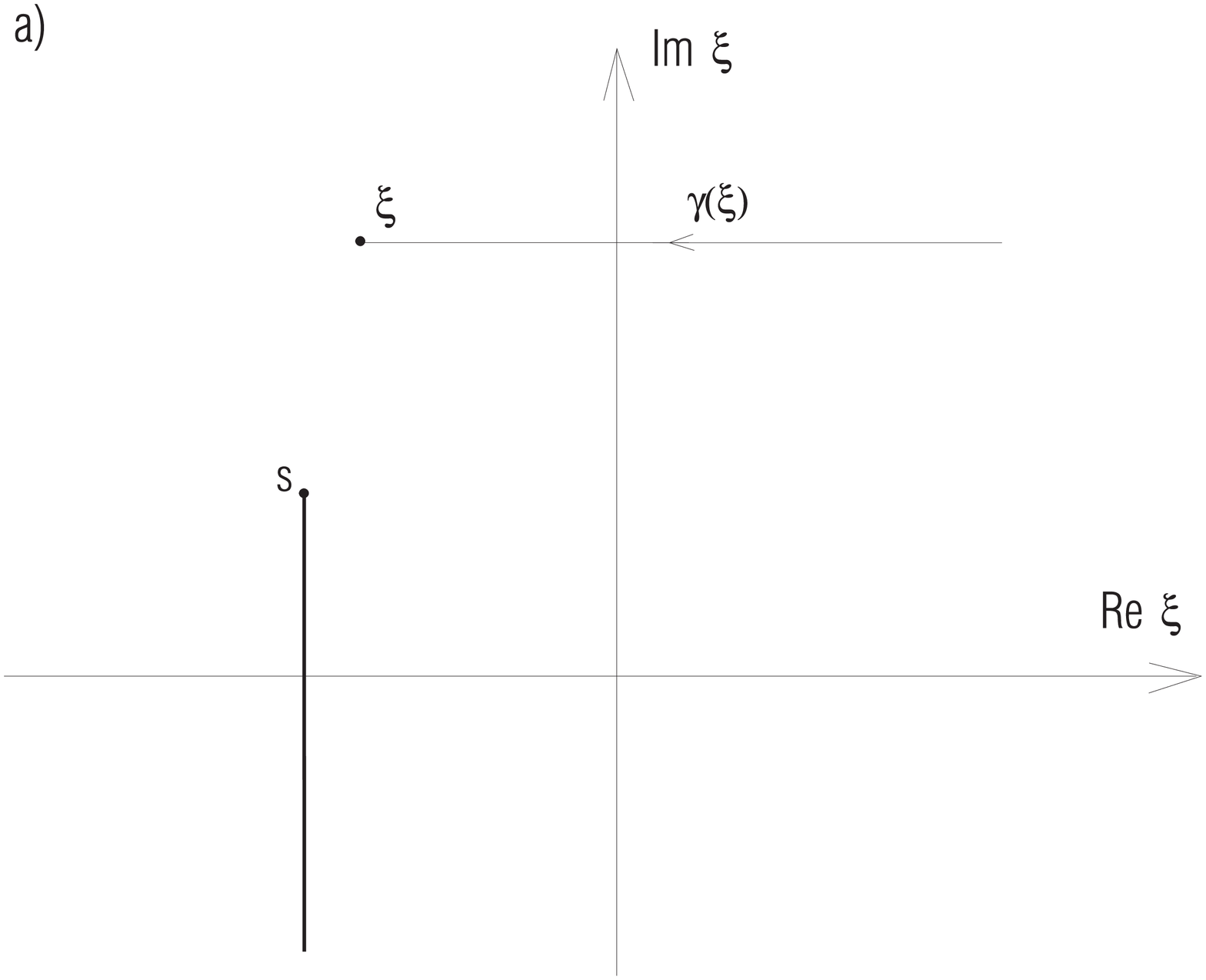,width=7cm} & \psfig{figure=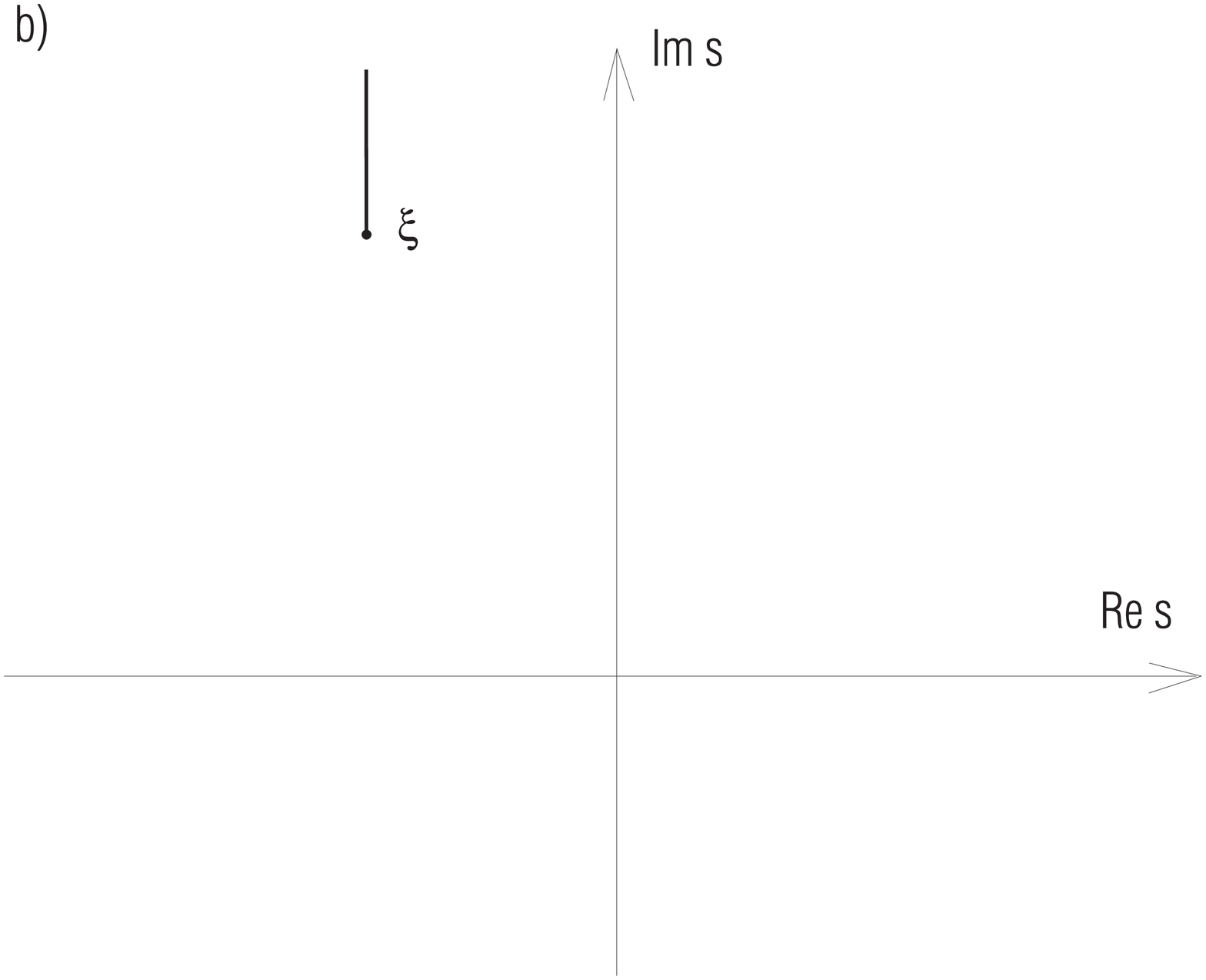,width=7cm} 
\end{tabular} \\
\parbox{14cm}{Fig. 10. $\;\;\;$ The $\xi$- and $s$-plane 
singularities corresponding to $\tilde{\Phi}_1^{(0)}(\xi,s)$}
\end{eqnarray*}
\vskip 18pt

  Nevertheless, one general
conclusion valid at least for all the polynomial potentials can
be drawn from the above consideration. Namely, if for a general
polynomial potential we consider any pair of neighbour sectors
joined by the analytic continuation in $\lambda$ when 
$\lambda \to e^{\pm i \pi}\lambda$ and we
continue a fundamental solution defined in one of the sectors to
the second along the canonical path then the corresponding
'$s$-plane' singularity structure of the first sheet of the
respective Borel function $\tilde{\Phi}_1 (\xi,s)$ is exactly the same as for
the 'simplest' case of the linear potential described above.

\vskip 12pt
{\it 4.2. An alternative non-standard Borel representation for the
linear potential wave function}
\vskip 12pt

 In the previous subsection we have made a disappointed note
that even in such a simple case as the linear potential one the
corresponding Borel function properties which follow from the
topological expansion are quite complicated. We have however
shown also that the actual structure of the linear potential Borel
plane should be rather simple. Below, we want to show that
indeed this complication is apparent and changing a little bit
the definition of the Borel function one can simplified the
latter enormously for the case considered. Namely, let us replace the
definition (\ref{2.7}) of the Borel function by the following one

\be\label{4.4}
\tilde{\chi}_1^{alt} (\xi,\sigma) =
\sum_{n\geq 0} \frac{(-\sigma)^{n+\fr}}{\Gamma(n+\frac{3}{2})}
\kappa_{1,n}(\xi)
\ee
which corresponds to the following representation of 
$\tilde{\chi}_1^{alt} (\xi,\sigma)$ by the Laplace transformation

\be\label{4.5}
 \tilde{\chi}_1^{alt} (\xi,\sigma) =
\frac{1}{\pi i} {\int\limits_{-i\infty+\lambda_0}^{+i\infty+\lambda_0} }
e^{-2\lambda \sigma}
\frac{\chi_1 (\xi,\lambda)}{(2\lambda)^{\frac{3}{2}}}d\lambda \\
 0<\lambda_0 <1 ,\;\;\;\;\;\;\;\;\;\;\;\;\;\; \sigma<0 \nn
\ee
 so that the invers Borel transformation is given by

\be\label{4.6}
\chi_1 (\xi,\lambda)=
(2\lambda)^{\frac{3}{2}} \int_{-\infty}^{0} e^{2\lambda \sigma}
\tilde{\chi}_1^{alt} (\xi,\sigma) d\sigma
\ee

Let us now make use of the fact that the fundamental solution
$\Psi_1 (x,\lambda)$ can be given the following integral representation (see
\cite{30}, Mathematical appendix)

\be\label{4.7}
\Psi_1 (x,\lambda)=
\frac{i}{\sqrt{\pi}}(2\lambda)^{\fr} \int_{C} e^{\lambda(xy-\frac{y^3}{3})} dy
\ee
 where we put $x$ real and positive and the contour $C$ is shown in Fig. 9.  

Changing in (\ref{4.7}) the
integration variable $y$ into $x^{-1/4}y$ and next putting 
$2\sigma = x^{3/4}y-x^{-3/4}y^3/3+2x^{2/3}/3$ we can bring the integral 
to the following form 

 \be\label{4.8}
 \Psi_1 (x,\lambda)=x^{-\frac{1}{4}}e^{-\frac{2}{3}\lambda x^{\frac{3}{2}}}
\sqrt{\frac{3}{\pi}} (2\lambda)^{\frac{3}{2}} \int\limits_{-\infty}^{0} 
e^{2\lambda \sigma} \;\;\;\;\;\;\;\;\;\;\; \\ 
 \left[ \left(-3(\sigma-\frac{1}{3}x^{\frac{3}{2}})
x^{\frac{3}{4}} + 
3x^{\frac{3}{4}}\sqrt{\sigma(\sigma-\frac{2}{3}x^{\frac{3}{2}})} 
\right)^{\frac{1}{3}} - \left(-3(\sigma-\frac{1}{3}x^{\frac{3}{2}})
x^{\frac{3}{4}} - 
3x^{\frac{3}{4}}\sqrt{\sigma(\sigma-\frac{2}{3}x^{\frac{3}{2}})} 
\right)^{\frac{1}{3}} \right] d\sigma \nn
\ee

Hence for $\tilde{\chi}_1^{alt} (\xi,\sigma)$ we get finally 

 \be\label{4.9}
\tilde{\chi}_1^{alt} (\xi,\sigma)=& \sqrt{\frac{3}{\pi}} 
\left[ 
\left(-3(\sigma-\fr \xi)(\frac{3}{2}\xi)^{\fr} + 
3(\frac{3}{2}\xi)^{\fr} \sqrt{\sigma(\sigma-\xi)} \right)^{\frac{1}{3}} 
\right. \\
& \left. - \left(-3(\sigma-\fr \xi)(\frac{3}{2}\xi)^{\fr} 
- 3(\frac{3}{2}\xi)^{\fr}\sqrt{\sigma(\sigma-\xi)} \right)^{\frac{1}{3}}
\right] \nn
\ee

It follows from (\ref{4.9}) that $\tilde{\chi}_1^{alt} (\xi,\sigma)$ is
defined on the $two$ sheeted Riemann surface having the branch
points $\sigma=0$ and $\sigma=\xi$ as its unique singularities.  

The non standard representation (\ref{4.5})-(\ref{4.6}) of the Borel function
considered above shows that the complicated form (\ref{2.7}) of the
standard one depends on the representation itself and it can be
simplified greatly by the proper choice of such a representation.

\vskip 12pt
{\it 4.3. The singularity structure of the Borel function for the
harmonic oscillator }
\vskip 12pt

Making, if necessary, a suitable rescaling
we can put in this case $q(x)=x^2 + 1$ (assuming the energy to be
negative). The corresponding Stokes graph is then shown in Fig.
11 and we choose as usually the sector $1$ to provide us with the
fundamental solution $\Psi_1 (x,\lambda)$ and its Borel function 
$\tilde{\Phi}_1 (\xi,s)$. Because of the last 
conclusion of Section \ref{4.1} we consider now a case of the Riemann surface 
structure corresponding to $\tilde{\Phi}_1 (\xi,s)$
when $\xi$($\equiv\int_{-i}^x \sqrt{y^2 +1} dy$) is continued to the sector 3 
of Fig. 11 (along a canonical path). Then the first sheets of 
$\tilde{\Phi}_1^{(1)} (\xi,s)$ and $\tilde{\Phi}_1^{(2)} (\xi,s)$
 are shown in the figures 12a,b and 13a,b respectively.  

Using again the formulae (\ref{A1.14}) we can show
inductively that the first sheets of $\tilde{\Phi}_1^{(2q)} (\xi,s)$ and
$\tilde{\Phi}_1^{(2q+1)} (\xi,s)$ look as in the figures 14 and 15. All the
detailed considerations establishing these can be found in
Appendix 3.2.

\vskip 12pt
{\it 4.4  Borel plane structure of harmonic oscillator Joos function}
\vskip 12pt

When in the consideration of the previous subsection we shall
push $\Re\xi$ to minus infinity (this corresponds to push $x$ to the
infinite point $\infty_3$ of the sector $3$ of Fig. 11) than we get the
'Borel plane' singularity structure of the so called Joos
function for the harmonic oscillator. The last name is given to
the coefficient $\chi_{1\to 3}(\lambda) \equiv \lim_{\xi\to \infty_3}
\chi_1(\xi,\lambda)$ \cite{19} so that the energy spectrum of the
harmonic oscillator is given by $\chi_{1\to 3}(\lambda)=0$. 
Note that in the limit $\xi\to \infty_3$
 all the functions $\tilde{\Phi}_1^{(2q+1)} (\xi,s)$ vanish so that the
corresponding limiting functions $\tilde{\Phi}_{1\to 3}^{(2q)}(s)$ 
contribute only to $\tilde{\chi}_{1\to 3}(s)$.

\begin{eqnarray*}
\psfig{figure=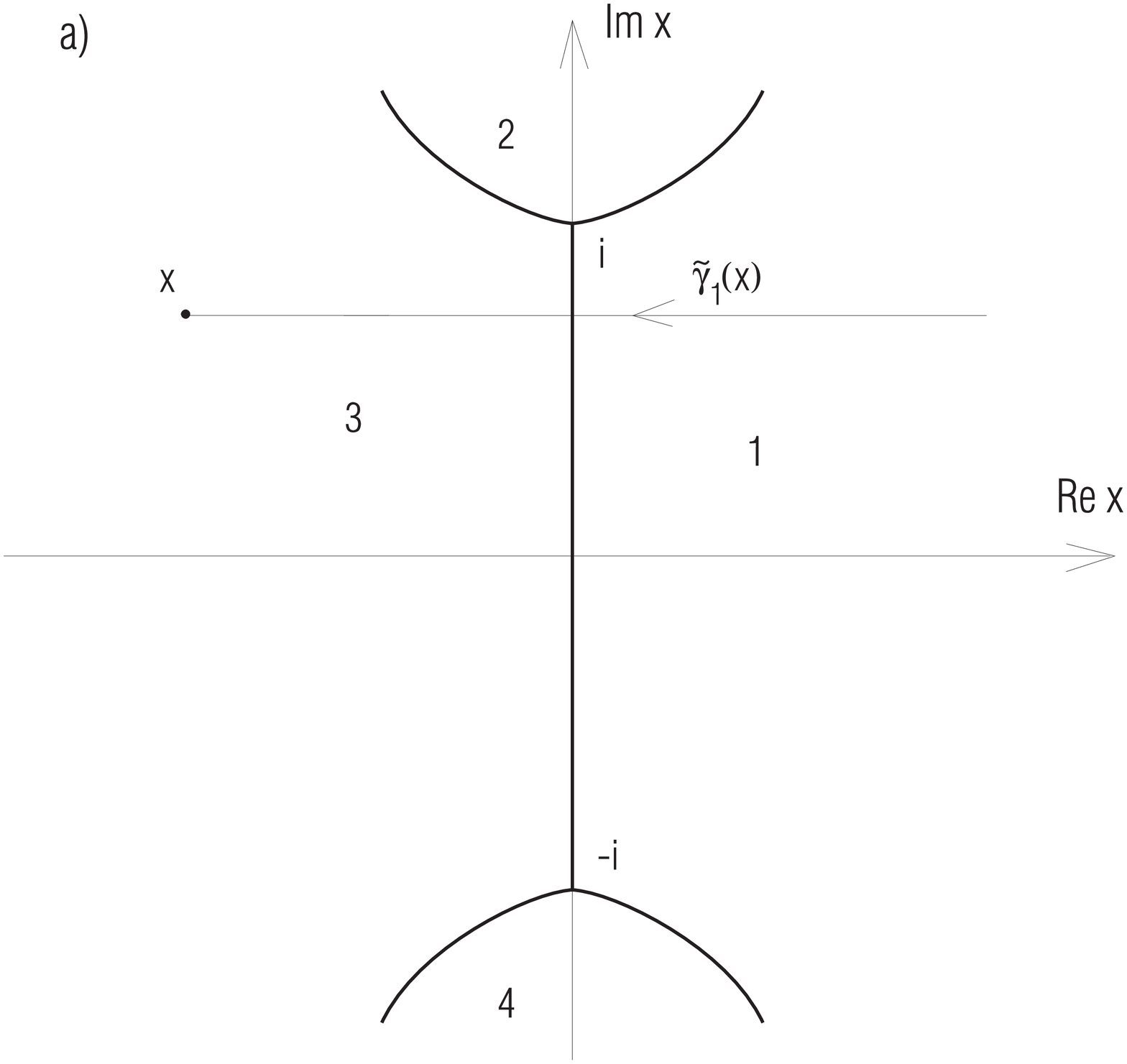,width=7cm} & \psfig{figure=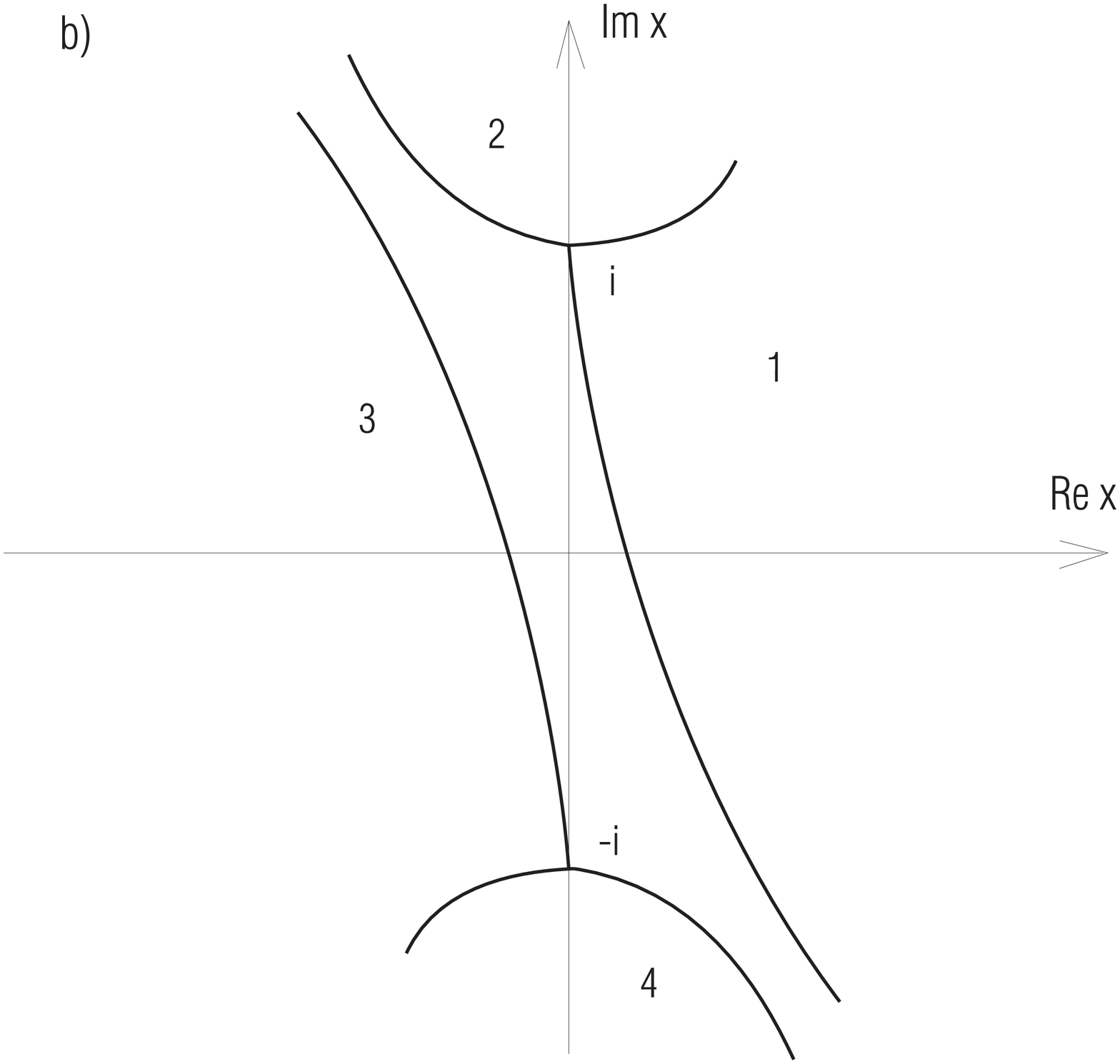,width=7cm} \\
\psfig{figure=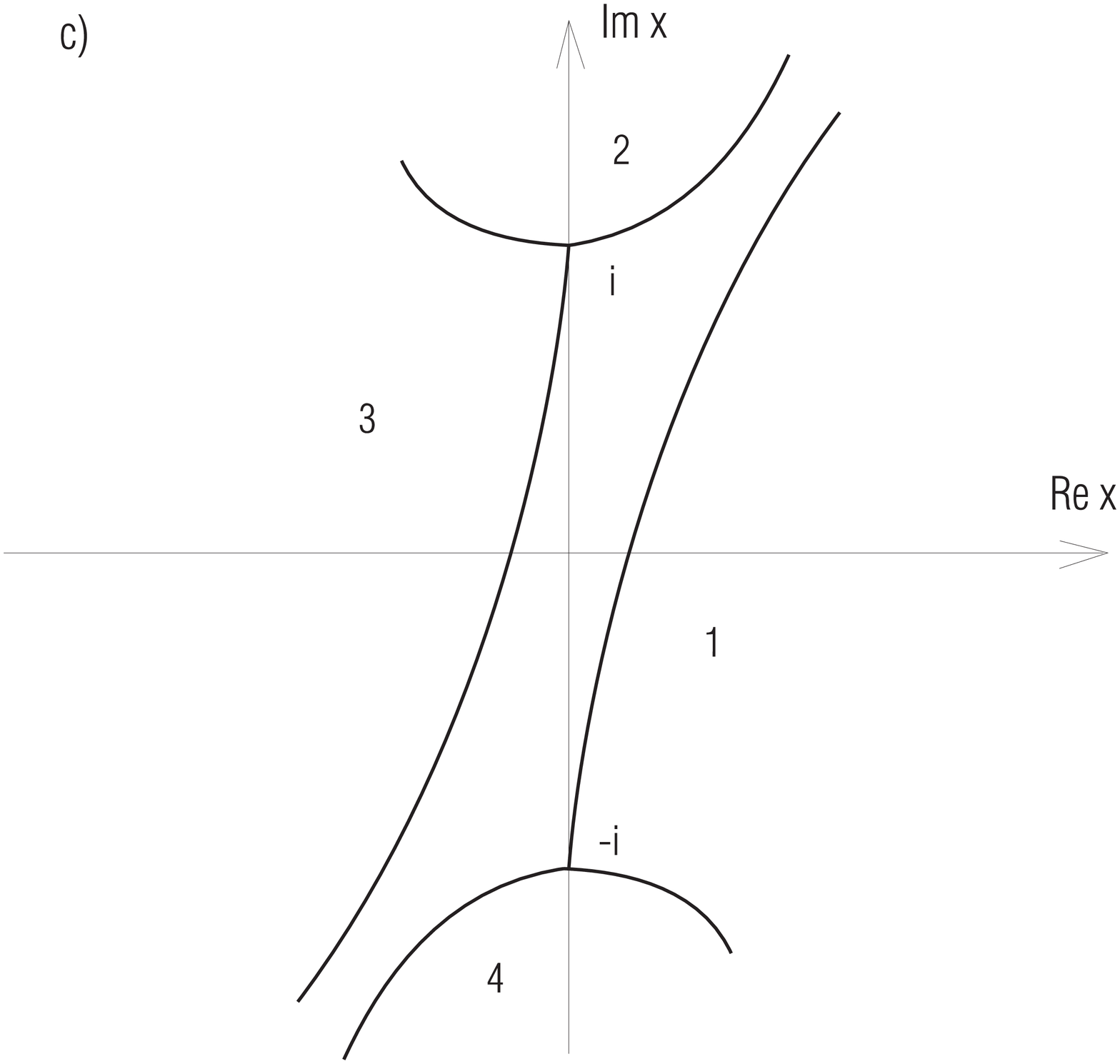,width=7cm} &
\parbox[b]{7cm}{Fig. 11. $\;\;\;$ The Stokes graphs corresponding to 
the harmonic oscillator}
\end{eqnarray*}
\vskip 18pt

As it follows from the considerations of the previous subsection
the singularity structure of the latter function is determind by
the branch points distributed along the imaginary axes of the
$s$-Riemann surface. This distribution can be discribed completely if instead 
of the Borel function $\tilde{\chi}_{1\to 3}(s)$ we shal consider the one
corresponding to $\log\chi_{1\to 3}(\lambda)$. To this end let us note 
that as it follows from Fig. 11a the normal sector of 
$\chi_{1\to 3}(\lambda)$ (i.e. the one
where $\chi_{1\to 3}(\lambda)$ is holomorphic and can be expanded into the
semiclassical series (\ref{2.5})) is defined by $|\arg \lambda|< \pi$.  
One can easily find also (by analytic continuation in $\lambda$) that 

\be\label{4.10}
\chi_{2\to 4}^{\sigma}(\lambda)= \chi_{1\to 3}(\lambda e^{i\sigma \pi}),
& 0<|\arg \lambda|< \pi
\ee
where $\sigma =
\arg \lambda/|\arg \lambda|$ and $\chi_{1\to 3}^{\mp}(\lambda)$ are the 
canonical coefficients corresponding to the graphs of Fig. 11b and 11c 
respectively. Despite (\ref{4.10}) these two canonical coefficients obey the
following two other relations 

\be\label{4.11}
 \chi_{1\to 3}(\lambda)\chi_{2\to 4}^{-}(\lambda)= 1+ e^{\pi i\lambda} , 
\;\;\;\;\;\;\;\;\;\;\;  0<\arg \lambda< \pi  \;\;\;\;\;\;\;\;\;\;\; \\
 \chi_{1\to 3}(\lambda)\chi_{2\to 4}^{+}(\lambda)= 1+ e^{-\pi i\lambda} ,
\;\;\;\;\;\;\;\;\;\;\; -\pi<\arg \lambda< 0 \;\;\;\;\;\;\;\;\;\;\; \nn
\ee
in which the fact that $2 \int^i_{-i}\sqrt{y^2+1} dy= \pi i$ has been used.

The relations (\ref{4.11}) follows as a
result of an identity which the four fundamental solutions $\Psi_k,\;\; k
= 1, ..., 4$, corresponding to the Stokes graphs of Fig. 11 have
to satisfy since only two of them can be linearly independent.

Using (\ref{4.10}) we get from (\ref{4.11}) 
\be\label{4.12}
 \chi_{1\to 3}(\lambda)\chi_{1\to 3}(\lambda e^{-i\sigma\pi})= 
1+ e^{\pi i \sigma\lambda} 
\ee
where $\sigma=\arg \lambda/|\arg \lambda|$ and $0 <|\arg \lambda|< \pi$.  

The formula (\ref{2.6}) can now be used directly to
define the Laplace transform $\tilde{\chi}_{1\to 3}(s)$  of the Joos 
function $\chi_{1\to 3}(\lambda)$
with the integration contour $C_{13}$ running around the negative half of 
the real axis of the $\lambda$-plane. In this way $\tilde{\chi}_{1\to 3}(s)$ 
is defined by (\ref{2.6}) as the holomorphic function in the half-plane
$\Re s < 0$.  

\begin{eqnarray*}
\begin{tabular}{cc}
\psfig{figure=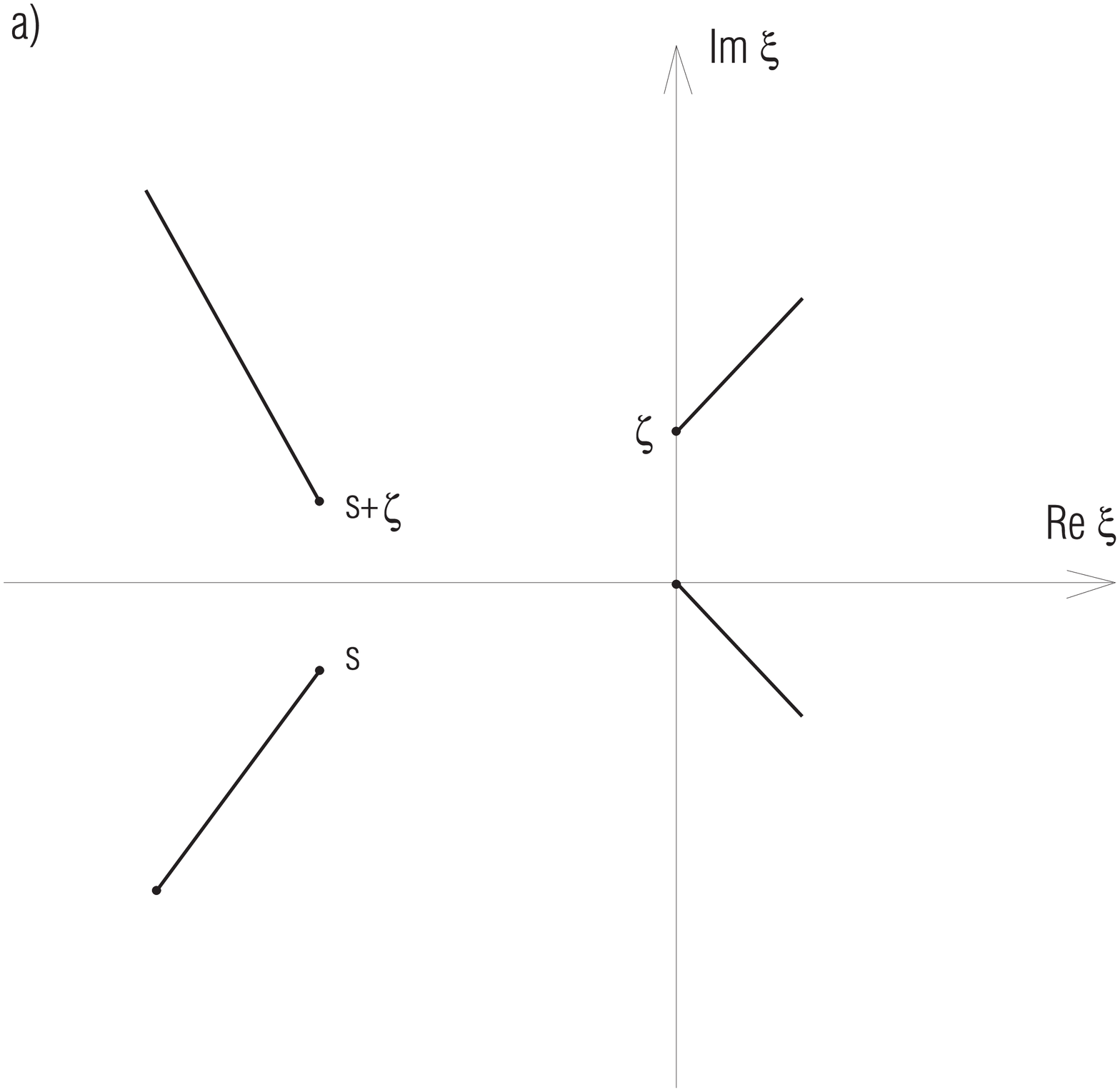,width=7cm} & \psfig{figure=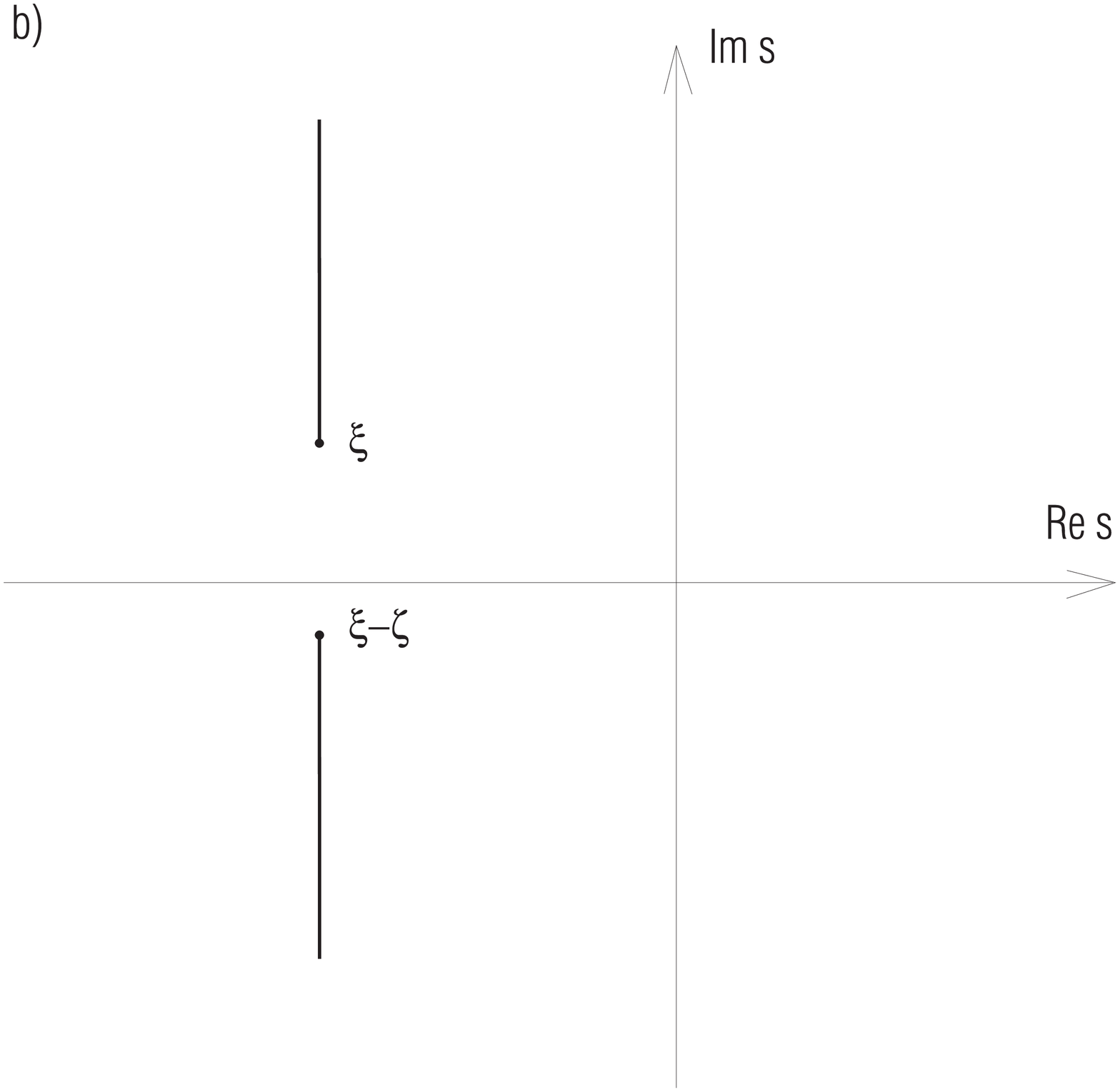,width=7cm} 
\end{tabular} \\
\parbox{14cm}{Fig. 12. $\;\;\;$ The  "first sheets" singularities 
of $\tilde{\Phi}_1^{(1)}(\xi,s)$ for the harmonic potential}
\end{eqnarray*}
\vskip 18pt

The analytic structure of $\tilde{\chi}_{1\to 3}(s)$ can, however, be best 
handled if we consider $\log^{*}\tilde{\chi}_{1\to 3}(s)$ rather than the 
function itself \cite{19}. Namely, we have: 
\be\label{4.13}
\log^{*}\tilde{\chi}_{1\to 3}(s) = \frac{1}{2\pi i}
\int_{C_{13}}e^{-2\lambda s}
\log \chi_{1\to 3}(\lambda) d\lambda
\ee
and we can use (\ref{4.12}) to calculate (\ref{4.13}) exactly. (Note, that 
$\chi_{1\to 3}(\lambda)$ does not vanish in the $\lambda$-plane cut
along the negative half of the real axis). Using (\ref{4.12}) we have:

\be\label{4.14}
 \log^{*}\tilde{\chi}_{1\to 3}(s) =
 \frac{1}{2\pi i} {\int\limits_{C^u(\lambda_0)}}e^{-2\lambda s}
\log (1 + e^{\pi i\lambda})d\lambda  \\
  + \frac{1}{2\pi i} {\int\limits_{C^d(\lambda_0)}}e^{-2\lambda s}
\log (1 + e^{-\pi i\lambda})d\lambda 
+ \frac{1}{2\pi i} {\int\limits_{C_{13}(\lambda_0)}}e^{-2\lambda s}
\log \chi_{1 \to 3}(-\lambda)d\lambda \nn
\ee
where $C_{13}(\lambda)$ is one of the contours $C_{13}$ crossing the real axis
at $\lambda_0 > 0$ and $C^u(\lambda_0)$, $C^d(\lambda_0)$ are parts of it 
lying above and below of the real axis correspondingly.

\begin{eqnarray*}
\begin{tabular}{cc}
\psfig{figure=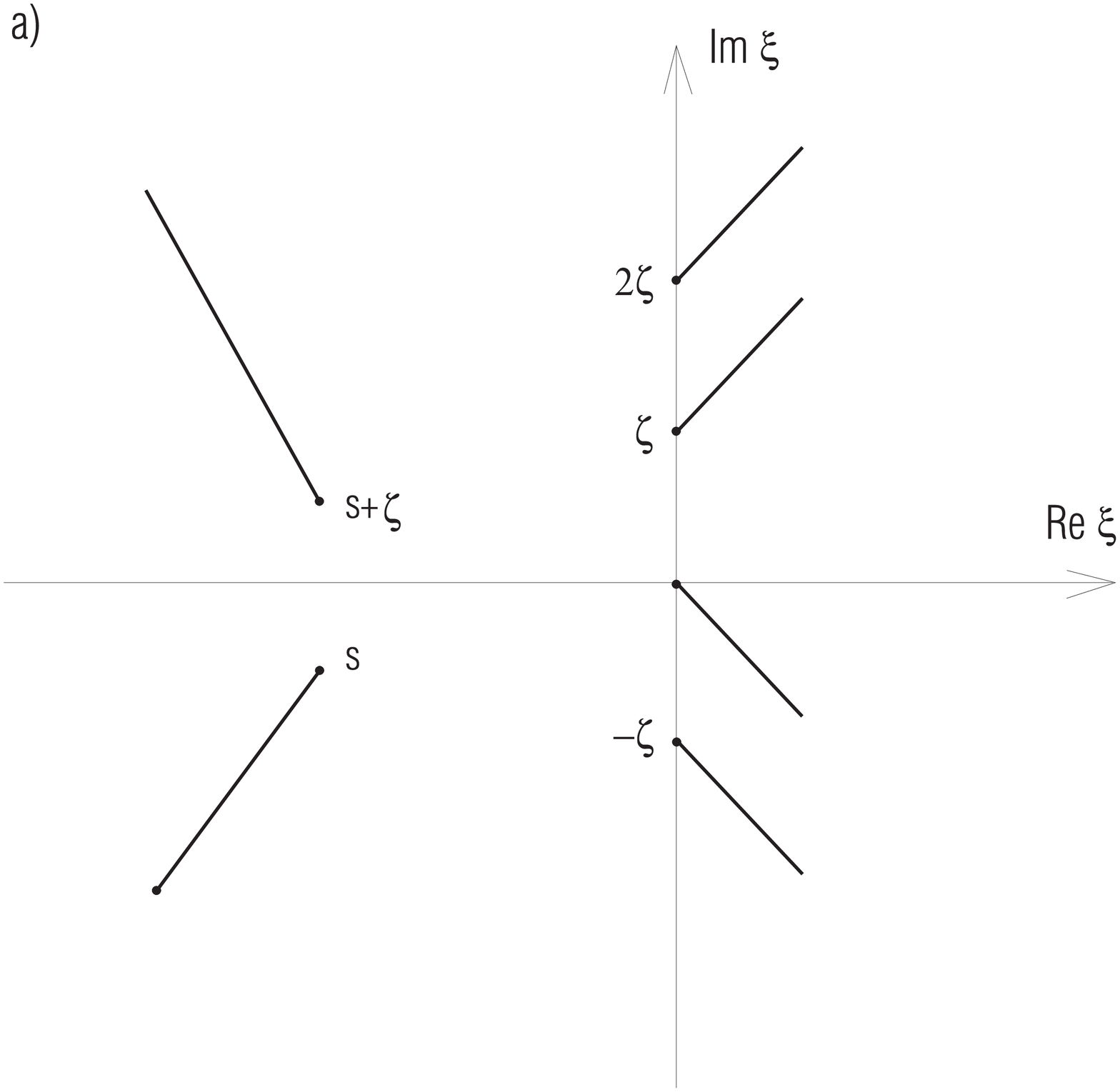,width=7cm} & \psfig{figure=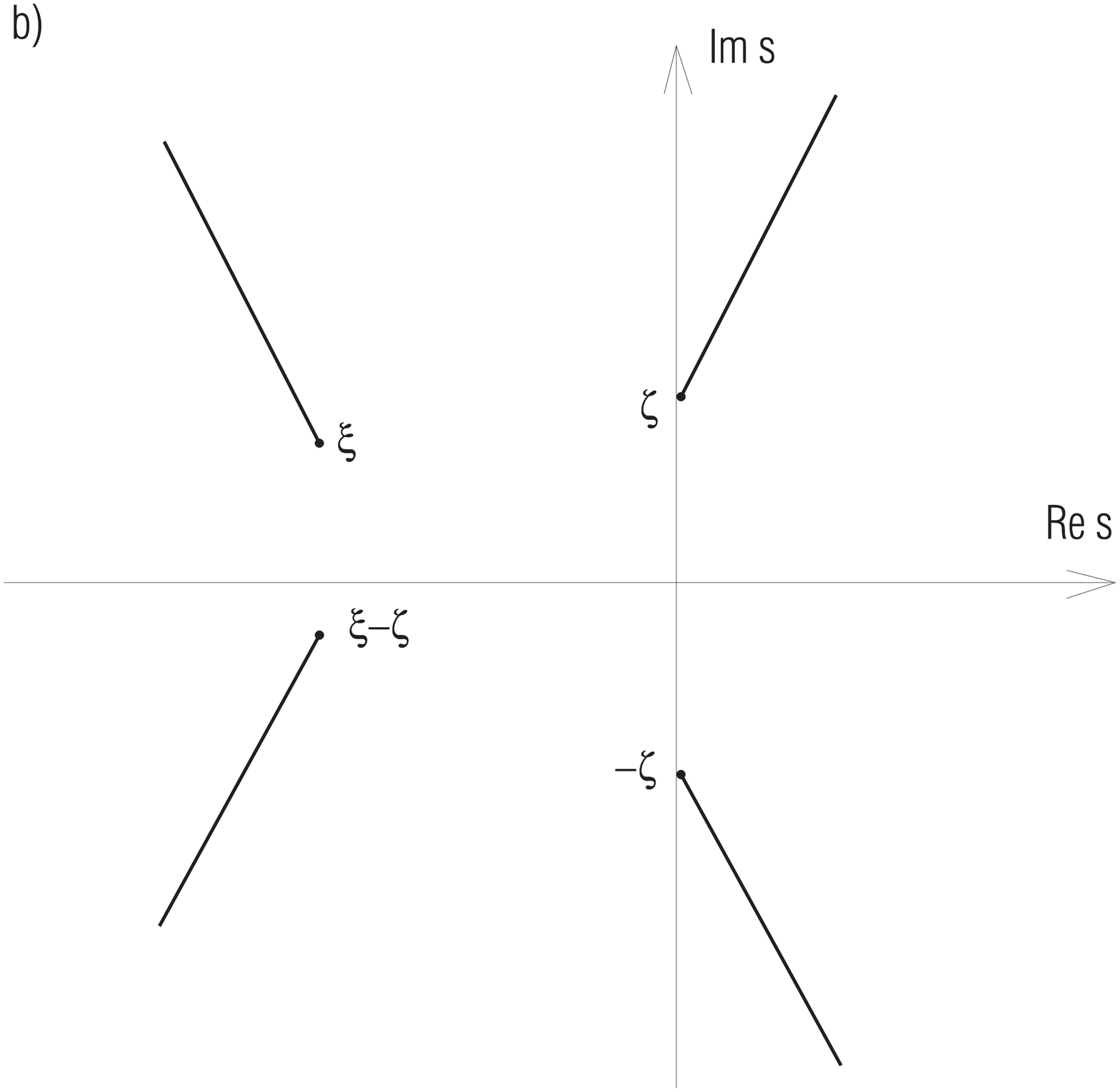,width=7cm} 
\end{tabular} \\
\parbox{14cm}{Fig. 13. $\;\;\;$ The  "first sheets" singularities 
of $\tilde{\Phi}_1^{(2)}(\xi,s)$ for the harmonic potential}
\end{eqnarray*}
\vskip 18pt

Performing the integrations in the first two integrals in (\ref{4.14}) 
(by expanding the logarithms) and changing $\lambda$ into -$\lambda$ in 
the third one we get:

\be\label{4.15}
 \log^{*}\tilde{\chi}_{1\to 3}(s) =
 \frac{1}{4\pi i} \sum_{n\leq1} \frac{(-1)^{n+1}}{n}
\left\{ \frac{e^{\pi i n \lambda_0}}{s - \frac{i\pi n}{2}} -
\frac{e^{-2\pi i n \lambda_0}}{s +  \frac{i\pi n}{2}} \right\} 
e^{-2 \lambda_0 s} \\
 \;\;\;\;\;\;\;\;\;\;\;\;\;\;\;\;\;\; +  
\frac{1}{2\pi i} \int\limits_{C'(\lambda_0)}e^{2\lambda s}
\log \chi_{1 \to 3}(\lambda)d\lambda \nn
\ee
where $C'(\lambda_0)$ is the contour encircling (anticlockwise) the point
$\lambda= 0$ and starting and ending at the point $\lambda = -\lambda_0$ 
of the real axis. Since the right hand side of (\ref{4.15}) is independent of 
$\lambda_0$ it can be calculated at $\lambda_0 \to 0$. It can be shown 
(see Appendix 5)
that the integral in (\ref{4.15}) vanishes in this limit and therefore
we finally get: 

\be\label{4.16}
 \log^{*}\tilde{\chi}_{1\to 3}(s) =
 \frac{1}{2} \sum_{n\leq1} \frac{(-1)^{n+1}}{s^2 +  \frac{n^2 \pi^2}{4}} =
\frac{1}{2is} \left( \frac{1}{2is} - \frac{1}{\sin{(2is)}}
\right)
\ee

The result (\ref{4.16}) was established essentially by Voros \cite{19} 
but here it is obtained directly by the definition (\ref{4.13}) of the 
Laplace transform for $\log\chi_{1\to 3}(\lambda)$. The inverse
transformation can be also performed to give the known expression for 
$\chi_{1\to 3}(\lambda)$ \cite{19}.  

\begin{eqnarray*}
\begin{tabular}{cc}
\psfig{figure=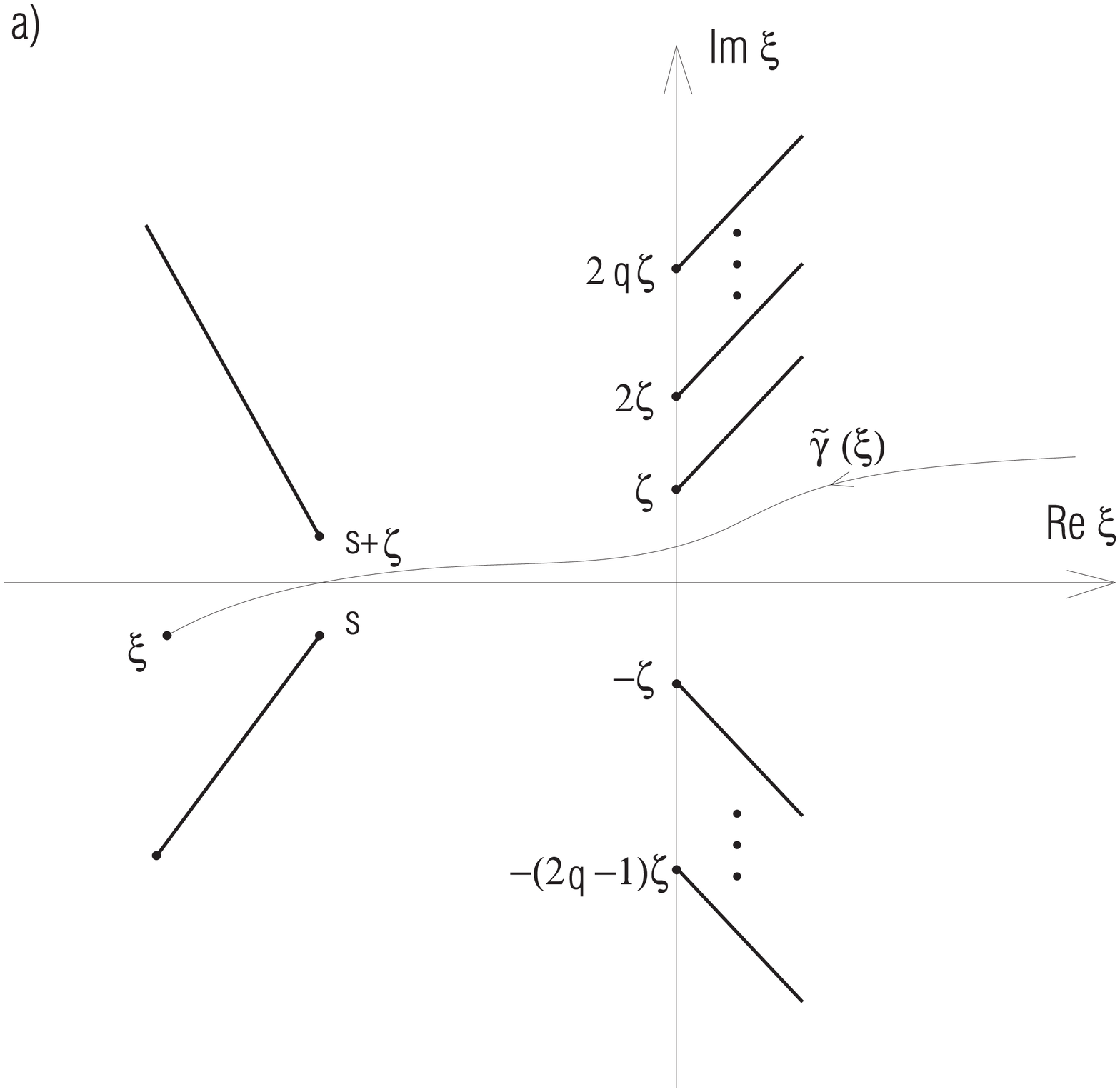,width=7cm} & \psfig{figure=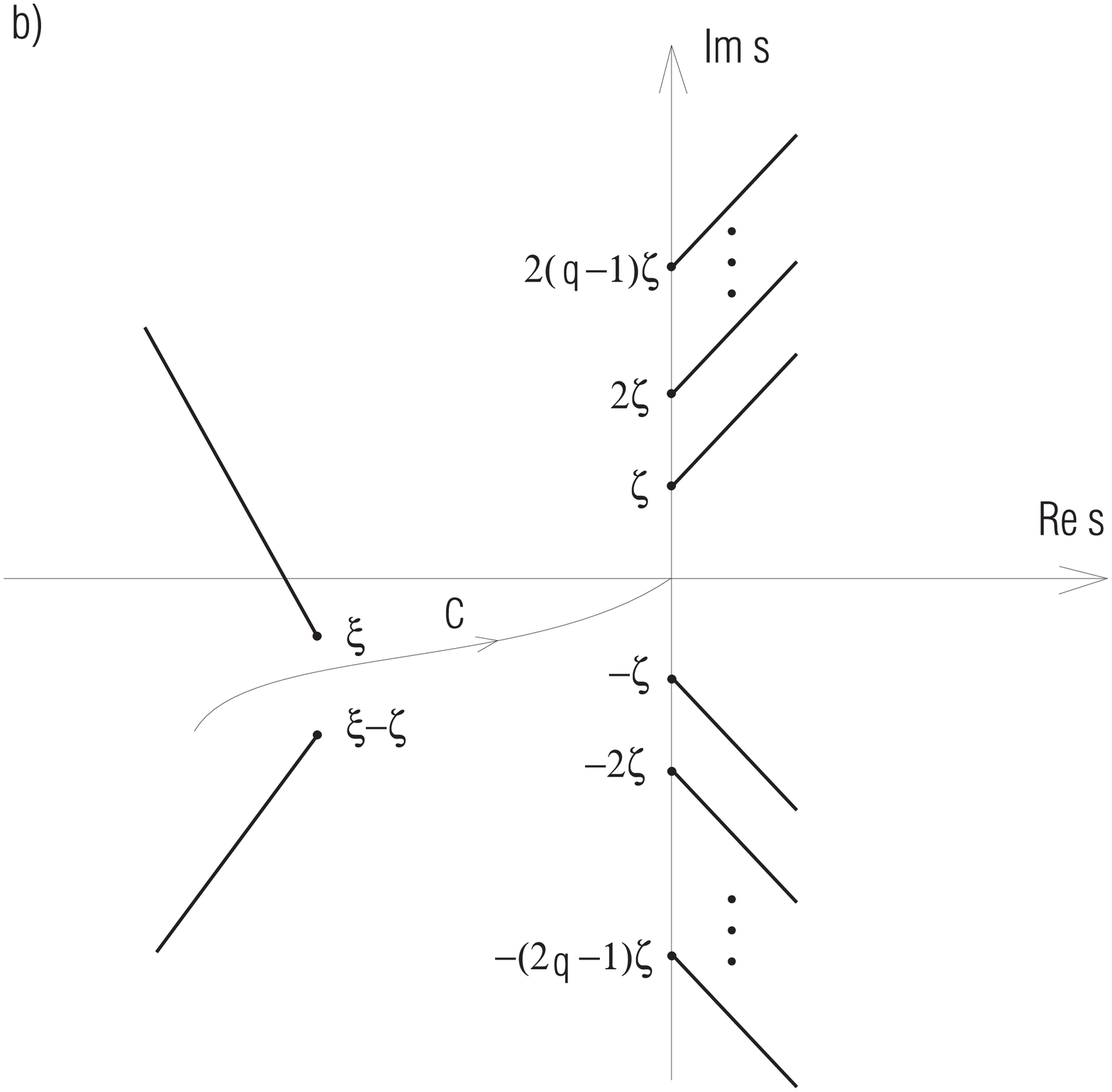,width=7cm} 
\end{tabular} \\
\parbox{14cm}{Fig. 14. $\;\;\;$ The  "first sheets" singularities 
of $\tilde{\Phi}_1^{(2q)}(\xi,s)$ for the harmonic potential}
\end{eqnarray*}
\vskip 18pt

Summarizing the above analyses one
can see that despite the clear way of obtaining corresponding
singularity patterns and the underlying structures of the
Riemann surfaces both they become still more and more
complicated with increasing $q$. The following main observations
follow, however, from this analyses: 

\begin{enumerate}
\item The set $S_{q+1}$ of singularities corresponding to 
$\tilde{\Phi}_1^{(q+1)} (\xi,s)$ contains the set
$S_{q}$ of these corresponding to $\tilde{\Phi}_1^{(q)} (\xi,s)$.  
\item The new singularities which belong to $S_{q+1}\setminus S_q$ 
are generated on the
sheets originated by the singularities of $S_q$; the latter is true
both on the $\xi$- and on the $s$-planes. 
\end{enumerate}

\begin{eqnarray*}
\begin{tabular}{cc}
\psfig{figure=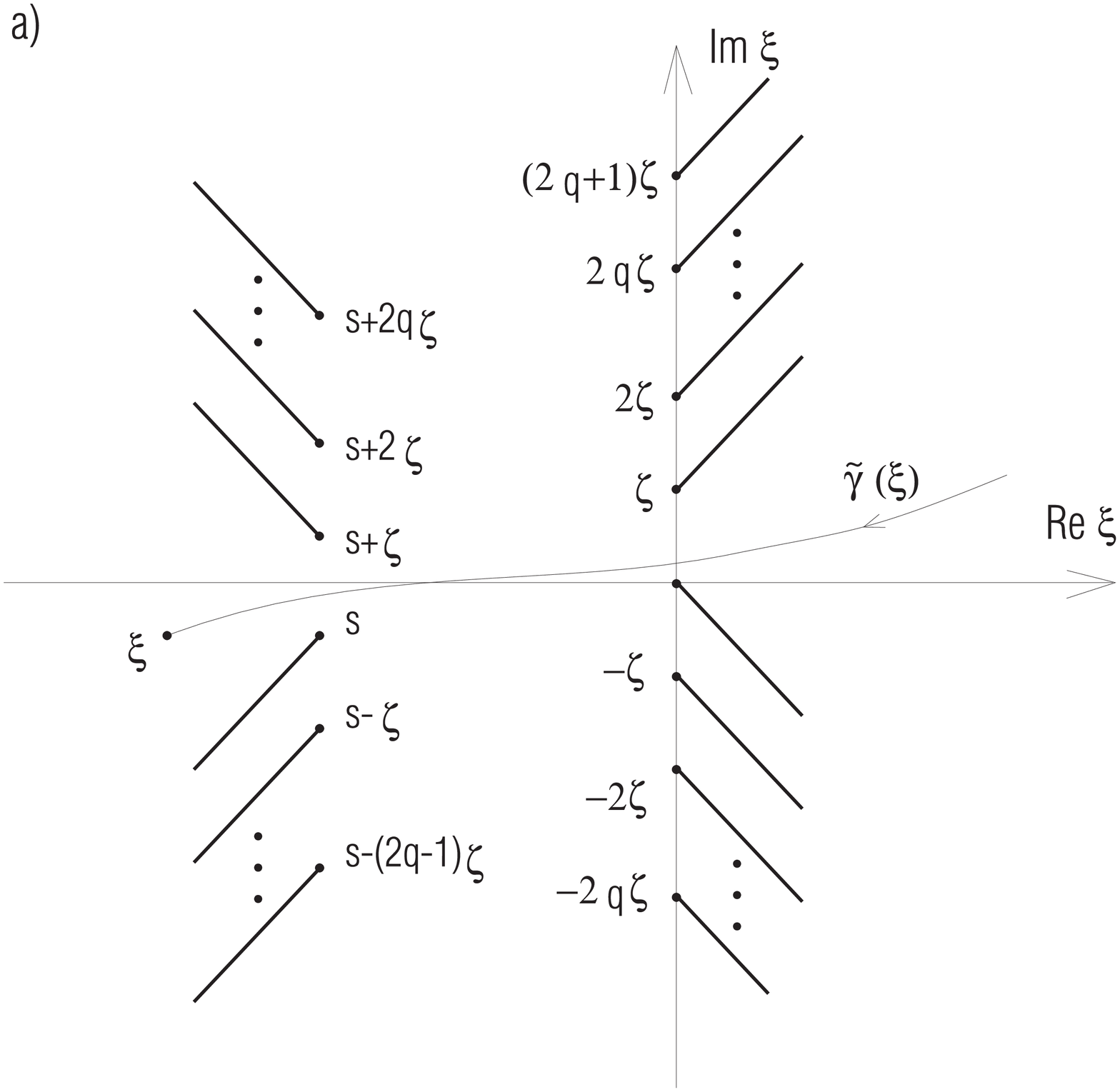,width=7cm} & \psfig{figure=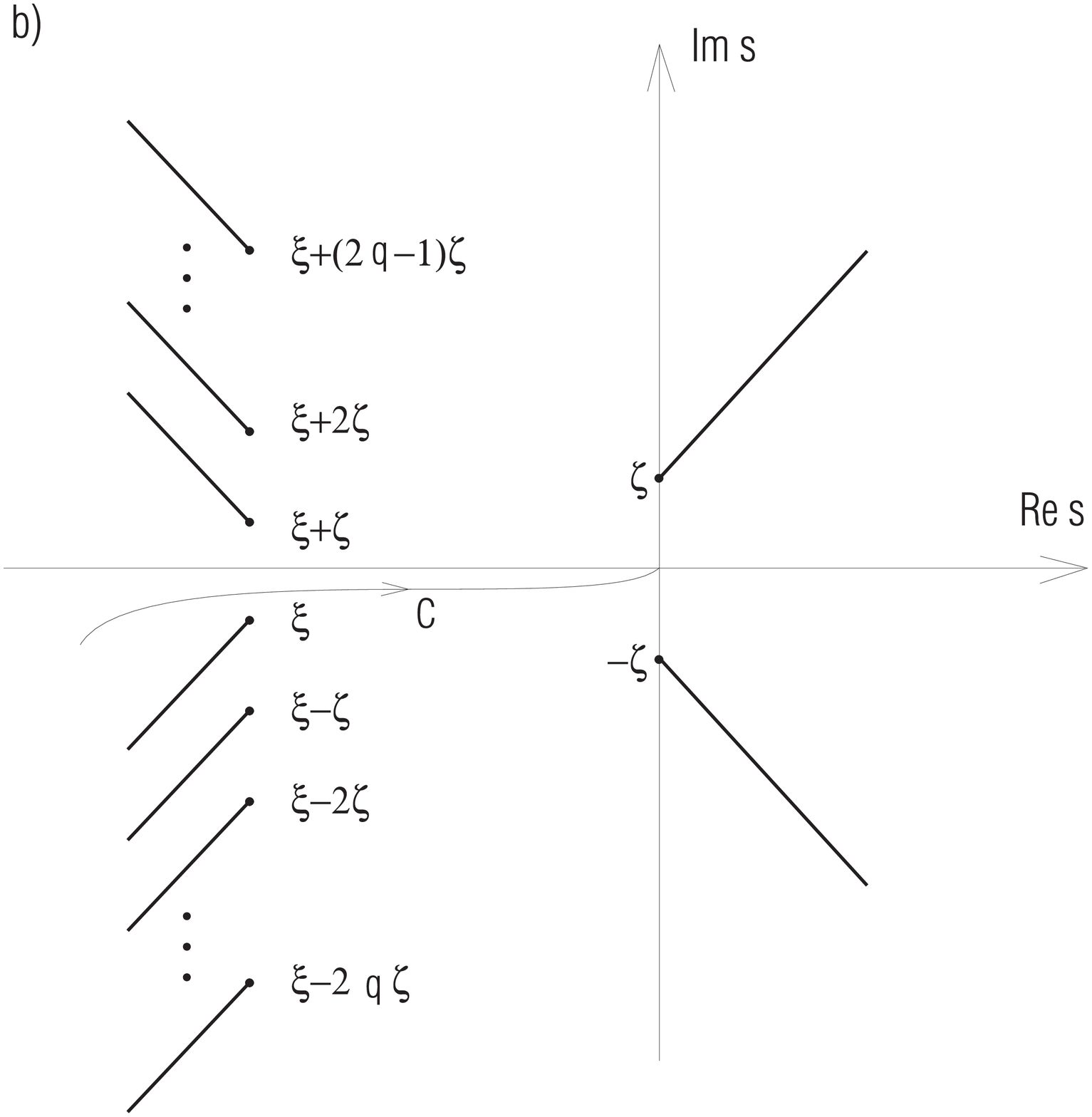,width=7cm} 
\end{tabular} \\
\parbox{14cm}{Fig. 15. $\;\;\;$ The  "first sheets" singularity structure
of $\tilde{\Phi}_1^{(2q+1)}(\xi,s)$ for the harmonic potential}
\end{eqnarray*}
\vskip 18pt

 The following comment
concerning the positions of the singularities themselves and
their relation to the Feynman path integral is in order. From
the above discussion it is seen that these positions are
determined by the values of the classical action the latter
takes on along suitable classical paths corresponding to the
case considered. The paths are real as well as complex (i.e.
they are real or complex solutions to the classical equation of motion). They 
contribute to calculated quantities $\tilde{\Phi}_1^{(q)} (\xi,s)$,
$q \geq 0$, in a hierarchical way described above so that a path with
greater absolute value of the real part of the corresponding
action contributes to the later term $\tilde{\Phi}_1^{(q)} (\xi,s)$ of the
topological expansion. In this way the latter expansion reflects
its close relation to the semiclassical expansion based on the
Feynman path integral and the saddle-point technique as well as
it confirms the role of complex classical paths in such
calculations \cite{15,16,17,18}.

\section{ An application: the connection problem\label{s5}}
\zero

\hskip+2em The connection problem
is an old problem of the JWKB theory which in the context of the
Balian-Bloch representation was considered first by Voros \cite{19}.
We shall discuss again this problem within the framework of our
formalism to show the equivalence of the solution it provides
with the corresponding method used in our earlier papers (see,
for example, \cite{12,22,28}).  

The main question is how the JWKB
formula, being a good approximation to a given solution in some
domain of the $x$-plane, should be changed (in order to remain
still the good approximation of the solution) when the solution
is continued analytically to another domain of the $x$-plane. The
problem can be solved in many different ways depending on the
type of the considered solutions (see, for example, \cite{20,21,22,28}).
In particular, it can be solved with the aid of fundamental
solutions (see \cite{12} for the relevant procedure in an application
to matrix element evaluations in JWKB approximation).

\begin{eqnarray*}
\begin{tabular}{cc}
\psfig{figure=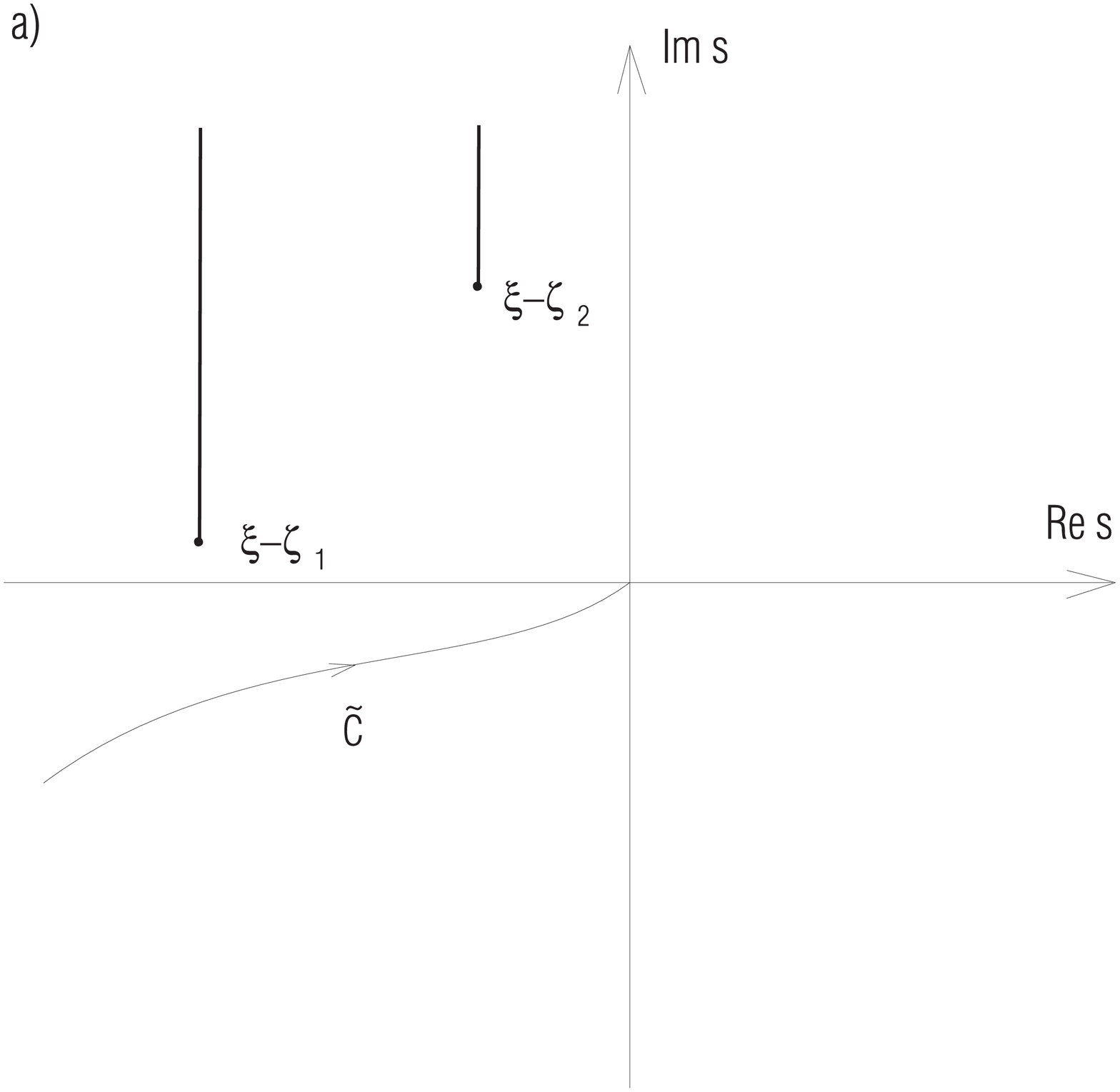,width=7cm} & \psfig{figure=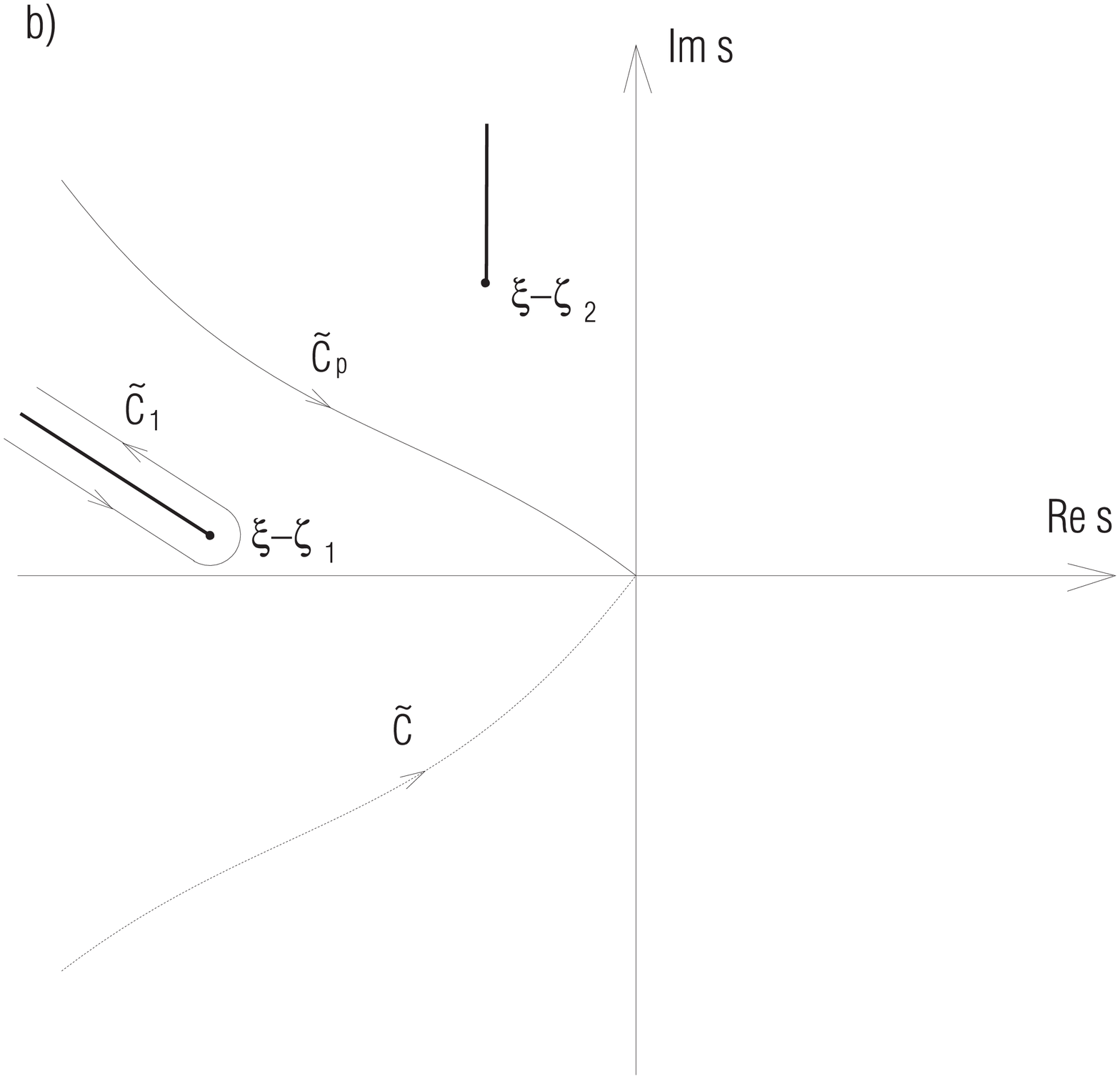,width=7cm} \\
\psfig{figure=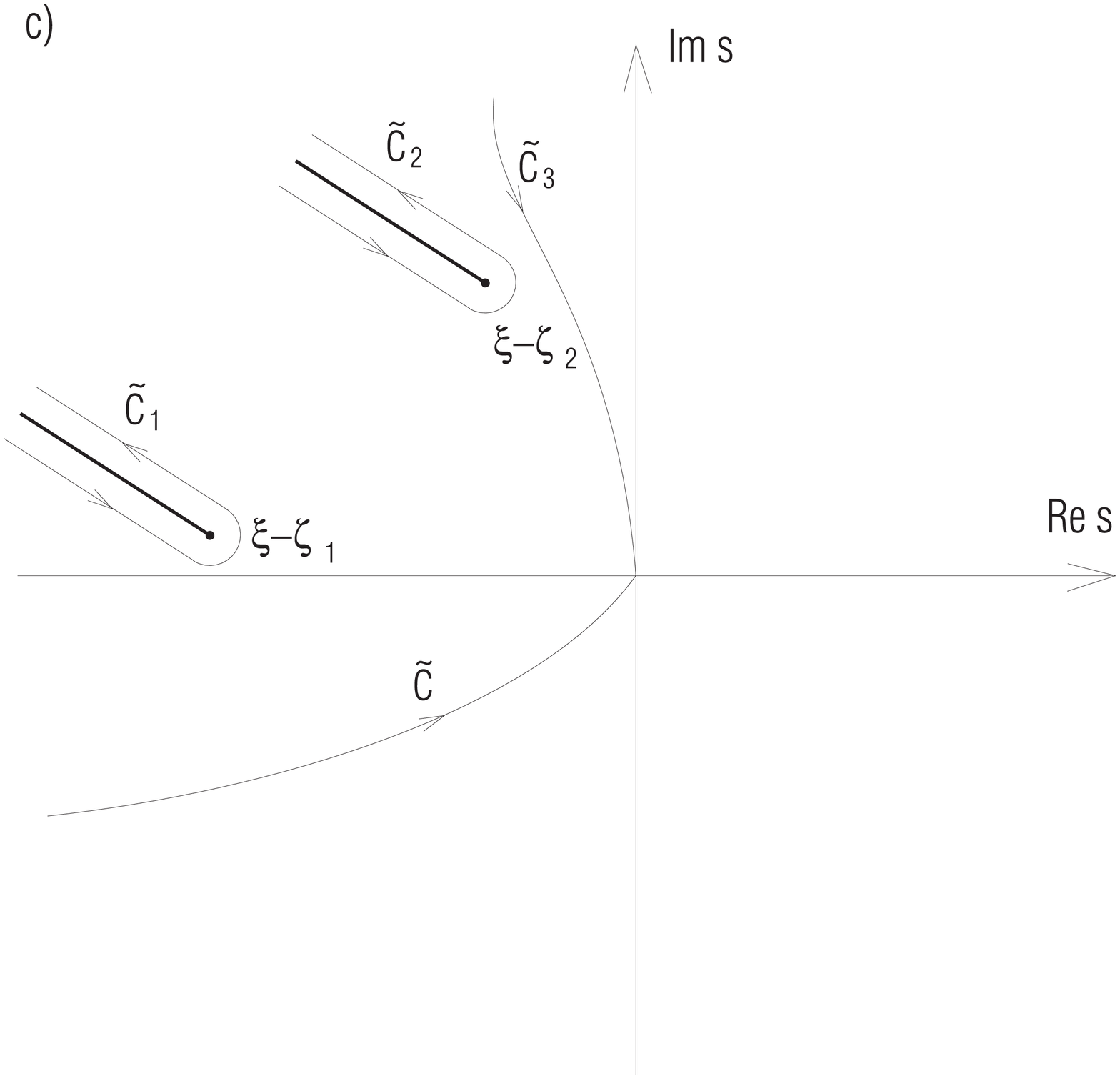,width=7cm} & \psfig{figure=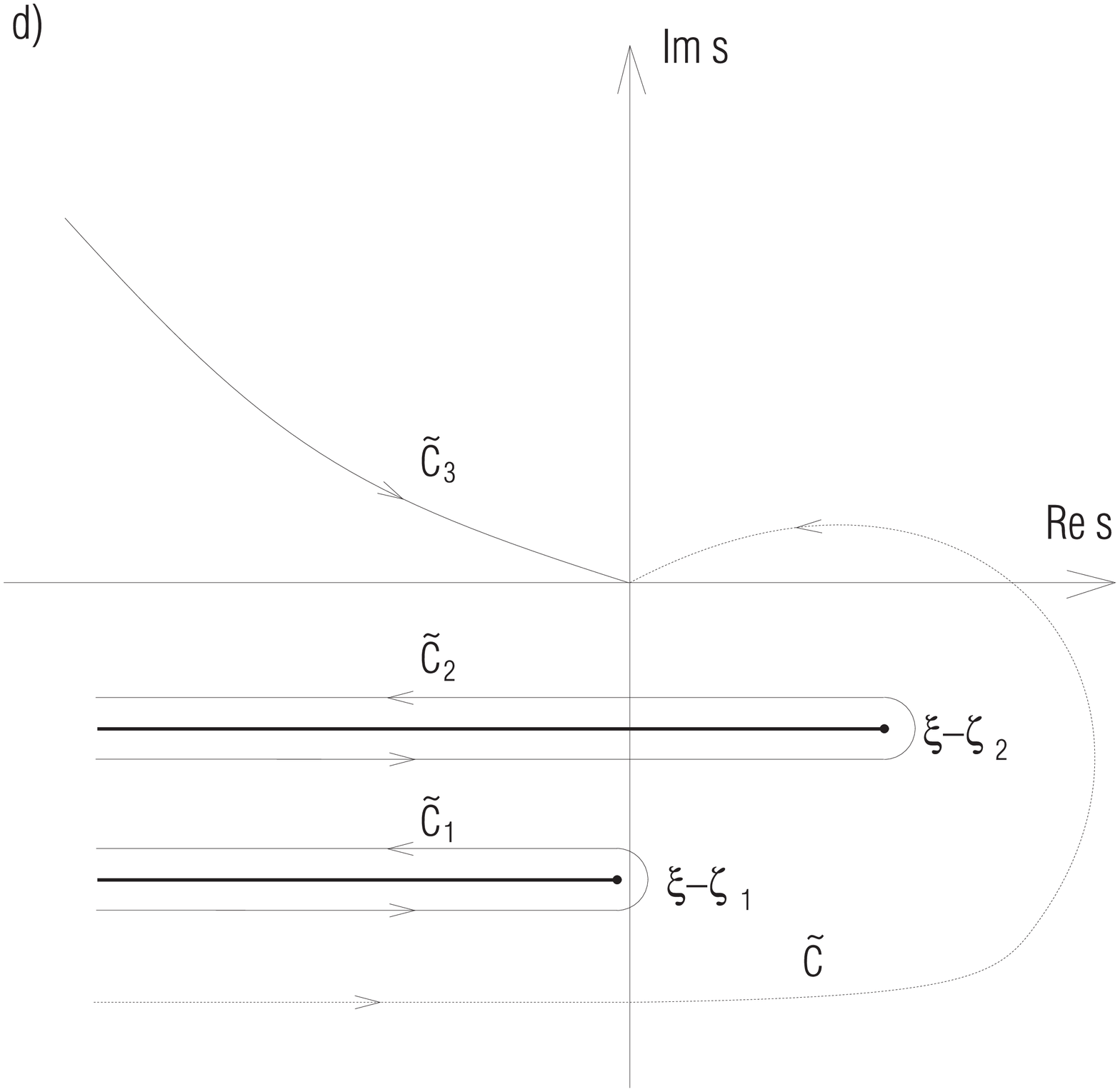,width=7cm} 
\end{tabular} \\
\parbox{14cm}{Fig. 16. $\;\;\;$ The  Borel plane  singularities corresponding
to $\tilde{\chi}_1^{-}(\xi,s)$}
\end{eqnarray*}
\vskip 18pt

  In the framework of the Balian-Bloch representation the solution of the
problem is the following. Consider the fundamental solution to \mref{2.1} 
given by \mref{2.2}-\mref{2.4} and continued along a path $\gamma_1"$ to
sector $2$ (see Fig. 2). As it follows from the previous section
analysis continuing analytically along the considered path we can not 
meet singularities above the corresponding path $\tilde{\gamma}_1(\xi)$
in the $\xi$-plane and, therefore, the singularity pattern of 
$\tilde{\chi}_1 (\xi,s)$
in the $s$-plane looks like in Fig. 16a i.e. there are only two
cuts on the relevant sheet. Since the integration along $\tilde{C}$ is
limited only to lie in the left half-plane we can deform it
freely in this half-plane to the position $\tilde{C}_3$, for example (see
Fig. 16c). But doing this we have to integrate also along the
cuts starting at the points $s = \xi - \zeta_1$ and $s = \xi - \zeta_1$,
respectively. Thus, $\psi_1 (\xi,\lambda)$ is represented in this way as the
following sum: 
\be\label{5.1}
\psi_1 (\xi,\lambda) = 2 q^{-\frac{1}{4}}(\xi) e^{-\lambda\xi}
\int_{C} e^{2\lambda s} \tilde{\chi}_1 (\xi,s)ds \\
  2 q^{-\frac{1}{4}}(\xi) e^{-\lambda\xi}
\left(\int_{C_1} + \int_{C_2} + \int_{C_3} \right) e^{2\lambda s} 
\tilde{\chi}_1 (\xi,s)ds \nn
\ee
where $\xi=\int^x_{x_0} q^{\fr} dy$. 
It should be noticed now that each term
of the sum in (5.1) is a solution to the Schr\"odinger equation
\mref{2.1} (see, for example, \cite{3}). It is also not difficult to see
that each of the solutions generated by the integrations along
$\tilde{C}_1$ and $\tilde{C}_2$ is proportional to the fundamental solution 
defined in the sector 2 of Fig. 2, whilst the remaining third solutions
generated by the integration along $\tilde{C}_3$ - to $\psi_3 (\xi,\lambda)$
- the fundamental solution defined in the sector 3. (An easy way to
establish these facts is to investigate the behaviour of these
solutions when $\xi$ goes to $\infty_2$ and $\infty_3$ correspondingly 
($\infty_k$ being the infinity point in the sector $k$)). In this way a linear
combination of the fundamental solutions $\psi_2$ and 
$\psi_3$ to form the solution $\psi_1 (\xi,\lambda)$ 
is realized simply by moving the contour $\tilde{C}$ in the $s$-plane.

  The connection problem arises when $\psi_1 (\xi,\lambda)$ is
continued to the sector 3 by crossing the sector 2 i.e. along
some non-canonical path $\gamma_1"$ in Fig. 2. At the end of such a
continuation the dominant character of the JWKB factor
$q^{-1/4}\exp(-\lambda\xi)$ is lost in favour of the amplitude factor 
$\chi_1 (\xi,\lambda)$ but the series (\ref{2.3}) does not give 
then an easy answer to what actually happens when  
$\Re \xi \to \infty$ along such a path. (In fact $\chi_1 (\xi,\lambda)$
behaves then as $\exp(2\lambda\xi)$).  

In the $s$-plane the analytic
continuation just described results in a deformation of the
contour $\tilde{C}$ to the form shown in Fig. 16d (broken line). It
follows obviously from the figure that the dominant contribution
to $\psi_1 (\xi,\lambda)$ comes now from the integration along 
$\tilde{C}_2$ i.e. from
the solution proportional to $\psi_2 (\xi,\lambda)$ and this is the way by
which the connection problem is solved within the framework of
the Balian-Bloch representation. (Note, that both the solutions defined by 
the integrations along $\tilde{C}_1$ and $\tilde{C}_3$ are subdominant when 
$\lambda\to\infty$ and $\xi$ stays in the sector 3 or when 
$\Re \xi \to \infty_3$  and $\lambda$ is fixed).

  It is easy to see further, that the linear combination
in the RHS of (\ref{5.1}) can be explicitly reconstructed with the aid
of the canonical coefficients 
$\alpha_{i/j \to k}$ ($\alpha_{i/j \to k} = \lim_{x\to\infty_k}
\frac{\psi_i (x)}{\psi_j (x)}$, see \cite{2}) as follows: 

\be\label{5.2}
\psi_1 (\xi,\lambda) = \alpha_{1/2 \to p}\psi_2 (\xi,\lambda) + 
\alpha_{1/p \to 2} \psi_p (\xi,\lambda) \\
= \alpha_{1/2 \to p} \psi_2 (\xi,\lambda) +
\alpha_{1/p \to 2}  \alpha_{p/2 \to 3}\psi_2 (\xi,\lambda) + 
\alpha_{1/p \to 2} \alpha_{p/3 \to 2}\psi_3 (\xi,\lambda) \nn
\ee
where the sequence of terms of the last sum in (\ref{5.2})
corresponds strictly to the sequence of integrations along $\tilde{C}_1$, 
$\tilde{C}_2$ and $\tilde{C}_3$ in (\ref{5.1}). The first linear combination 
appears in (\ref{5.2}) when the contour $\tilde{C}_1$ is deformed 
to the position  $\tilde{C}_p$ shown in Fig. 16b.  

It is also worthwhile to note that the formula (\ref{5.2}) giving us 
the continuation of $\psi_1 (\xi,\lambda)$ to the sector 3
along the noncanonical path $\gamma_1"$ can be also used to obtain in a
simple way the improved connection formula of Silverstone \cite{22}
(see also the recent work of Fr\"oman and Fr\"oman \cite{28}) with $\psi_3$
playing the role of the subdominant contribution. It is enough
to this end to substitute each term in the sums in (\ref{5.2}) by its
corresponding JWKB approximation (i.e. none cumbersome Borel
resummation used by Silverstone is necessary).

\section{ Exponential asymptotics \label{s6}}
\zero

\hskip+2em  The problem of the semiclassical
expansions for physical quantities is strictly related to the
problem of so called exponentially small contributions absent
(by definition) when only the bare semiclassical expansions of
these quantities are considered \cite{23,24,25,26,27}. The exponentially small
contributions become important if the accuracy of the best
semiclassical approximation is considered to be insufficient.
There are however two aspects of this problem a difference
between which was, according to our knowledge, not discussed properly.  

The first one appears when the corresponding
semiclassical series if Borel summed does not reproduce
correctly the quantity considered. A discrepancy has to be of
course exponentially small not contributing to the semiclassical
limit. It can however be incorporated into the resurgent
quantity by choosing another integration path in the Borel
'plane' i.e. the original path has had to be chosen incorrectly
if the quantity considered was to be Borel summable (see \cite{3} for
the corresponding discussion). If it is the case then the Borel
summation along the correct path has to incorporate $all$ the
exponentially small contributions to the quantity considered.

The second aspect appears when the Borel transform reproduces
completely the quantity considered and the semiclassical series
is used as a source of the best approximation. As it is well
known (see for example \cite{29}) the latter is obtained in this case
by abbreviating the series on its least term (since the series
is divergent) the order of which is proportional to actual value
of $\lambda(\hbar^{-1})$ (in fact $n$ should be equal to the integer part of
$\lambda|s_0|$ where $s_0$ is a singularity of the Borel function closest to
the origin). The remainder (i.e. the difference between the
quantity and its approximation) is then exponentially small quantity.  

To improve this approximation using still the
semiclassical tools we have to be able to identify the
exponential factor of the remainder and to multiply the last
factor again by some optimal abbreviation of a new semiclassical
expansion of the remainder. Next, we should be able to repeat
this procedure to the remainder of the remainder constructing in
this way still more and more accurate semiclassical
approximation which includes as many exponentially small
contributions as we need to make the approximation as good as we
wish. In Appendix 4 we show how to do it. According to the
beginning of this discussion the exponentially small
contributions obtained in this way have to be determind by the
singularity structure of the corresponding Borel functions
provided for example by the topological expansions. The results
of Appendix 4 confirm these expectations.

\section{ Exponential asymptotics of energy levels \label{s7}}
\zero

\hskip+2em  Using the
approximation scheme which follows obviously from the
topological expansion and from the results of App. 4 we shall
determine in this section the way of obtaining the semiclassical
exponential asymptotic for energy levels of the anharmonic
osciallator corresponding to the potential $V(x)=x^2 + x^4$ with the
Stokes graph shown in Fig. 17 and drawn for $E>0$. Taking into
account the symmetry of the potential we can write the
quantization condition for the energy levels in the form 

\be\label{7.1}
\exp\left(\frac{\lambda}{2} \oint_K \sqrt{V(x)-E^{\pm}(\lambda)} dx
\pm i \frac{\pi}{2} \right) = \chi_{1\to 3}(E^{\pm}(\lambda),\lambda)
\ee
where
the contour $K$ is shown in Fig. 17 and "$\pm$" in (\ref{7.1}) correspond to
the even and odd parities of the levels respectively.

\begin{tabular}{c}
\psfig{figure=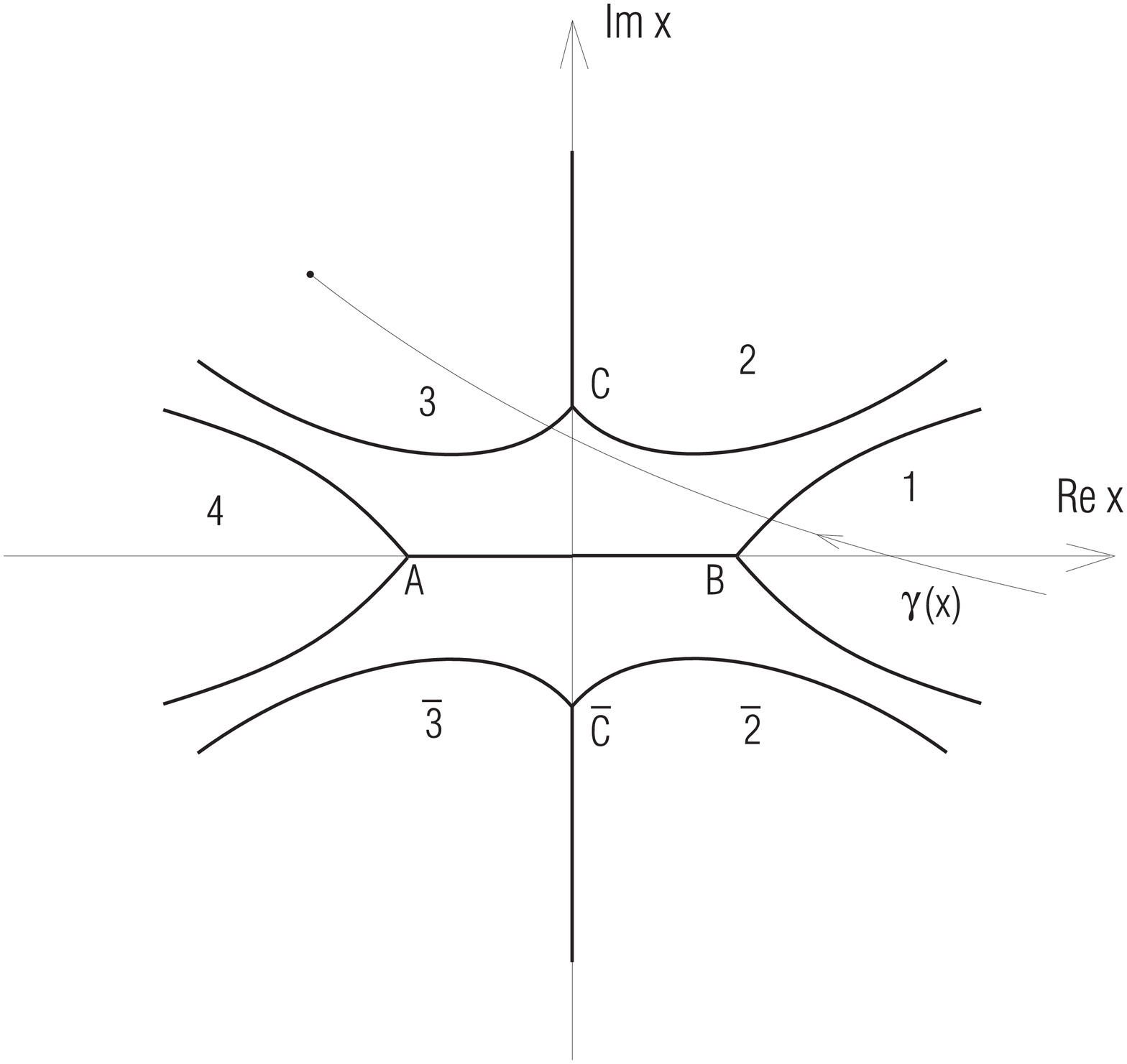,width=8cm} \\
\parbox{14cm}{Fig. 17. $\;\;\;$ The  Stokes graph for the harmonic potential
$V(x) = x^2 + x^4$ for $E>0$}
\end{tabular}
\vskip 18pt

Making the complex conjugation of both the sides of (\ref{7.1}) we get an
alternative condition for the energy level quantization. Both
the versions are important since they determine the normal
sector of $E(\lambda)$ to be $0<|\arg\lambda|<3\pi/2$ for $\lambda$
sufficiently large \cite{2}. Because of this the semiclassical series of 
$E(\lambda)$ is, as we
have shown in our earlier paper \cite{2} (see also \cite{5,6,7,8,23}), Borel
summable to $E(\lambda)$ itself and the singularity structure of 
$\tilde{E}(s)$ on its
Borel plane is also determined by (\ref{7.1}) and its complex
conjugation. As it follows from (\ref{7.1}) this structure is
symmetric with respect to the real axis of the $s$-plane and $E(\lambda)$
can be recovered by integrating $\tilde{E}(s)$ along the negative halfaxis.
Of course to apply to the last integral the procedure of App. 4 we
have to know a detailed distribution of singularities of $\tilde{E}(s)$ on its
Borel plane. But instead of this we can use (\ref{7.1}) directly to
establish the respective exponentially small contributions to
$E(\lambda)$.  Namely ordering these contributions according to their
exponential smallness we can treat each such a contribution as a
correction to its predecessors and use the Taylor series
expansion to take into account the corresponding contribution.
So we can write 

\be\label{7.2}
E(\lambda)=E_0(\lambda) + E_1(\lambda) \ldots
\ee
with $E_1(\lambda)$ being polynomially dependent on $\lambda^{-1}$
and with the contributions $E_1(\lambda)$, $E_2(\lambda)$,..., being each
exponentially small with respect to their predecessors. Of
course, $E_0(\lambda)$ is constructed in the standard way (see \cite{11} and
App. 4) and, for a given $\lambda$, it has some well defined numerical value.
  
The corresponding Taylor expansion of $\chi_{1\to 3}(E(\lambda),\lambda)$ with
respect to the dependence of the latter on the energy is the following
\be\label{7.3}
\chi_{1\to 3}(E(\lambda),\lambda)=
\chi_{1\to 3}(E_0(\lambda),\lambda) +
 \frac{\partial\chi_{1\to 3}(E_0(\lambda),\lambda)}{\partial E}
(E_1(\lambda)+ E_2(\lambda) + \ldots) + \ldots
\ee

Since the quantities in (\ref{7.2}) are real we shall write the
corresponding quantization condition in its real form, too, to get

\be\label{7.4}
\sin\left(\frac{\lambda}{2 i}{\oint\limits_K} 
\sqrt{V(x)-E_0^{\pm}(\lambda)}dx -
\frac{\lambda}{4 i}(E_1^{\pm}(\lambda)+E_2^{\pm}(\lambda)+\ldots)
{\oint\limits_K} \frac{dx}{\sqrt{V(x)-E_0^{\pm}(\lambda)}}+\ldots \right) \\
=\mp \Re \left(\chi_{1\to 3}(E^{\pm}(\lambda),\lambda)
 + (E^{\pm}_1(\lambda)+ E^{\pm}_2(\lambda) + \ldots) 
 \frac{\partial\chi_{1\to 3}(E^{\pm}_0(\lambda),\lambda)}{\partial E} 
+ \ldots \right) \nn
\ee

It is now clear that we can apply to the coefficient 
$\chi_{1\to 3}(E_0(\lambda),\lambda)$ and its
derivatives the procedure of App. 4 considering $E_0(\lambda)$ as having
well defined value so that the singularity structure of the
corresponding Borel function $\tilde{\chi}_{1\to 3}(E,s)$ 
is determined just by the value of $E$ equal to $E_0(\lambda)$.  

Assuming the exponential asymptotics for $\chi_{1\to 3}(E_0(\lambda),\lambda)$
to be ordered in a way analogous to (\ref{7.2}) we get for the first two
terms of (\ref{7.2}) 

\be\label{7.5}
\sin\left(\frac{\lambda}{2 i}\oint_K \sqrt{V(x)-E_0^{\pm}(\lambda)}dx \right)
=\mp \Re \left(\chi_{1\to 3}^{(0)}(E_0^{\pm}(\lambda),\lambda) \right)
\ee
and

\be\label{7.6}
E_1^{\pm}(\lambda)=
\frac{\pm \Re \left(\chi_{1\to 3}^{(1)}(E_0^{\pm}(\lambda),\lambda) \right)}
{\frac{\lambda}{4 i} \oint\limits_K \frac{dx}{\sqrt{V(x)-E_0^{\pm}(\lambda)}}
\cos\left( \frac{\lambda}{2 i}\oint\limits_K 
\sqrt{V(x)-E_0^{\pm}(\lambda)}dx \right) \mp \Re \left(
 \frac{\partial\chi^{(0)}_{1\to 3}(E^{\pm}_0(\lambda),\lambda)}{\partial E} 
\right)}
\ee
where $\chi_{1\to 3}^{(0)}(E_0(\lambda),\lambda)$
(we shall suppress the parity indices as unimportant for
our further considerations) is given by the respective number of
the first terms of the series (\ref{2.5}) (i.e.  abreviated at its
corresponding least term; note also that the integrations in (\ref{2.5}) go 
from $\infty_1$ to $\infty_3$ along the canonical path) whilst the
exponential contribution $\chi_{1\to 3}^{(1)}(E_0(\lambda),\lambda)$
is determined according to App. 4 by the singularities of 
$\tilde{\chi}_{1\to 3}(E_0(\lambda),s)$ in its Borel plane closest 
to the origin.

Applying now the approximations following from the topological
expansion (\ref{A1.11}) for $\tilde{\chi}_{1\to 3}(E_0(\lambda),s)$
we can write (keeping only the first two terms of this expansion) 

\be\label{7.7}
\tilde{\chi}_{1\to 3}(E_0(\lambda),s)=
\tilde{\Phi}^{(0)}_{1\to 3}(E_0(\lambda),s)
+ \tilde{\Phi}^{(2)}_{1\to 3}(E_0(\lambda),s)
\ee
where 

\be\label{7.8}
\tilde{\Phi}^{(0)}_{1\to 3}(E_0(\lambda),s) = I_0 \left(
\sqrt{4s\int\limits_{\infty_1}^{\infty_3}\tilde{\omega}(\xi,E_0(\lambda))d\xi}
\right) \nn \\
\tilde{\Phi}^{(2)}_{1\to 3}(E_0(\lambda,s))= \int_s^0 d\eta
\int\limits_{\infty_1}^{\infty_3}d\xi \tilde{\omega}(\xi+\eta,E_0(\lambda))
\tilde{\omega}(\xi,E_0(\lambda))\frac{I_2(\sqrt{z})}{z} \\
z= 4(s-\eta)
\int\limits_{\infty_1}^{\infty_3}\tilde{\omega}(\zeta,E_0(\lambda))d\zeta
+ 8(s-\eta)\left(\int\limits_{\infty_1}^{\xi}-
\int\limits_{\infty_1}^{\xi+\eta}\right)
\tilde{\omega}(\zeta,E_0(\lambda))d\zeta \nn
\ee

Assuming the same order of approximation for  
$\chi_1(x,E_0(\lambda),\lambda)$ we can see that when $x$ stays in
the sector 3 as it is shown in Fig. 18a then its Borel plane
looks as in Fig. 18b on which $C_1$ is the path of the Borel integration to 
recover $\chi_1(x,E_0(\lambda),\lambda)$. The distribution of the
singularities on the figure follows now from (\ref{7.7}). Fig. 18c
shows the Borel plane for $\tilde{\chi}_{1\to 3}(E_0(\lambda),s)$
i.e. when $x\to \infty_3$. The singular points are
$\zeta_C=\int\limits_{B(E_0(\lambda))}^{C(E_0(\lambda))} 
\sqrt{V(x)-E_0(\lambda)}dx$, $-\zeta_C$, $\zeta_C -\zeta_A=
\int\limits_{A(E_0(\lambda))}^{C(E_0(\lambda))} 
\sqrt{V(x)-E_0(\lambda)}dx$ and $\zeta_A-\zeta_C$. Therefore, 
$\chi_{1\to 3}(E_0(\lambda),\lambda)$ can be given as 

\be\label{7.9}
{\chi}_{1\to 3}(E_0(\lambda),\lambda) =
2\lambda\int\limits_C e^{2\lambda s}
\tilde{\chi}_{1\to 3}(E_0(\lambda),s)ds
\ee
where the integration path $C$ is shown in Fig.18c.

\begin{eqnarray*}
\psfig{figure=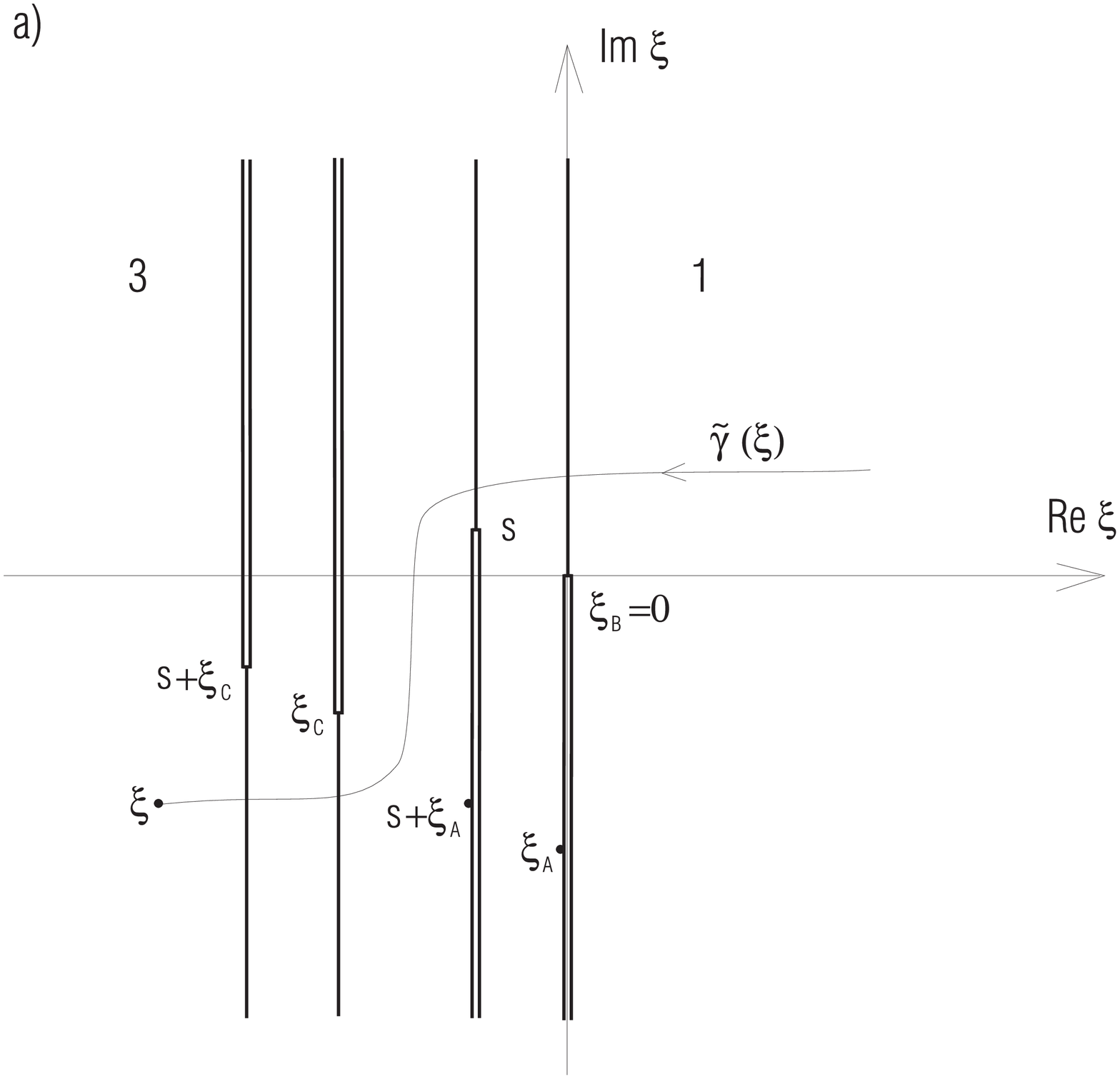,width=7cm} & \psfig{figure=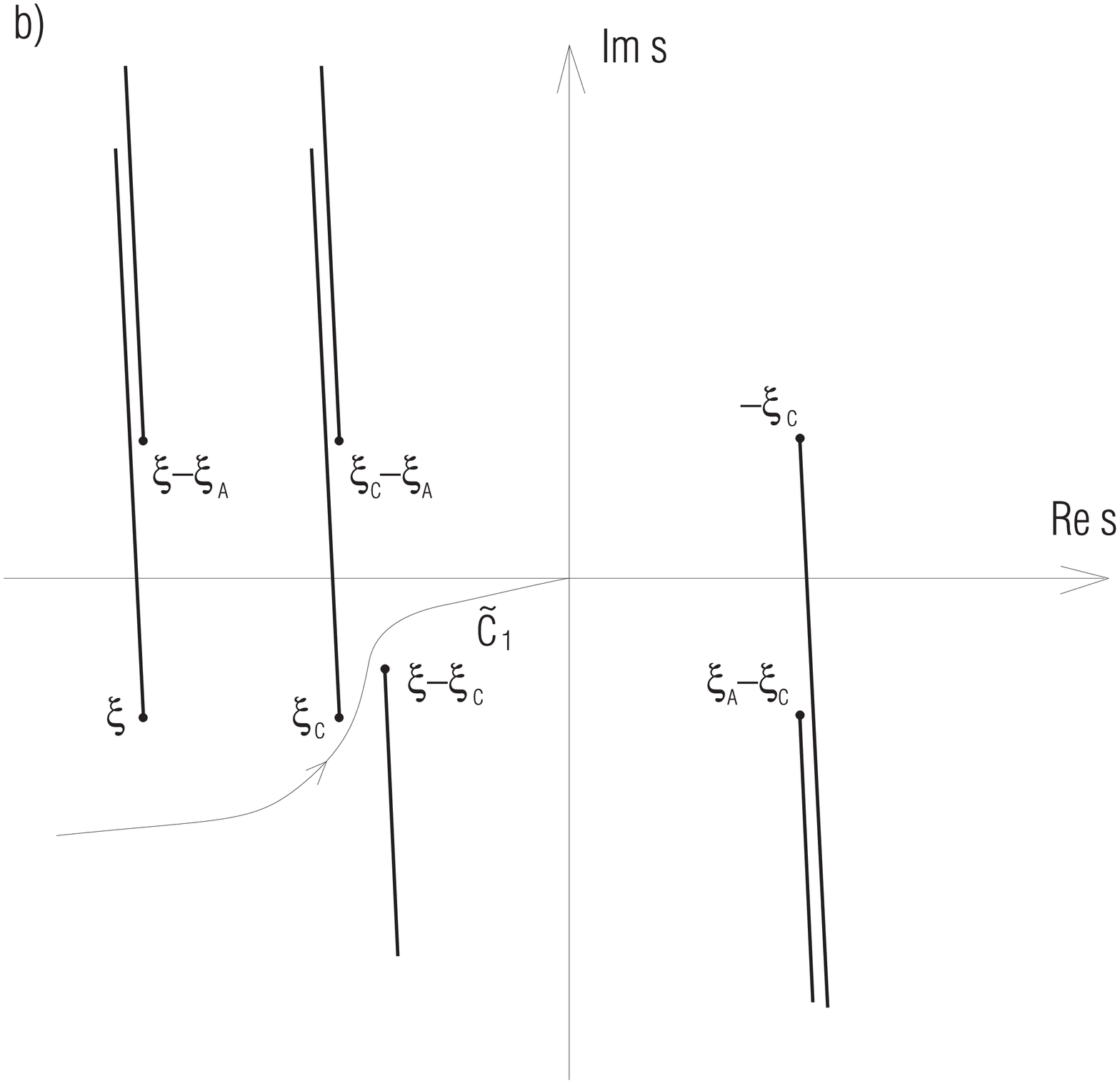,width=7.2cm} \\
\psfig{figure=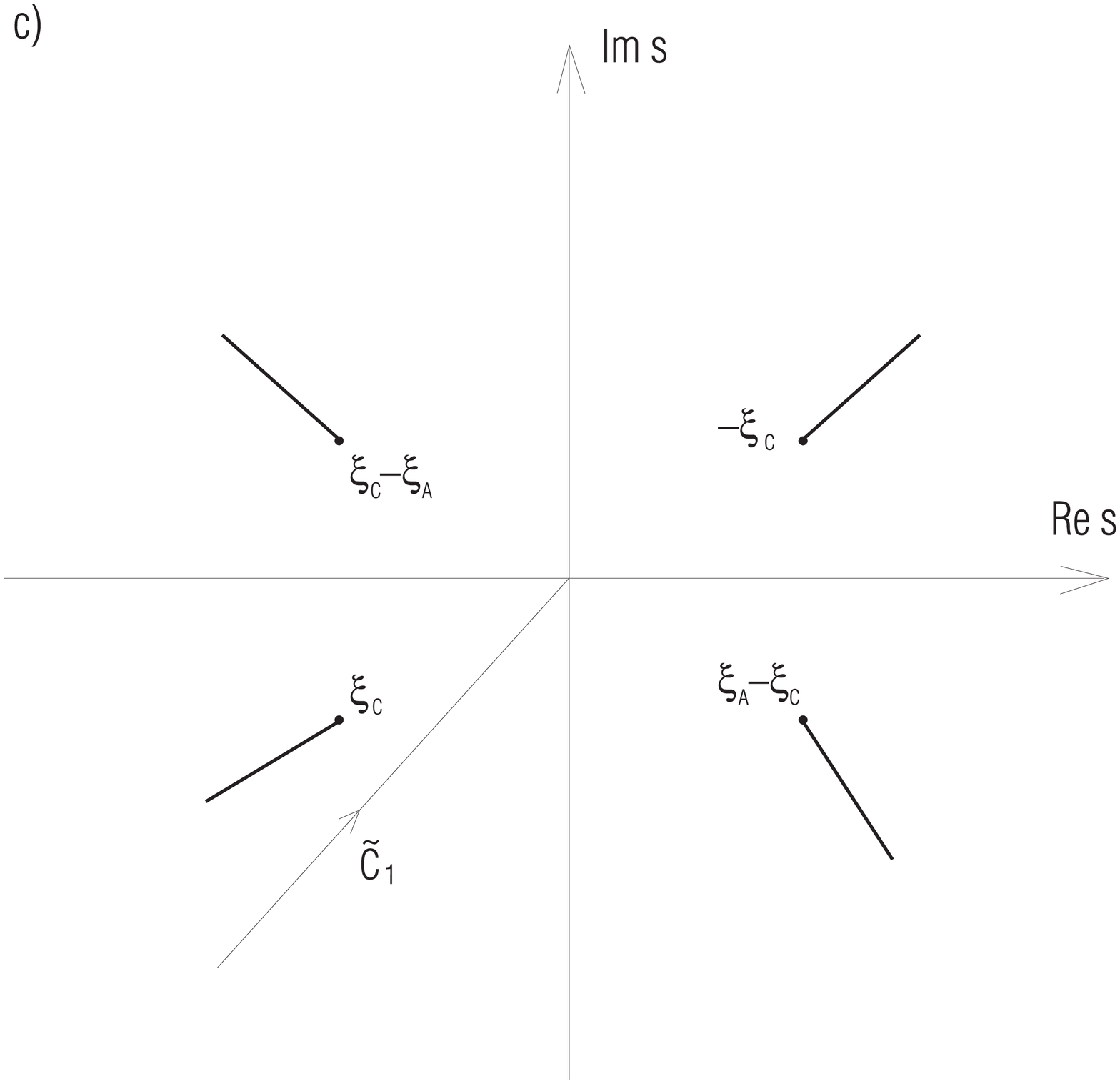,width=7cm} & 
\parbox[b]{7cm}{Fig. 18. $\;\;\;$ The singularity patterns of 
$\chi_1(x,E_0(\lambda),\lambda)$ for the anharmonic oscillator (figure (a))
and $\tilde{\chi}_1,(x,E_0(\lambda),s)$ (figure (b)). Figure (c) shows
the latter pattern for $x(\xi) \to \infty_3$}
\end{eqnarray*}

\vskip 12pt

The path $C$ differs from the one considered in App. 4 but this does not
prevent us to apply the procedure of this appendix. Therefore
according to the approximation (\ref{7.7}) we have 

\be\label{7.10}
{\chi}_{1\to 3}(E_0(\lambda),\lambda) =
2\lambda\int\limits_C e^{2\lambda s}
\left( \tilde{\Phi}^{(0)}_{1\to 3}(E_0(\lambda),s)
+ \tilde{\Phi}^{(2)}_{1\to 3}(E_0(\lambda),s) \right)ds
\ee
so that 

\be\label{7.11}
{\chi}_{1\to 3}^{(0)}(E_0(\lambda),\lambda) =
\sum\limits_{k=0}^{n_0} \frac{(-1)^k}{(2\lambda)^k}
\frac{\partial^k}{\partial s^k}
\left( \tilde{\Phi}^{(0)}_{1\to 3}(E_0(\lambda),s)
+ \tilde{\Phi}^{(2)}_{1\to 3}(E_0(\lambda),s) \right)|_{s=0}
\ee
where $n_0=[|\lambda\zeta_C|]$.

Using the formulae (\ref{A4.5}) and (\ref{A4.6}) of App. 4 for
${\chi}_{1\to 3}^{(1)}(E_0(\lambda),\lambda)$ we get 

\be\label{7.12}
{\chi}_{1\to 3}^{(0)}(E_0(\lambda),\lambda) =
- \sum\limits_{j=C,-C,A-C,C-A}
\frac{(n_0+1)!}{(2\lambda)^{n_0}\zeta^{n_0}} \sum\limits_{m=0}^{n_1} 
\frac{(-1)^m \kappa^{(m)}_j (E_0(\lambda),0)}{(2\lambda)^{m+1}}
\ee
where $\zeta_{-C}=-\zeta_{C}$, $\zeta_{A-C} =\zeta_{A} -\zeta_{C}$,
$\zeta_{C-A} =\zeta_{C}-\zeta_{A}$, 
$n_1=[|\lambda\zeta_A|]$ (with $|\zeta_A|$ determining the common distance of
singularities of $\kappa_j (E_0(\lambda),s)$ closest to the origin) and 
$\kappa_j$'s are given by

\be\label{7.13}
{\kappa}_{j}(E_0(\lambda),s) =
\frac{1}{2\pi i} \int\limits_{K_j} dt
\frac{ \tilde{\Phi}^{(2)}_{1\to 3}(E_0(\lambda),\zeta_j +t)}
{(1+\frac{t}{\zeta_j})^{n_0} } \\
 \cdot \frac{1}
{\left(t+\zeta_j +\frac{n_0}{\lambda} \ln(1+\frac{t}{\zeta_j}) -s\right)
\left(t+\zeta_j +\frac{n_0}{\lambda} +\frac{n_0}{\lambda} 
\ln(1+\frac{t}{\zeta_j}) -s\right)} \nn
\ee
where the contours $K_j$ surround the cuts originated by the
singularities of $\tilde{\Phi}^{(2)}_{1\to 3}(E_0(\lambda),s)$
 at $\zeta_j$'s each shifted to the origin $s=0$ of the Borel plane.

\section{ Summary \label{s8}}
\zero

\hskip+2em  In this paper we have found the representation for
the Borel functions of the quantities relevant for the 1D
quantum mechanics. The representation takes the form of the
topological expansion. This expansion provides us with an
algorithm determining in a systematic way the singularity
structure of the Borel plane for the relevant quantities and
orders the appearing of the Borel plane singularity structures
in a hierarchical way allowing for the formulation of the
approximation scheme of the semiclassical calculations
alternative to the other ones \cite{8,9,10,11,12}.  

We have also remarked limitations of our method in the
description of the proper nature of singularities of the
quantities represented by the expansion.

We have formulated also
the scheme of the semiclassical approximations including the
exponentially small contributions to the desired order of
accuracy. It makes use of the Borel plane singularity structure
in the most natural and effective way, particularly, if it is
accompanied by the topological expansion method of
approximations of the Borel functions.  

We have demonstrated the
action of both the methods considering some simple (but not
quite trivial) examples of their applications in Sections 4-5
and 7. However, it was not our aim in this paper to perform some
numerical tests of the method presented. Rather we have limited
ourselves to test both the methods as theoretical tools for
better understanding of the mutual relations between the
semiclassical expansions, Borel plane singularity structure and
the exponential asymptotics. For the latter goal both the
expansions (i.e. the topological and the exponential ones)
appeared to be very useful. Nevertheless, their test as a
practical method of extended semiclassical approximations is
certainly desired.

\section*{Acknowledgment}

 The idea of this paper originated during my
participation in the workshop on the Exponential Asymptotics
organized by the Isaac Newton Institute for Mathematical
Sciences (Cambridge, June 1995). I am greatly indebted to the
organizers of the program, particularly to M.V. Berry and C.J.
Howls for inviting me to the workshop and for the inspiring
atmosphere of the Institute.  

\vskip 12pt 

\section*{ Appendix 1}
\renewcommand{\theequation}{A1.\arabic{equation}}
\zero
\vskip 12pt
\hskip+2em {\it A1.1. The Laplace transforms 
$\tilde{Y}_{n;r_1,...,r_q}(\xi,s)$}

We shall determine below
the Laplace transforms $\tilde{Y}_{n;r_1,...,r_q}(\xi,s)$, $n \geq q \geq 0$, 
as defined by (\ref{3.5}). To begin with consider first the case 
$r_1 = 0$ and $q = 1$. We have:

\be\label{A1.1}
 Y_{n;r}^{(1)}(\xi,\lambda)=
\int\limits_{\tilde{\gamma}_1(\xi)} d\xi_1 \ldots 
\int\limits_{\tilde{\gamma}_1(\xi_{n-1})} d\xi_n
\tilde{\omega}(\xi_1) \ldots \tilde{\omega}(\xi_r)
\ldots \tilde{\omega}(\xi_n) e^{2\lambda(\xi - \xi_r)} \\
 r = 1,..., n, \;\;\;\;\;\;\;\;\;\;\;  n = 1, 2, ...\;\; etc. \nn
\ee

The multiple integral in (\ref{A1.1}) can be rewritten further as
follows: 

\be\label{A1.2}
Y_{n;r}^{(1)}(\xi,\lambda)=
\int\limits_{\tilde{\gamma}_1(\xi)} d\xi_r
e^{2\lambda(\xi - \xi_r)}
\tilde{\omega}(\xi_r)\Omega_{r-1}(\xi,\xi_r)
Y_{n-r;}^{(0)}(\xi_r)  \\
 r = 1,..., n, \;\;\;\;\;\;\;\;\; n = 1, 2, ...\;\; etc. \nn
\ee

where $Y_{n-r;}^{(0)}(\xi_r)\equiv Y_{n-r;}(\xi_r)$ is defined by (\ref{3.3}) 
and 

\be\label{A1.3}
\Omega_{r-1}(\xi,\xi_r)= ((r-1)!)^{-1}
(\Omega(\xi) -\Omega(\xi_r))^{r-1}
\ee

Making further in (\ref{A1.2}) a
change $\xi_r \to \xi_r - s$ of the integration variable we get finally: 

\be\label{A1.4}
Y_{n;r}^{(1)}(\xi,\lambda)=
\int\limits_{\tilde{C}} ds
e^{2\lambda s}
\tilde{Y}_{n;r}^{(1)}(\xi_r) \\
 r = 1,..., n, \;\;\;\;\;\;\;\;\;\;\;\;\; n = 1, 2, ...\;\; etc. \nn
\ee
where 

\be\label{A1.5}
\tilde{Y}_{n;r}^{(1)}(\xi,s)=
-\tilde{\omega}(\xi -s)\Omega_{r-1}(\xi,\xi-s)
Y_{n-r;}^{(0)}(\xi-s)
\ee
and the contour $\tilde{C}$ runs
from the infinity $\Re s = -\infty$ to the origin $s = 0$. Note, that
because of $\Re \xi > 0$ (by assumption) the contour  $\tilde{C}$ is 
independent of $r$ and $n$ and also, as it follows from (\ref{A1.5}), 
$\tilde{Y}_{n;r}^{(1)}(\xi,s)$ is holomorphic for  $\Re s < \Re \xi$.  

Resoning in the completely similar
way we get for $\tilde{Y}_{n;r_1 r_2}^{(2)}(\xi,s)$: 

\be\label{A1.6}
\tilde{Y}_{n;r_1 r_2}^{(2)}(\xi,s)=
-\int\limits_{\tilde{\gamma}(\xi-s)} d\xi_1
\tilde{\omega}(\xi_1+s)\tilde{\omega}(\xi_1)\Omega_{r_1 -1}(\xi,\xi_1+s)
\Omega_{r_2 - r_1 -1}(\xi_1+s,\xi_1)
Y_{n-r_2;}^{(0)}(\xi_1)
\ee  

The remaining Laplace transforms 
$\tilde{Y}_{n;r_1...r_{2q+1}}^{(2q+1)}(\xi,s)$ and 
$\tilde{Y}_{n;r_1...r_{2q}}^{(2q)}(\xi,s)$, $q = 1, 2,...$ etc. can
be defined recurrently as follows: 

\be\label{A1.7}
\tilde{Y}_{n;r_1 r_2 r_3...r_{2q}}^{(2q)}(\xi,s)=
- \int\limits_{\tilde{C}(s)} d\eta \int\limits_{\tilde{\gamma}(\xi-s)} d\xi_1
\tilde{\omega}(\xi_1 +s)\tilde{\omega}(\xi_1 +\eta)
\Omega_{r_1 -1}(\xi,\xi_1 +s)  \\
\Omega_{r_2 - r_1 -1}(\xi_1 +s,\xi_1 +\eta)
\tilde{Y}_{n-r_2;r_3-r_2...r_{2q}-r_2}^{(2q-2)}(\xi_1 +\eta,\eta),
 q=2,3,... \nn
\ee  
and 

\be\label{A1.8}
\tilde{Y}_{n;r_1 r_2 r_3...r_{2q+1}}^{(2q+1)}(\xi,s)=
- \int\limits_{\tilde{C}(s)} d\eta 
\tilde{\omega}(\xi -s+ \eta)\Omega_{r_1 -1}(\xi,\xi_1 -s +\eta) \\
\cdot \tilde{Y}_{n-r_1;r_2-r_1...r_{2q+1}-r_1}^{(2q)}(\xi_1 -s +\eta,\eta)
\;\;\;\;\;\;\;\;\;\;\;\;\;\; q=1,2,... \nn
\ee  

Every of them is
holomorphic in $\xi$ and $s$ for  $\Re \xi > 0$ and $\Re s < \Re \xi$. 
The contour $\tilde{C}(s)$ in (A1.7)-(A1.8) starts at the point $s$ 
with $\Re s < \Re \xi$ and ends at $s = 0$.

\vskip 12pt
{\it A1.2. Topological expansion}
\vskip 12pt

 A further step we can do is to fix $q$
and to take sums with respect to $n,\; r_1, ..., r_{2q}$. It can be done
as follows. First, we consider rather $\chi_1(\xi,\lambda)/(2\lambda)$ than 
$\chi_1(\xi,\lambda)$ itself. Next we note that to each term 
$-(-2\lambda)^{-n-1}\tilde{Y}_{n;r_1...r_{q}}^{(q)}(\xi,s)$ there 
corresponds the following Laplace transform: 

\be\label{A1.9}
 \frac{1}{n!}s^n * \tilde{Y}_{n;r_1...r_{q}}^{(q)}(\xi,s)
\ee
where the star means the convolution of the factors.  

The sums we are now looking for are the following: 

\be\label{A1.10}
\tilde{\Phi}_1^{(q)}(\xi,s)= \sum\limits_{1\leq r_1< ... <r_q \leq n}
\frac{(-1)^{r_1 +r_2+ ... +r_q}}{n!} 
s^n * \tilde{Y}_{n;r_1...r_{q}}^{(q)}(\xi,s)
\ee
so that the series: 

\be\label{A1.11}
\tilde{\Phi}_1(\xi,s)= \sum_{q\geq0} \tilde{\Phi}_1^{(q)}(\xi,s)
\ee
(its convergence is discussed
below) represents a function $\tilde{\Phi}_1(\xi,s)$ such that 
$\partial\tilde{\Phi}_1(\xi,s)/\partial s$
is the Laplace transform of $\chi_1(\xi,\lambda)$.  

The sums in (\ref{A1.10}) can be performed explicitly to give: 

\be\label{A1.12}
\tilde{\Phi}_1^{(0)}(\xi,s)= I_0 \left(\sqrt{4s\Omega(\xi)}\right)  \nn \\
\tilde{\Phi}_1^{(2q)}(\xi,s)= 
 \int\limits_{\tilde{C}(s)} d\eta_1 
 \int\limits_{\tilde{C}(\eta_1)} d\eta_2  \ldots
 \int\limits_{\tilde{C}(\eta_{q-1})} d\eta_q 
 \int\limits_{\infty}^{\xi-\eta_1} d\xi_1
\int\limits_{\infty}^{\xi_1} d\xi_2 \ldots
\int\limits_{\infty}^{\xi_{q-1}} d\xi_q  \nn \\
 \tilde{\omega}(\xi_1 +\eta_1)
\tilde{\omega}(\xi_1 +\eta_2) \cdot \ldots \cdot
\tilde{\omega}(\xi_q +\eta_q)\tilde{\omega}(\xi_q)(2s-2\eta_1)^{2q}
\frac{I_{2q}(z_{2q}^{\fr})}{z_{2q}^q}  \nn \\
z_{2q}= 
4(s-\eta_1)\Omega(\xi) + 8(s-\eta_1) \sum_{p=1}^q
\left( \Omega(\xi_p + \eta_{p+1}) - \Omega(\xi_q + \eta_p) \right), 
\;\;\;\;\;\;\;\;\;\;\;\;\;\;\;  \nn \\
\eta_{q+1} \equiv 0, \;\;\;\;\;\;\;\;\;\; q=1,2,\ldots  \\
\tilde{\Phi}_1^{(2q+1)}(\xi,s)= 
 \int\limits_{\tilde{C}(s)} d\eta_1  \ldots
 \int\limits_{\tilde{C}(\eta_{q})} d\eta_{q+1} 
\tilde{\omega}(\xi-\eta_1+\eta_2)
 \int\limits_{\infty}^{\xi-\eta_1} d\xi_1 \ldots
\int\limits_{\infty}^{\xi_{q-1}} d\xi_q  \nn \\
 \tilde{\omega}(\xi_1 +\eta_2)
\tilde{\omega}(\xi_1 +\eta_3) \cdot \ldots \cdot
\tilde{\omega}(\xi_q +\eta_{q+1})\tilde{\omega}(\xi_q)(2s-2\eta_1)^{2q+1}
\frac{I_{2q+1}(z_{2q+1}^{\fr})}{z_{2q+1}^{\frac{2q+1}{2}}}  \nn \\
z_{2q+1}= 
4(s-\eta_1)\Omega(\xi) + 8(s-\eta_1) \sum_{p=0}^q
\left( \Omega(\xi_p + \eta_{p+2}) - \Omega(\xi_p + \eta_{p+1}) \right), 
 \nn \\
\xi_0 \equiv \xi ,\;\;\;\;\;
\eta_{q+2} \equiv 0 , \;\;\;\;\; q=0,1,2,...\;\;\;\;\;\;\;\;\;\; \nn 
\ee
 
The functions $I_q(x)$, $q \geq 0$, in
(\ref{A1.12}) are the modified Bessel functions (of the first kind,
see \cite{31} p.5, formula (12)). The results (\ref{A1.12}) have been
obtained from (\ref{A1.10}) by using repeatedly the following sum rule
\cite{31}: 

\be\label{A1.13}
\sum_{k\geq0}\frac{1}{k!}(\frac{t}{2})^k z^{-\frac{\nu +k}{2}}I_{\nu+k}
(z^{\fr}) =
(z+t)^{-\frac{\nu}{2}}I_{\nu}\left((z+t)^{\fr}\right)
\ee
valid for any $\nu$.  

The formulae (\ref{A1.12}) provide us with the general forms of 
$\tilde{\Phi}_1^{(q)}(\xi,s)$'s. However, for the singularity
analysis of the latter the more convenient representation for
them is the following recurrent one: 

\be\label{A1.14}
\tilde{\Phi}_1^{(2q+2)}(\xi,s)=
- \int\limits_{\tilde{C}(s)} d\eta
 \int\limits_{\tilde{C}(\eta)} d\eta' 
 \int\limits_{\tilde{\gamma}(\eta)} d\eta_1
 \tilde{\omega}(\xi_1)\tilde{\omega}(\xi_1 -\eta')(2s-2\eta) \nn \\
\tilde{\Phi}_1^{(2q)}(\xi_1 -\eta',\eta- \eta')
\frac{I_{1}\left(
\sqrt{4(s-\eta)(\Omega(\xi)-2\Omega(\xi_1)+\Omega(\xi_1-\eta'))}\right)}
{\sqrt{4(s-\eta)(\Omega(\xi)-2\Omega(\xi_1)+\Omega(\xi_1-\eta'))}} \nn \\
\tilde{\Phi}_1^{(2q+1)}(\xi,s)= 
- \int\limits_{\tilde{C}(s)} d\eta
 \int\limits_{\tilde{C}(\eta)} d\eta' 
\tilde{\omega}(\xi-\eta') \\
 \tilde{\Phi}_1^{(2q)}(\xi -\eta',\eta- \eta')
I_{0}\left(
\sqrt{-4(s-\eta)(\Omega(\xi)-\Omega(\xi-\eta'))}\right) 
\;\;\;\;\;\;\;\;\;\;\;\;\;\;\;  q=0,1,2,... \nn
\ee
where $\tilde{\Phi}_1^{(0)}(\xi,s)$ is given by (\ref{A1.12}).  

Note that (\ref{A1.14}) can be obtained from (\ref{A1.12}) and
vice versa by applying the following relations:

\be\label{A1.15}
 \int\limits_0^1 dx I_m(\sqrt{\alpha x})I_m(\sqrt{\beta(1- x)})
(\alpha x)^{\fr m}(\beta(1-x))^{\fr n}= 
2\alpha^m \beta^n \frac{I_{m+n+1}(\sqrt{\alpha+\beta})}
{(\sqrt{\alpha+\beta})^{m+n+1}} \\
\frac{(s-\eta)^n}{n!}= \frac{1}{(k-1)!(n-k)!}
\int\limits_{\eta}^s d\eta'(s-\eta')^{k-1}(\eta'-\eta)^{n-k} \nn
\ee

\vskip 12pt
{\it A1.3. Analytic properties of the functions 
$\tilde{\Phi}_1^{(0)}(\xi,s)$}
\vskip 12pt

 Since each of the functions 
$I_q(z_q^{1/2})/z_q^{q/2}$, $q \geq 0$, is an entire
function of its argument then it follows from (\ref{A1.12}) that
possible singularities of $\tilde{\Phi}_1^{(q)}(\xi,s)$ are generated by the
(known) singularities of the functions $\tilde{\omega}(\eta)$ and 
$\Omega(\eta)$ and their
integrations present in (\ref{A1.12}). However, it can be easily
checked that the conditions:

\be\label{A1.16}
\Re \xi >0 &\mbox{and} &\Re s < \Re \xi
\ee
 determine the domain where the
integrands in (\ref{A1.12}) are holomorphic. Therefore, this is also
the domain of holomorphicity of $\tilde{\Phi}_1^{(q)}(\xi,s)$ since all the
integration paths in (\ref{A1.1}), (\ref{A1.12}) can be chosen to lie
completely in this domain.

Let us note, however, that as it follows from (\ref{A1.12}) each
$\tilde{\Phi}_1^{(q)}(\xi,s)$, $q \geq 1$ can be continued
analytically from the domain (\ref{A1.16}) to any point $\xi$ of the
$\xi$-Riemann surface of $\tilde{\Phi}_1^{(q)}(\xi,s)$  (obtained for 
fixed $s$) if the distribution of branch points of $\tilde{\omega}(\xi)$
along a path of the corresponding analytical continuation is such that the 
distance of any of them from the path is greater than $|s|$. This statement
is the direct conclusion from the corresponding formulae in
(\ref{A1.12}) since all the integrations on the $\xi$-Riemann surface
present there are performed inside a strip no wider than $|s|$.

Let us note further that for the polynomial potentials the
branch points of $\tilde{\omega}(\eta)$ are isolated and on each sheet of the
$\xi$-Riemann surface of $\tilde{\omega}(\eta)$ their numbers are finite. The
distances between them on each sheet are nothing but the
corresponding distances between turning points measured by the
action. Therefore, there is the smallest distance $d$ among them.
If we take, therefore, $s$ in (\ref{A1.12}) such that $|s|<d'<d/2$ then we
can penetrate by paths of the analytical continuations the whole
$\xi$-Riemann surface of $\tilde{\omega}(\xi)$ if the former is 
deprived all the
circular vicinities of radius $d''$, $d'<d"<d/2$, centered at each
branch point of the surface. We shall denote the corresponding
part of the $\xi$-Riemann surface as {\bf R}(d").  

Consider now a question
of convergence of the series in (\ref{A1.11}). We shall show below
that the series is convergent absolutely and uniformly in the
domain {\bf R}(d"). It means that the series (\ref{A1.11}) determines 
$\tilde{\Phi}_1(\xi,s)$ as the holomorphic function in these domains.

To this end
let us note that if $|s|$ is chosen to satisfy the condition
$|s|<d'<d"$ all the integration paths $\tilde{\gamma}(\xi-\eta_1)$ can be 
deformed then to lie inside an infinite strip $S(\xi,s)$ bounded by the paths
$\tilde{\gamma}(\xi)$ and $\tilde{\gamma}(\xi-s)$ so having the width $|s|$
with the one end
of the strip being placed at the infinity $\infty_1$ and the other one
being a segment $(\xi,\xi-s)$. The latter bound can be chosen as such
because the path $\tilde{C}(s)$ can be deformed to a segment (with its
ends anchored at the origin and at $s$). Introducing now the
following functions: 

\be\label{A1.17}
|\tilde{\omega}|(\xi_r,\eta_1)= \limsup\limits_{\eta\in \tilde{C}(\eta_1)}
 |\tilde{\omega}(\xi_r +\eta_1)| \\
|\tilde{\rho}|(\xi,\eta_1)= \int\limits_{\tilde{\gamma}_1(\xi)}
|d\xi_r| |\tilde{\omega}|(\xi_r ,\eta) \nn
\ee
we have: 

\be\label{A1.18}
|\Omega(\xi_r+\eta)| <  \tilde{\rho}(\xi,\eta_1), \;\;\;\;\;\;\;\;\;\;
\eta\in \tilde{C}(\eta_1) , \;\;\;\;\;\;\; \xi_r \in \tilde{\gamma}_1(\xi)
\ee
and for $q$ large enough:

\be\label{A1.19}
|z_q| < 8(q+1)|s-\eta_1| \tilde{\rho}(\xi,\eta_1) \\
|2^q z_q^{-\frac{q}{2}}I_q| <  \frac{1}{q!}\exp(2|s-\eta_1| 
\tilde{\rho}(\xi,\eta_1)) \nn
\ee
so that: 

\be\label{A1.20}
|\tilde{\Phi}_1^{(2q)}(\xi,s)|< \left((2q)!q!(q-1)!\right)^{-1}
\int\limits_0^{|s|} dx x^{q-1}(|s|-x)^{2q}  \nn \\
 \cdot \left(
\int\limits_{\tilde{\gamma}_1(\xi-\eta_1)}
|d\eta| |\tilde{\omega}|^2 (\eta,\eta_1) \right)^q
\exp(2(|s|-x) \tilde{\rho}(\xi,\eta_1)) \\
|\tilde{\Phi}_1^{(2q+1)}(\xi,s)|< \left((2q+1)!(q!)^2 \right)^{-1}
\int\limits_0^{|s|} dx x^{q}(|s|-x)^{2q+1} |\tilde{\omega}|(\eta,\eta_1)
  \nn \\
 \cdot \left(
\int\limits_{\tilde{\gamma}_1(\xi-\eta_1)}
|d\eta| |\tilde{\omega}|^2 (\eta,\eta_1) \right)^q \exp(2(|s|-x)
\tilde{\rho}(\xi,\eta_1)) \nn
\ee
where $x = |\eta_1|$.

Introducing yet: 

\be\label{A1.21}
|{\omega}|(\xi,s)= \limsup\limits_{\eta_1 \in \tilde{C}(s)}
 |\tilde{\omega}|(\xi,\eta_1) \nn \\
\rho(\xi,s)= \limsup\limits_{\eta_1 \in \tilde{C}(s)}
\tilde{\rho}(\xi,\eta_1)  \\
Q(\xi,s)= \limsup\limits_{\eta_1 \in \tilde{C}(s)}
\int_{\tilde{\gamma}_1(\xi-\eta_1)}
|d\eta| |\tilde{\omega}|(\eta ,\eta_1) \nn
\ee
we obtain finally for $q \to \infty$:

\be\label{A1.22}
|\tilde{\Phi}_1^{(2q)}(\xi,s)|< \frac{|s|^{3q}}{(3q)! q!}
Q^{2q}(\xi,s) e^{2|s|\rho(\xi,s)} \\
|\tilde{\Phi}_1^{(2q+1)}(\xi,s)|< 
 \frac{|s|^{3q+2}}{(3q+2)! q!}
Q^{2q}(\xi,s)  |{\omega}|(\xi,s) e^{2|s|\rho(\xi,s)}  \nn
\ee

The bounds (\ref{A1.22}) show clearly that the series (\ref{A1.11}) is
convergent in the assumed domain {\bf R}(d") since 
$Q(\xi,s)$, $\rho(\xi,s)$ and $|\omega|(\eta,s)$ are finite there.

\section*{ Appendix 2}
\renewcommand{\theequation}{A2.\arabic{equation}}
\zero

If $x_p$ is a simple zero of $q(x)$ then the point $\xi_p =\xi(x_0, x_p)$
 is the branch point for the function $\tilde{\omega}(\xi)$ defined by
(\ref{2.5}) which can be expounded around the point $\xi_p$ into the
following series: 

\be\label{A2.1}
\tilde{\omega}(\xi) =\sum_{k\geq -3}\tilde{\omega}_k(\xi_p)(\xi-\xi_p)^{2k/3}
\ee
The coefficients $\tilde{\omega}_k(\xi_p)$ in (A2.1) are defined
by the identity: $\tilde{\omega}(\xi(x_0, x)) \equiv \omega(x) q^{-\fr}(x)$
 and the following expansions of 
$\xi(x_0, x)$ and $\omega(x) q^{-\fr}(x)$ (see \mref{2.4}) around $x_p$:

\be\label{A2.2}
\xi(x_0,x) -\xi_p =\sum_{k\geq 0}\xi_k(x_p)(x-x_p)^{k+\frac{3}{2}} \nn \\
\mbox{and} & \\
\omega(x) q^{-\fr}(x) =\sum_{k\geq 0}\omega_k(x_p)(x-x_p)^{k-3} \nn 
\ee
In particular, the coefficient at the most singular term in
(A2.1) $\tilde{\omega}_{-3}=- 5/36$ i.e. it is potential independent. It
depends, however, on the multiplicity of zero of $q(x)$ at $x_p$,
namely, $\tilde{\omega}_{-3} = -n(n + 4)/[4(n + 2)^2]$ for the $n$-fold zero.

\section*{ Appendix 3}
\renewcommand{\theequation}{A3.\arabic{equation}}
\zero

 We establish here the Riemann surface structure of
$\tilde{\Phi}_1^{(q)}(\xi,s)$ for the linear and harmonic potentials.
\vskip 12pt
{\it 3.1 The linear potential }
\vskip 12pt

According to Section 4.1 the Riemann surface structure of 
$\tilde{\Phi}_1^{(1)}(\xi,s)$ for this case is determined by

\be\label{A3.1}
\tilde{\Phi}_1^{(1)}(\xi,s) = -\frac{5}{8} {\int\limits_{\tilde{C}(s)}}
d\eta \frac{s-\eta}{(\xi-\eta)^2}
\frac{I_1 \left(\left(
\frac{5}{2}\frac{s-\eta}{\xi-\eta} - \frac{5}{4}\frac{s-\eta}{\xi}
\right)^{\fr}\right)}{\left(
\frac{5}{2}\frac{s-\eta}{\xi-\eta} - \frac{5}{4}\frac{s-\eta}{\xi}
\right)^{\fr}}
\ee

From (\ref{A3.1}) it follows that its subintegral function is singlar
at $\xi=\eta$ and at $\xi=0$ where it behaves as 
$e^{\pm(\xi-\eta)^{-\fr}}(\xi-\eta)^{-\frac{7}{4}}$ and 
$e^{\pm\xi^{-\fr}}\xi^{\fr}$ respectively. The
$\eta$-integration generates only a singularity at $s=\xi$ (by the
EP-mechanism) leaving the singularity at $\xi=0$ nad its character
unchanged. Therefore, assuming $\xi$ to be continued to the sector 2
(along the canonical path) the corresponding first sheets of the
Riemann surface look as in Fig. 10a,b.  

Consider now $\tilde{\Phi}_1^{(2)}(\xi,s)$.
Its Riemann surface structure is defined by: 

\be\label{A3.2}
\tilde{\Phi}_1^{(2)}(\xi,s) = -\frac{25}{64} {\int\limits_{\tilde{C}(s)}}
d\eta  {\int\limits_{\tilde{\gamma}(\xi)}}d\xi_1
\frac{(s-\eta)^2}{(\xi-\eta)^2\xi_1^2}
\frac{I_2 (z^{\fr}}{z} \\
z=
\frac{5}{4}(s-\eta)\left(\frac{1}{\xi} -\frac{2}{\xi_1}+ \frac{2}{\xi_1-\eta}
\right) \nn
\ee

As it follows from (A3.2) the subintegral function is singular at
$\xi_1=\eta$, $\xi=0$ and at $\xi_1=0$ behaving there as 
$e^{\pm(\xi_1-\eta)^{-\fr}}(\xi_1-\eta)^{-\frac{7}{4}}$, 
$e^{\pm \xi^{-\fr}} \xi^{\fr}$ and $e^{\pm \xi_1^{-\fr}}\xi_1^{-\frac{7}{4}}$ 
respectively.  

The $\xi_1$-integration in (\ref{A3.2}) generates the EP-singularities 
at $\xi=\eta$ and at $\xi=0$ but also the P-singularity at $\eta=0$ (when the
singularity at $\xi_1=\eta$ move around the end point of 
$\tilde{\gamma}(\xi)$ clockwise
pinching the latter against the singular point $\xi_1=0$).  

The final $\eta$-integration in (\ref{A3.2}) generates the EP-singularity
 at $s=\xi$ and $s=0$ and the P-singularity at $\xi=0$. Therefore, the 
'closest' singularities of $\tilde{\Phi}_1^{(2)}(\xi,s)$ are the following: 

\be\label{A3.3}
\xi=0, & \xi=\eta, & s=0
\ee

Let us make a general note that the EP-mechanism repeats the distribution of
the branch points and cuts whilst the P-one generates new branch
points on the Riemann surfaces obtained by the EP-mechanism
always however enforcing specific ways of moving around the
singularities generated by the EP-mechanism.  

All the singularities in (\ref{A3.3}) are the root branch points (of the forth
order) accompanied by essential singularities as we have
mentioned above. The singularity at $s=0$, however, to be reached
needs to round the branch point at $s=\xi$ moving clockwise i.e. it
lies on the sheet opened by the latter branch point. This
singularity is therefore a consequence of the singularity at $\xi=0$.  

A similar note concerns the singularity at $\xi=0$. In fact
there are two such singularities the one on the sheet shown in
Fig. 10a (arising by the EP-mechanism) and the second one at the
sheet opened by the branch point at $\xi=\eta$ i.e. to reach it one
needs to round this point clockwise.  

One can conclude therefore
that the P-mechanism applied once has generated a singularity at
$s=0$ and applied twice has generated a new singularity at þ$\xi=0$ (on
a different sheet) from the old one. It is clear that this
mechanism will proliferate the last singularity on all the
sheets of the Riemann surface of $\tilde{\Phi}_1(\xi,s)$ except the sheet we
have started with on which the point $s=0$ is regular for
$\tilde{\Phi}_1(\xi,s)$.  

Now we can use the formulae (\ref{A1.14}) to prove the form
of the first sheet of the Riemann surface as shown on Fig.10a,b. 
Namely, assuming for $\tilde{\Phi}_1^{(2q)}(\xi,s)$ the form of this sheet
shown in the last figure we deduce that it remains unchanged for
$\tilde{\Phi}_1^{(2q+2)}(\xi,s)$ whilst it is deprived of the singularity at 
$s=0$ for $\tilde{\Phi}_1^{(2q+1)}(\xi,s)$.  

Indeed, consider the subintegral function
in the first of the formulae (\ref{A1.14}) defining 
$\tilde{\Phi}_1^{(2q+2)}(\xi,s)$. It
has singularities at the following points: 

\be\label{A3.4}
\xi=0, \;\;\;\;\; \xi_1= 0, \;\;\;\;\; \xi_1- \eta'=0,\;\;\;\;\;\xi_1- \eta=0, \;\;\;\;\; \eta-\eta'=0
\ee
shown on Fig. 19a for
the $\xi_1$-Riemann surface (when the rest of the variables are fixed).  

\begin{eqnarray*}
\begin{tabular}{cc}
\psfig{figure=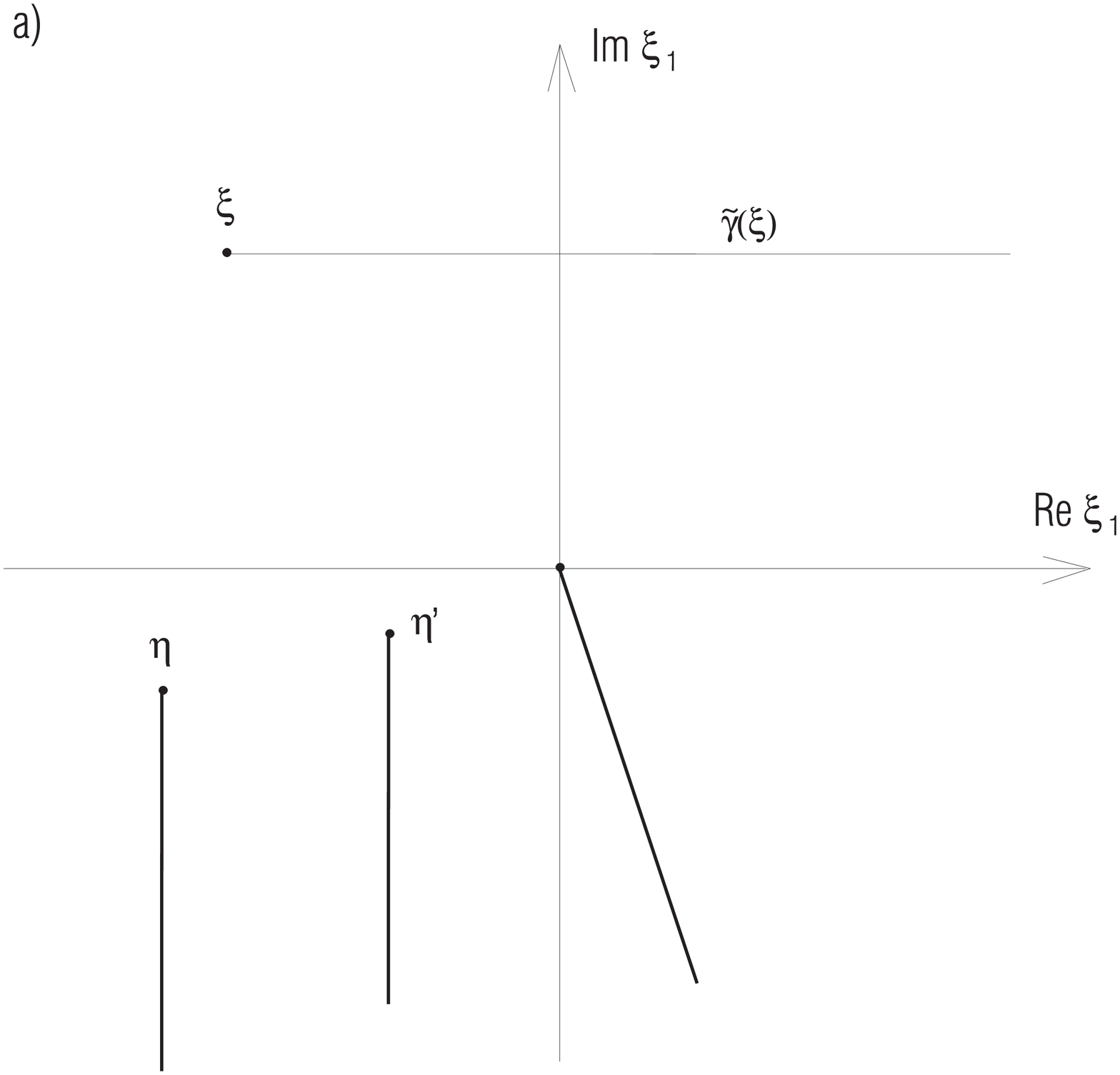,width=7cm} & \psfig{figure=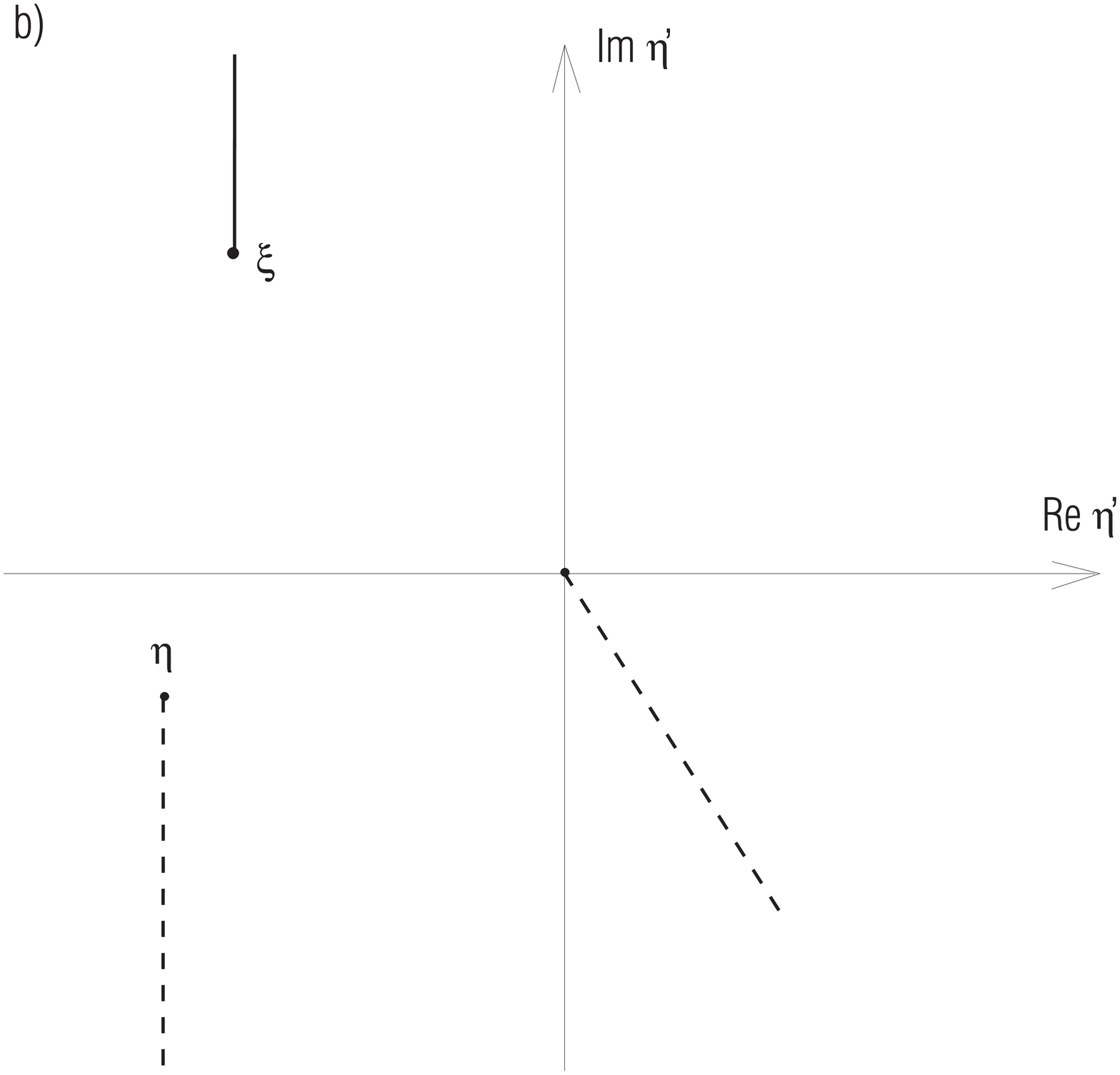,width=7cm} 
\end{tabular} \\
\parbox{14cm}{Fig. 19. $\;\;\;$ The  "first sheets" singularity pattern
of the subintegral function defining $\tilde{\Phi}_1^{(2q+1)}(\xi,s)$ 
for the linear potential before the $\xi_1$-integration (a) and after it (b)}
\end{eqnarray*}
\vskip 18pt

The $\xi_1$-integration provides us with the
EP-singularities at $\xi=0$, $\xi-\eta=0$ and $\xi-\eta'=0$ and with 
the P-ones at $\eta=0$, $\eta'=0$ (when the point $\xi$ is rounded by $\eta$ 
and $\eta'$ clockwise to touch the point $\xi=0$ by the latter) and at 
$\eta=\eta'$ (when the point $\eta(\eta')$ rounds $\xi$ clockwise 
(anticlockwise) to touch $\eta'(\eta)$). We get
in this way the singularity pattern before the $\eta'$-integration
shown in Fig. 19b where the singularities at $\eta'=0$ and at $\eta'=\eta$
are screened by the $\xi$-cut and to get them one has to go around $\xi$
clockwise or anticlockwise respectively.  

The $\eta'$-integration
therefore does not do much now providing us with the singularity
at $\eta=\xi$ and at $\eta=0$ (by the EP-mechanism) and at $\xi=0$ (by the
P-one) with the latter singularity placed on a sheet opened by
the singularity at $\xi=\eta$.  

The final $\eta$-integration repeat only the
singularity pattern described just above so we are left with the
distribution of the singularities as shown in Fig. 10a,b.

Consider now the second formula (\ref{A1.14}). There is no the
$\xi_1$-integration and therefore the singularity at $\eta'=0$ is not
generated and the other singularities at $\xi=0$, $\eta=0$ and $s=0$ can
not be generated as well by the further $\eta'$- and $\eta$-integrations.
Besides the generation of the singularities at $\xi=s$ goes exactly
in the same way so that the final picture of the corresponding
Rieman surface is the same as in Fig. 10a,b except of missing of
the respective singularities at $\xi=0$ and $s=0$ on the lower sheets.

\vskip 12pt
{\it 3.2 The harmonic potential}
\vskip 12pt

 We assume here $\tilde{\gamma}_1(\xi)$ to be continued
canonically to the sector 3 of Fig. 11 and we put 
$\xi(i)=\int_{-i}^i \sqrt{x^2+1}dx=\zeta$. Nor $\tilde{\omega}(\xi)$
nor ${\Omega}(\xi)$are now simple functions of $\xi$. $\tilde{\omega}(\xi)$ 
is periodic (with its period
$2\zeta$ acting between different sheets of the infinitely sheeted
Riemann surface on which this function is defined) whilst ${\Omega}(\xi)$
is not.  

\begin{eqnarray*}
\begin{tabular}{cc}
\psfig{figure=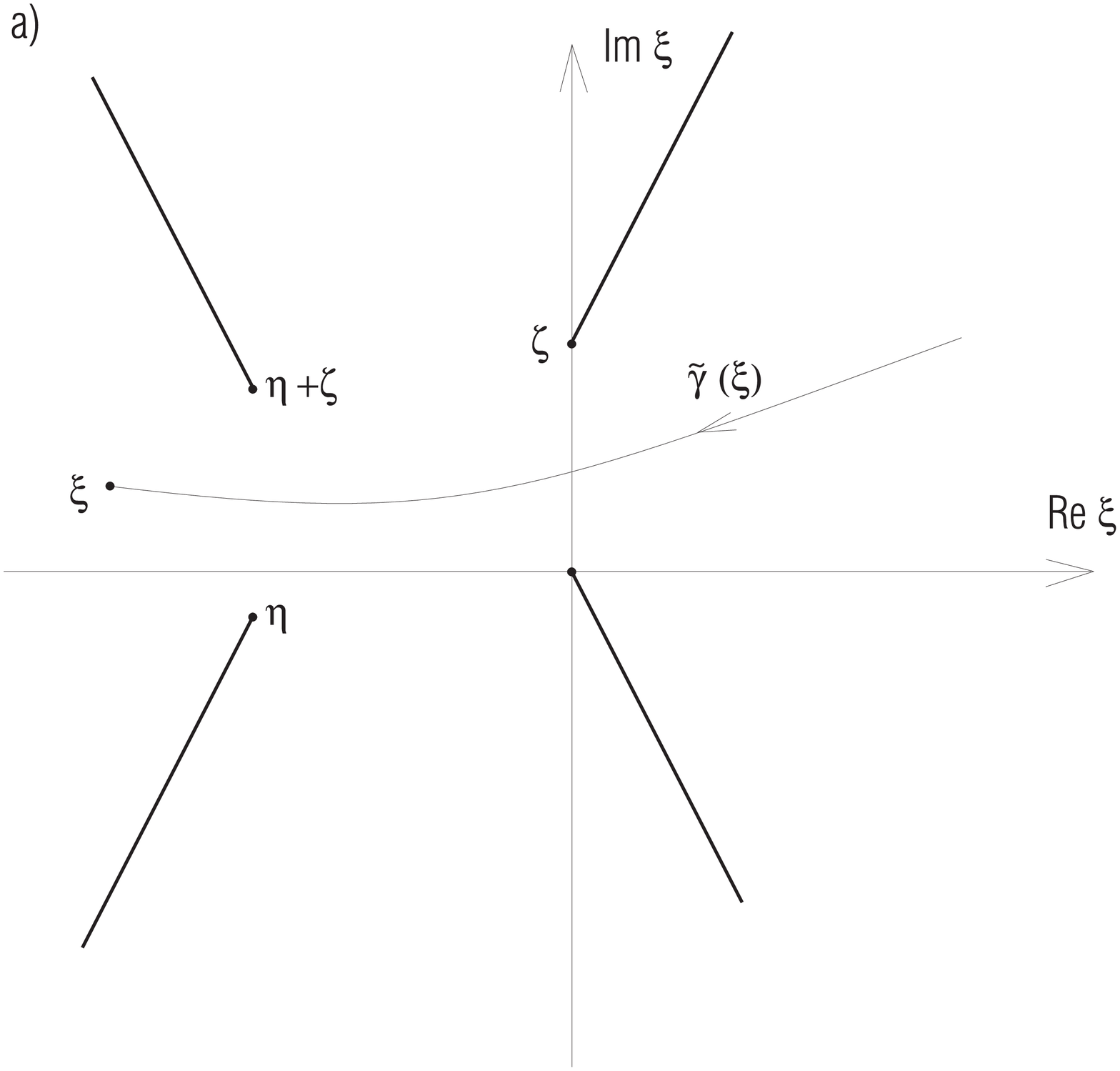,width=7cm} & \psfig{figure=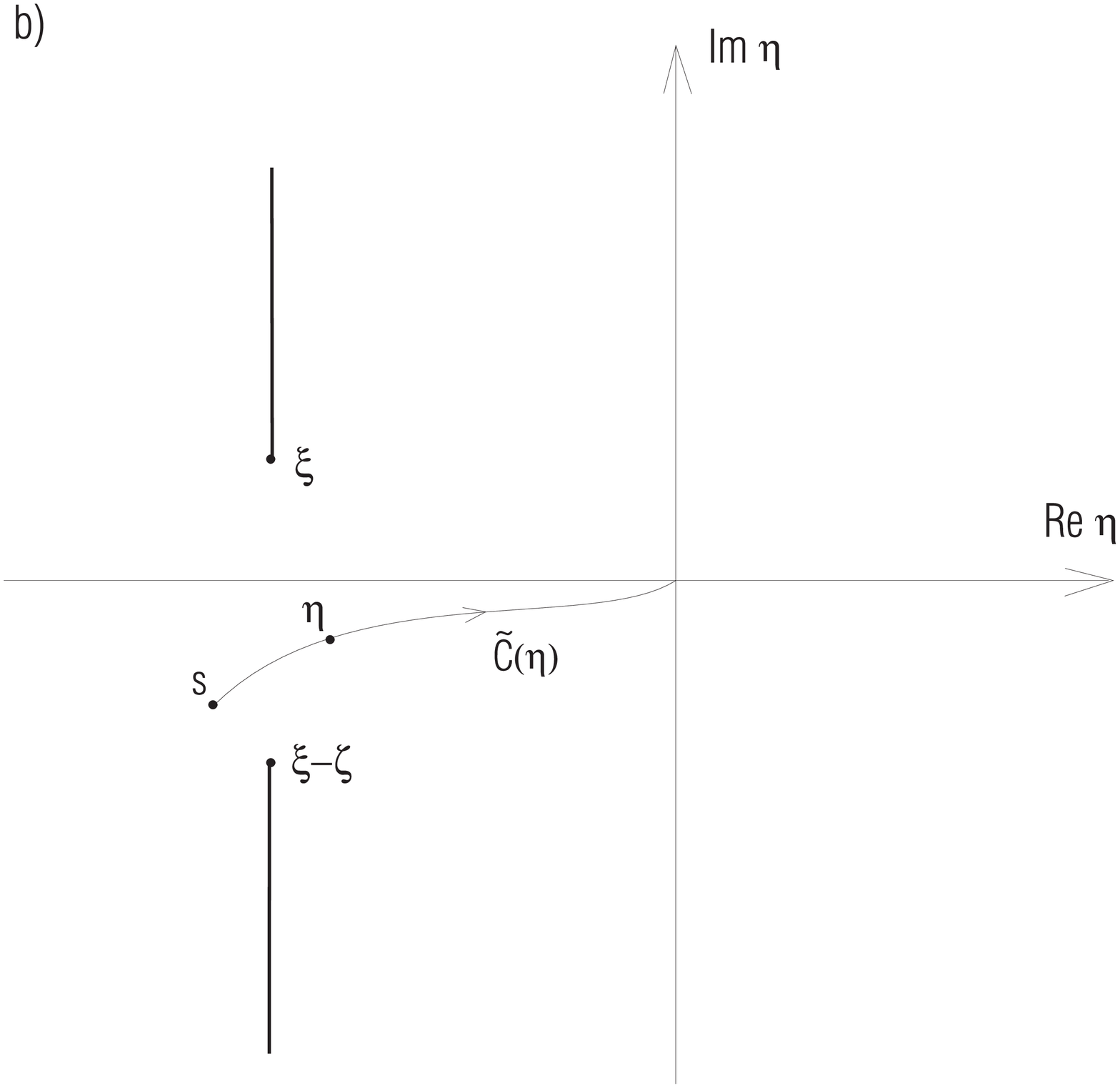,width=7cm} 
\end{tabular} \\
\parbox{14cm}{Fig. 20. 
$\;\;\;$ The singularities of the subintegral function 
in \mref{4.1} defining $\tilde{\Phi}_1^{(1)}(\xi,s)$ 
for the harmonic potential}
\end{eqnarray*}
\vskip 18pt

Consider however again $\tilde{\Phi}_1^{(1)}(\xi,s)$ as given by (\ref{4.1}).
The closest singularities of the subintegral function are shown
in Fig. 20a,b i.e. they are:

\be\label{A3.5}
\xi=0, \;\;\;\;\; \xi -\zeta = 0, \;\;\;\;\; \xi - \eta=0,\;\;\;\;\;
\xi- \eta  -\zeta=0 \;\;\;\;\;\;
\ee

Therefore the $\eta$-integration in (\ref{4.1}) provides us with the
following singularities of $\tilde{\Phi}_1^{(1)}(\xi,s)$ shown in Fig. 12a,b: 

\be\label{A3.6}
\xi=0, \;\;\;\;\; \xi -\zeta = 0, \;\;\;\;\; \xi - s=0 , \;\;\;\;\;
\xi- \zeta - s=0
\ee
i.e. none the P-singularity is generated.  

Consider next $\tilde{\Phi}_1^{(2)}(\xi,s)$.
According to (\ref{4.3}) singularities of the subintegral function in
this formula are now the following: 

\be\label{A3.7}
\xi=0, \;\;\;\;\; \xi -\zeta = 0, \;\;\;\;\; \xi_1=0,\;\;\;\;\; 
\xi_1 - \zeta=0 \\
\;\;\;\;\; \xi_1 - \eta,\;\;\;\;\; \xi_1  -\zeta - \eta =0  \;\;\;\;\; \nn
\ee
and we can use Fig. 20 to show the corresponding situation situation for the
$\xi_1$- and $\eta$-dependance by making the substitution $\xi \to \xi_1$ in
the figure.

The $\xi_1$-integration generates the following singularities: 

\be\label{A3.8}
\xi=0, \;\;\;\;\; \xi -\zeta = 0, \;\;\;\;\; \xi - \eta=0,\;\;\;\;\;
 \xi- \zeta  -\eta=0
\ee
by the EP-mechanism and 

\be\label{A3.9}
\eta = 0, & \eta -\zeta = 0,&  \eta + \zeta=0
\ee
by the P-mechanism. The latter singularities lie on the 'lower' sheets
of the $\eta$-Riemann surface.  

Finally, integrating in (\ref{4.3}) over $\eta$ we generate singularities 
of $\tilde{\Phi}_1^{(2)}(\xi,s)$ at 

\be\label{A3.10}
\xi -s =0, \;\;\;\;\; \xi -\zeta -s = 0, \;\;\;\;\; s - \zeta=0,\;\;\;\;\;
s + \zeta=0
\ee
by the EP-mechanism and at 

\be\label{A3.11}
\xi + \zeta =0, \;\;\;\;\; \xi = 0, \;\;\;\;\; \xi - \zeta = 0,\;\;\;\;\;
\xi - 2\zeta=0
\ee
by the P-mechanism. The proper distribution of these
singularities is shown on Fig. 13.

  Now we can proceed inductively assuming for 
$\tilde{\Phi}_1^{(2q)}(\xi,s)$ the singularity pattern
shown in Fig.14a,b where $\tilde{\gamma}(\xi)$ 
is the integration path in the formulae
(\ref{A1.12})-(\ref{A1.14}) and $C$ the corresponding path to recover from
${\chi}_1^{(2q)}(\xi,\lambda)$ by the Borel transformation (at $s=0$ 
$\tilde{\Phi}_1^{(2q)}(\xi,s)$ is then regular).  

Taking into account the second of the formulae (\ref{A1.14}) we see 
that the singularities of the subintegral function are determined 
mostly by its factor $\tilde{\Phi}_1^{(2q)}(\xi-\eta',\eta-\eta')$
according to which and Fig. 21a these singularity are at the
points

\be\label{A3.12}
 \xi -\eta = 0, \;\;\;\;\; \xi - \eta - \zeta =0,\;\;\;\;\;
\xi- \eta'  - k \zeta=0, \;\;\;\;\; k=-(2q-1),...,2q \\
\;\;\;\;\; \eta' -\eta +k\zeta =0 ,\;\;\;\;\; k=-(2q-1),...,(2q-1), 
\;\;\;\;\; k \neq 0 \;\;\;\;\;\;\;\;\;\;\;\;\;\;\; \nn 
\ee
 shown for the case of the corresponding $\eta'$-Riemann
surface on Fig. 21b.  

\begin{eqnarray*}
\psfig{figure=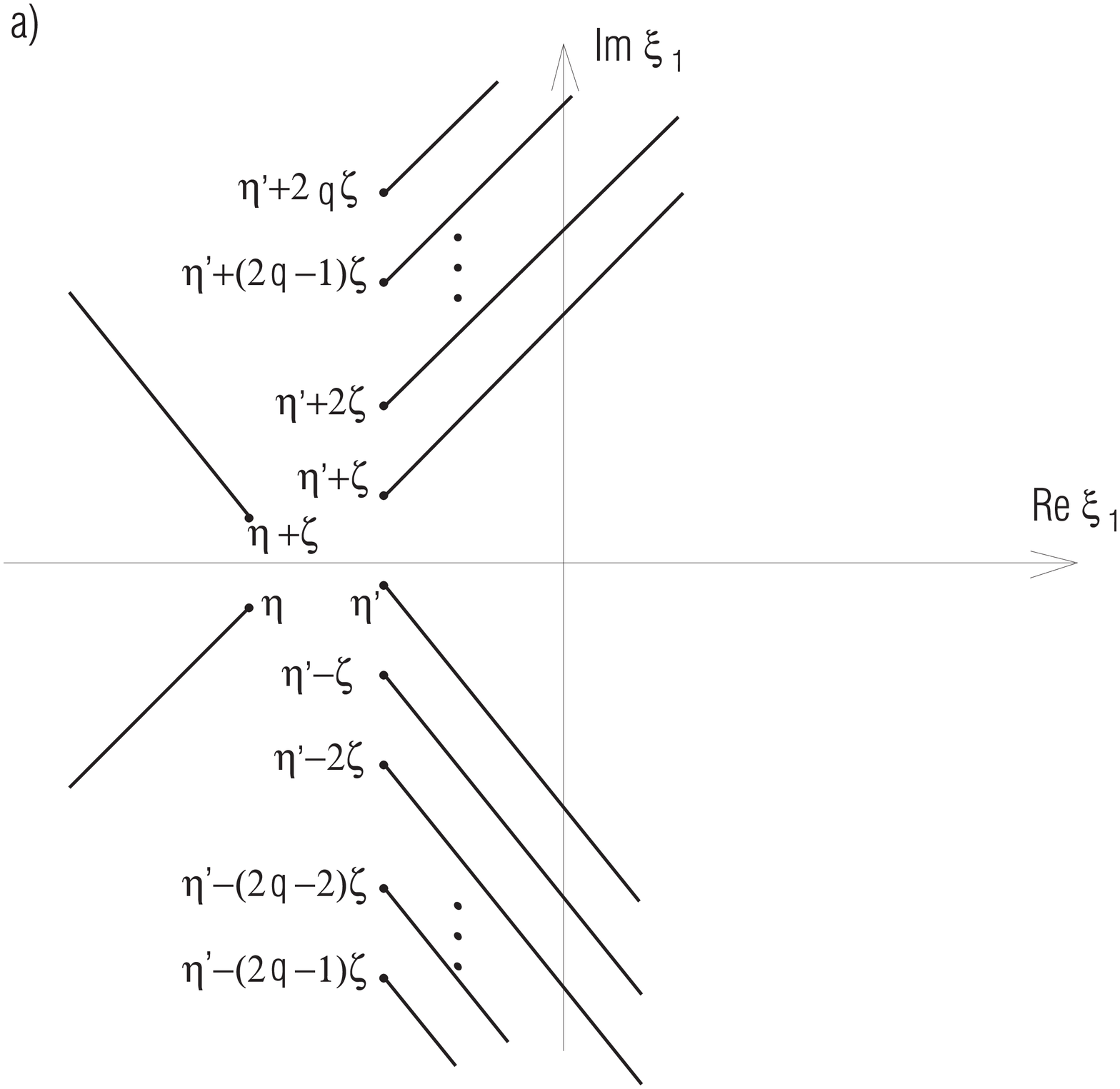,width=7cm} & \psfig{figure=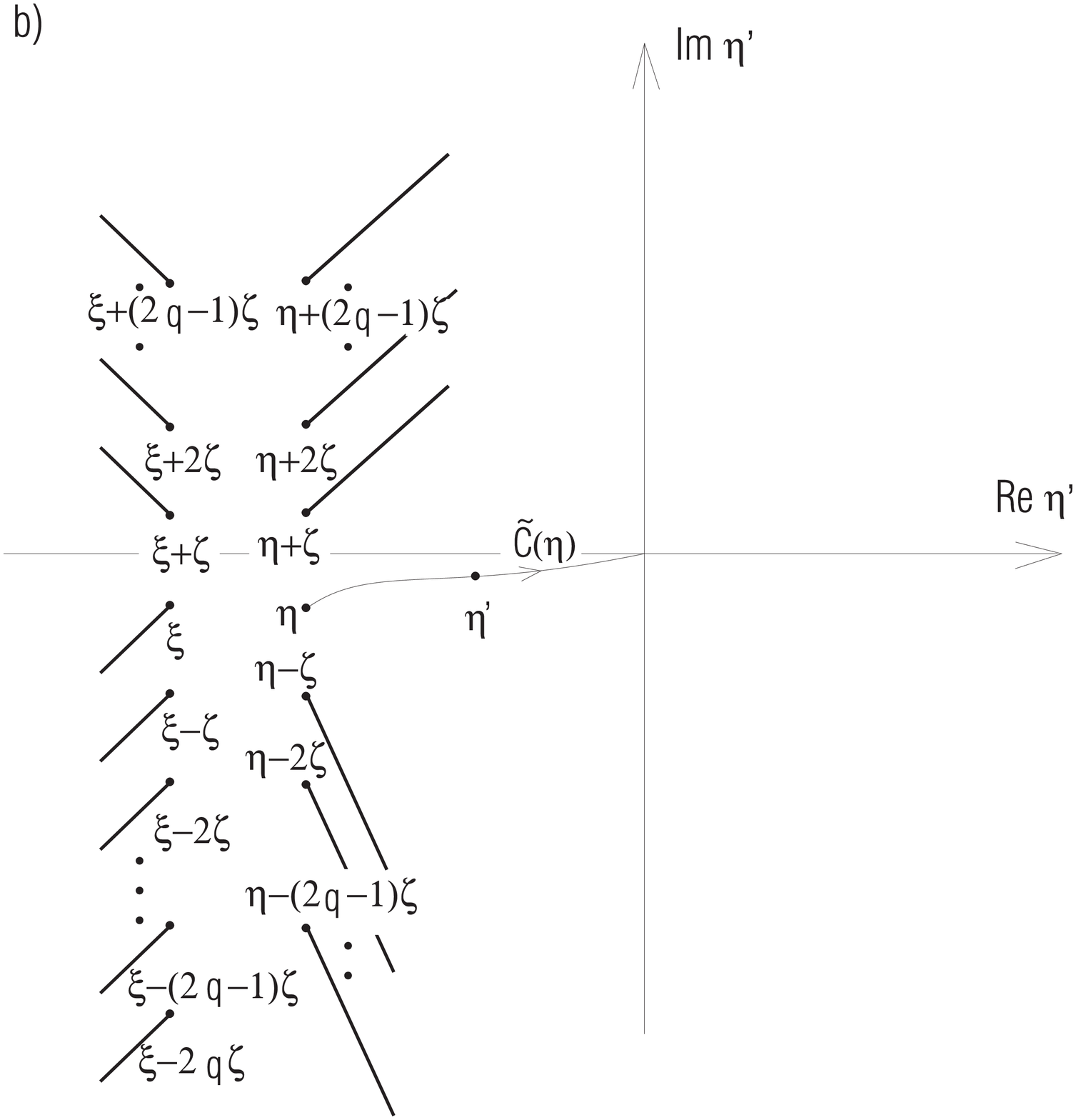,width=7cm} \\
\psfig{figure=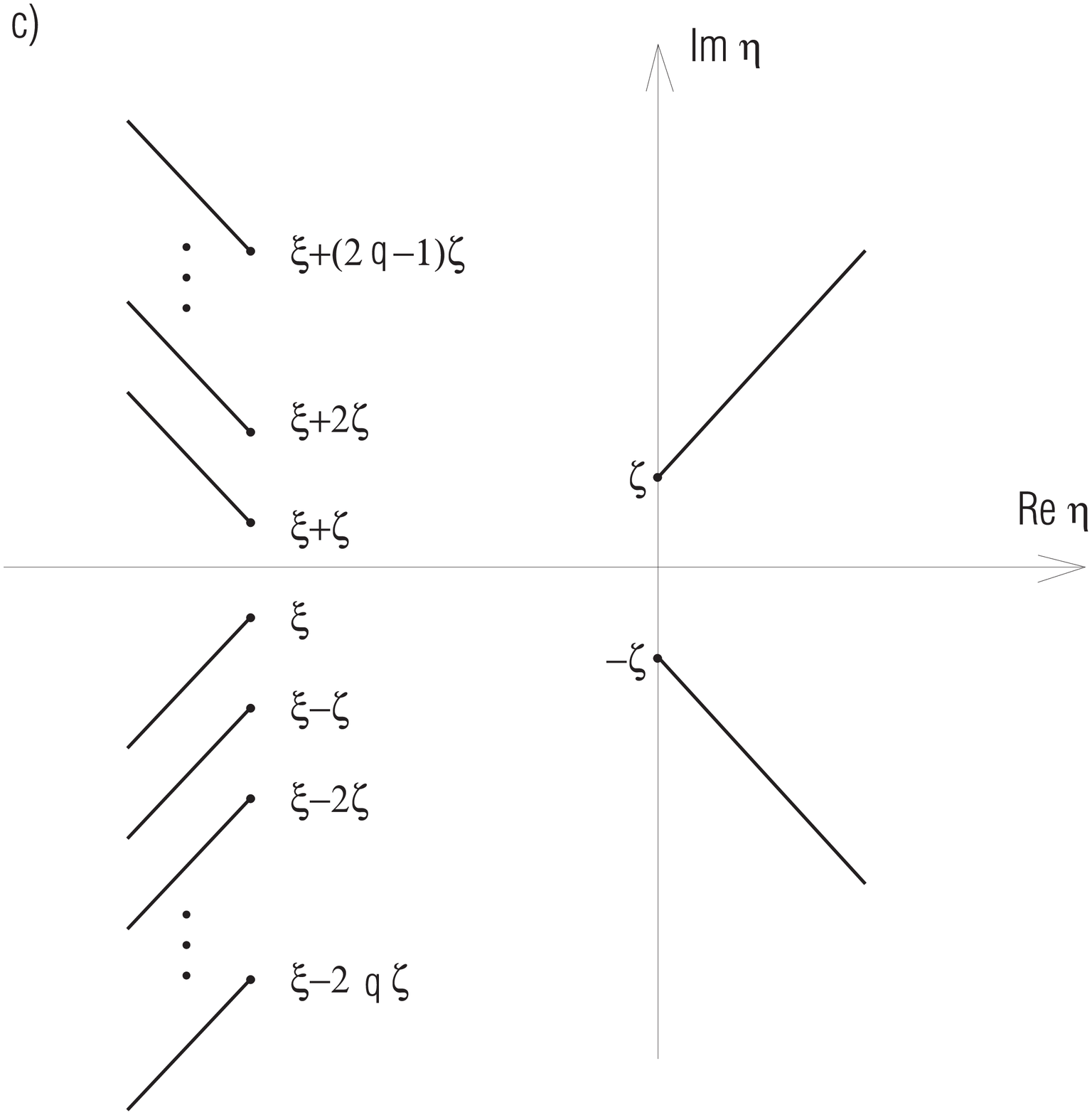,width=7cm} &
\parbox[b]{7cm}{Fig. 21. $\;\;\;$ The  singularity
structure of  $\tilde{\Phi}_1^{(2q)}(\xi-\eta',\eta-\eta')$ 
for the harmonic potential in the $\xi$-plane (a), in the $\eta'$-plane (b)
and in the $\eta$-plane after the $\eta'$-integration (c)}
\end{eqnarray*}
\vskip 18pt

Therefore, making the $\eta'$-integration we
obtain the '$\eta$-plane' singularity pattern shown in Fig. 21c on
which the $\xi$-dependent singularities are created by the
EP-mechanism whilst the two fixed ones on the imaginary axis by
the P-mechanism. Other singularities generated in the last way
appear on the lower sheets originated by the two singularities
at $\eta=\zeta$ and $\eta=\zeta'$.  

The successive $\eta$-integration does change
nothing in the $s$-variable singularity pattern (in comparison
with this on Fig. 21c) so providing us finally with its form
shown in Fig. 15b but seriously changes the original pattern of
Fig.14a. Namely, the EP-mechanism generates the $\xi$-singularities
at the points $\xi=s-(2q-1)\zeta,s-(2q-2)\zeta,...,s-\zeta,s,s+\zeta,...,
s+2q\zeta$ and by the P-mechanism at the points 
$\xi=-2q\zeta,...,-\zeta,0,\zeta,...,(2q+1)\zeta$. As the 
final result we have for $\tilde{\Phi}_1^{(2q+1)}(\xi,s)$ the picture of
Fig.15 for its both types of singularities.  

The corresponding analysis of the case 
$\tilde{\Phi}_1^{(2q+2)}(\xi,s)$ is still a little bit more
tedious but nevertheless direct due to the first of the formulae
(\ref{A1.14}). The valid singularities of the subintegral function in
this case are 

\be\label{A3.13}
\;\;\;\;\;\xi_1=0,\;\;\;\;\; \xi_1 -\eta = 0, \;\;\;\;\;\;\;\;\;
\;\;\;\;\;\;\;\;\;\;\;\;\;\;\;\;\;\;\; \nn \\ 
\xi_1 - \eta =0,\;\;\;\;\; \xi_1 - \eta - \zeta =0,\;\;\;\;\;
\xi_1- \eta'  - k \zeta=0, \;\;\;\;\; k=-(2q-1),...,2q \\
\;\;\;\;\; \eta' -\eta +k\zeta =0 ,\;\;\;\;\; k=-(2q-1),...,(2q-1), 
\;\;\;\;\; k \neq 0 \nn 
\ee

The first $\xi_1$-integration is performed on the sheet
shown in Fig. 22a. By the EP- and P-mechanism it generates $\eta$-
and $\eta'$-singularities.  Limiting ourselves to collect only these
singularities which appear on the $\eta'$-sheet on which the result
of this $\xi_1$-integration is regular at $\eta'=0$ we arrive at the
pattern shown in Fig. 22b.

\begin{eqnarray*}
\psfig{figure=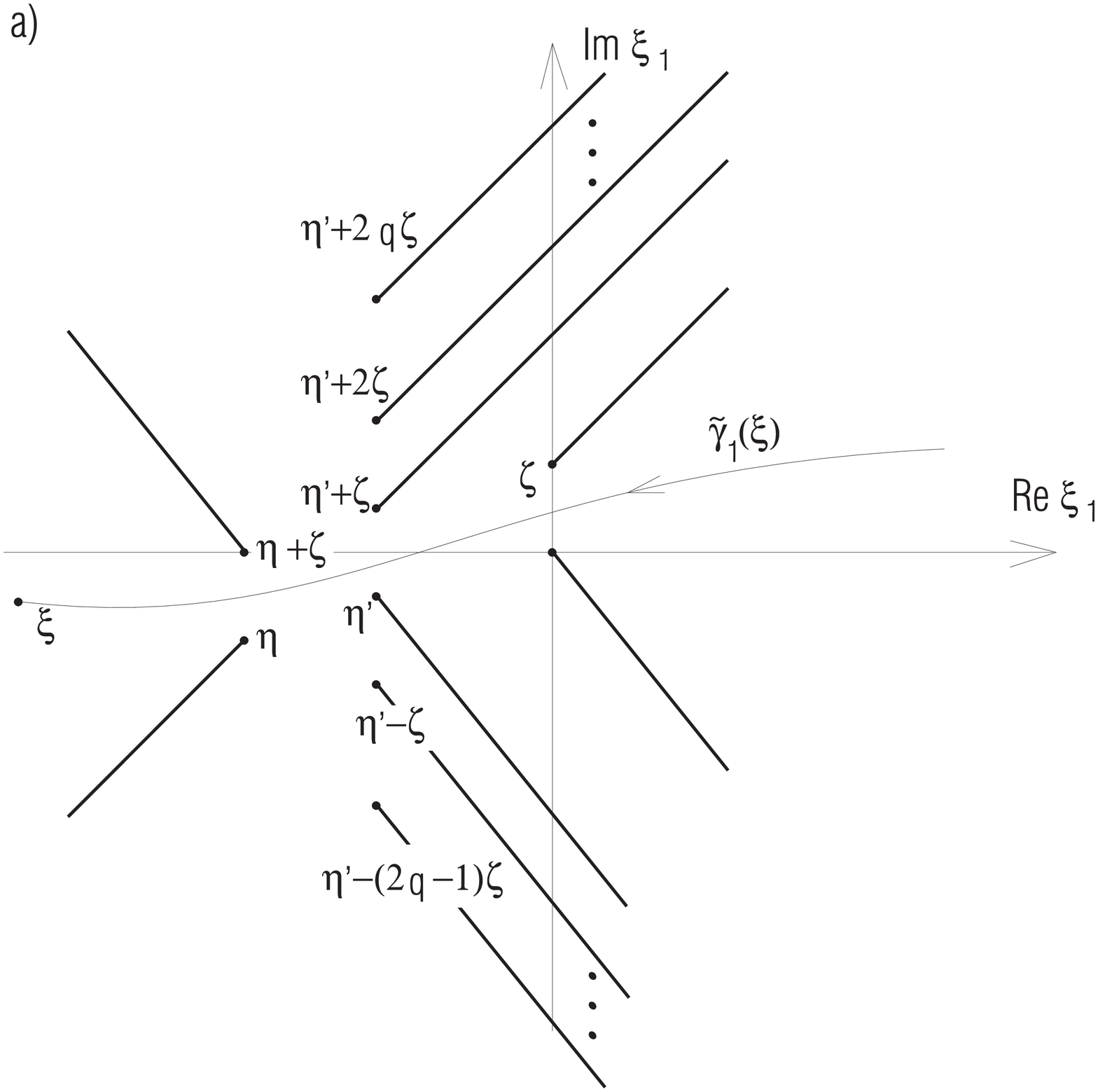,width=7cm} & \psfig{figure=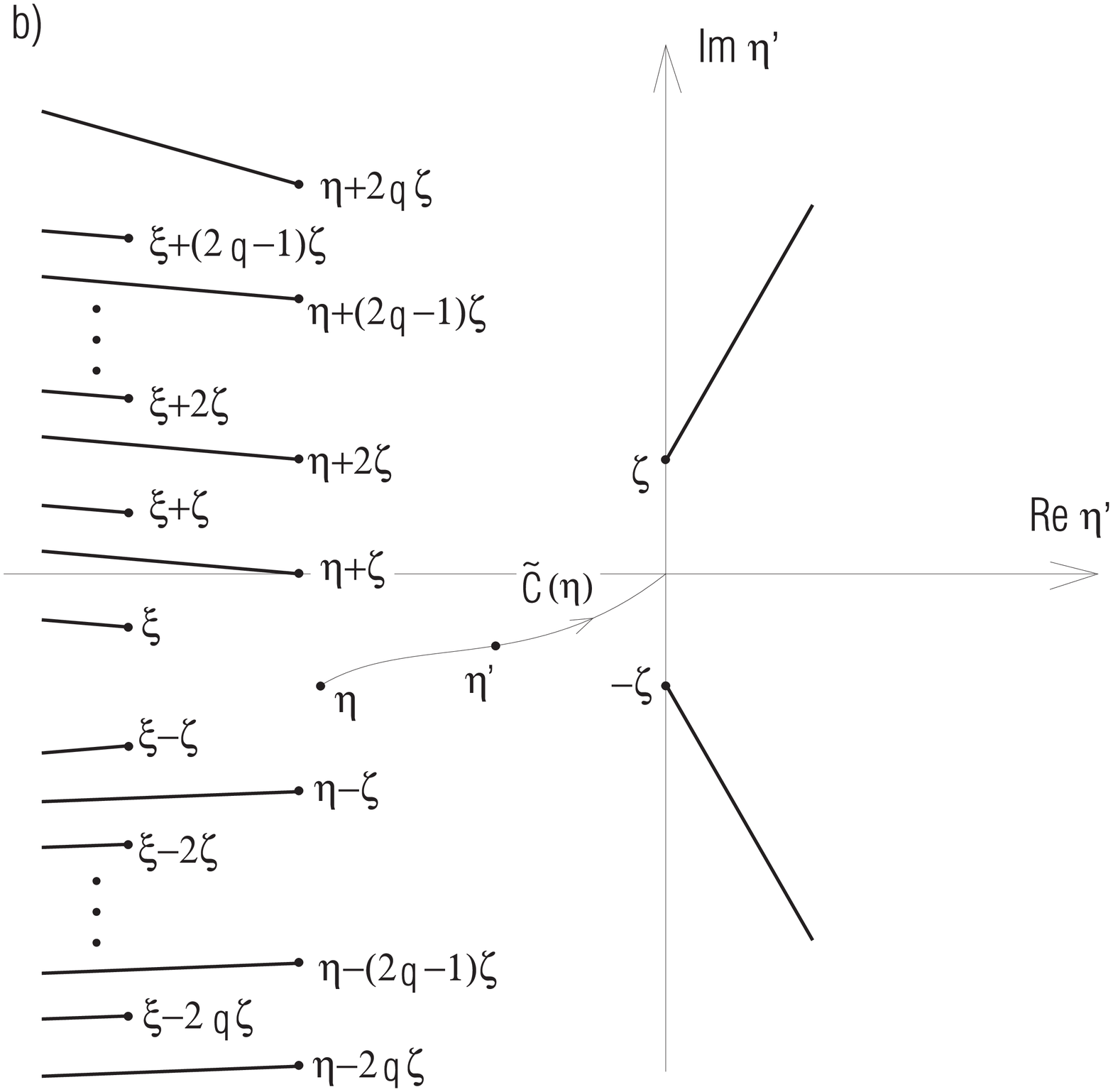,width=7cm} \\
\psfig{figure=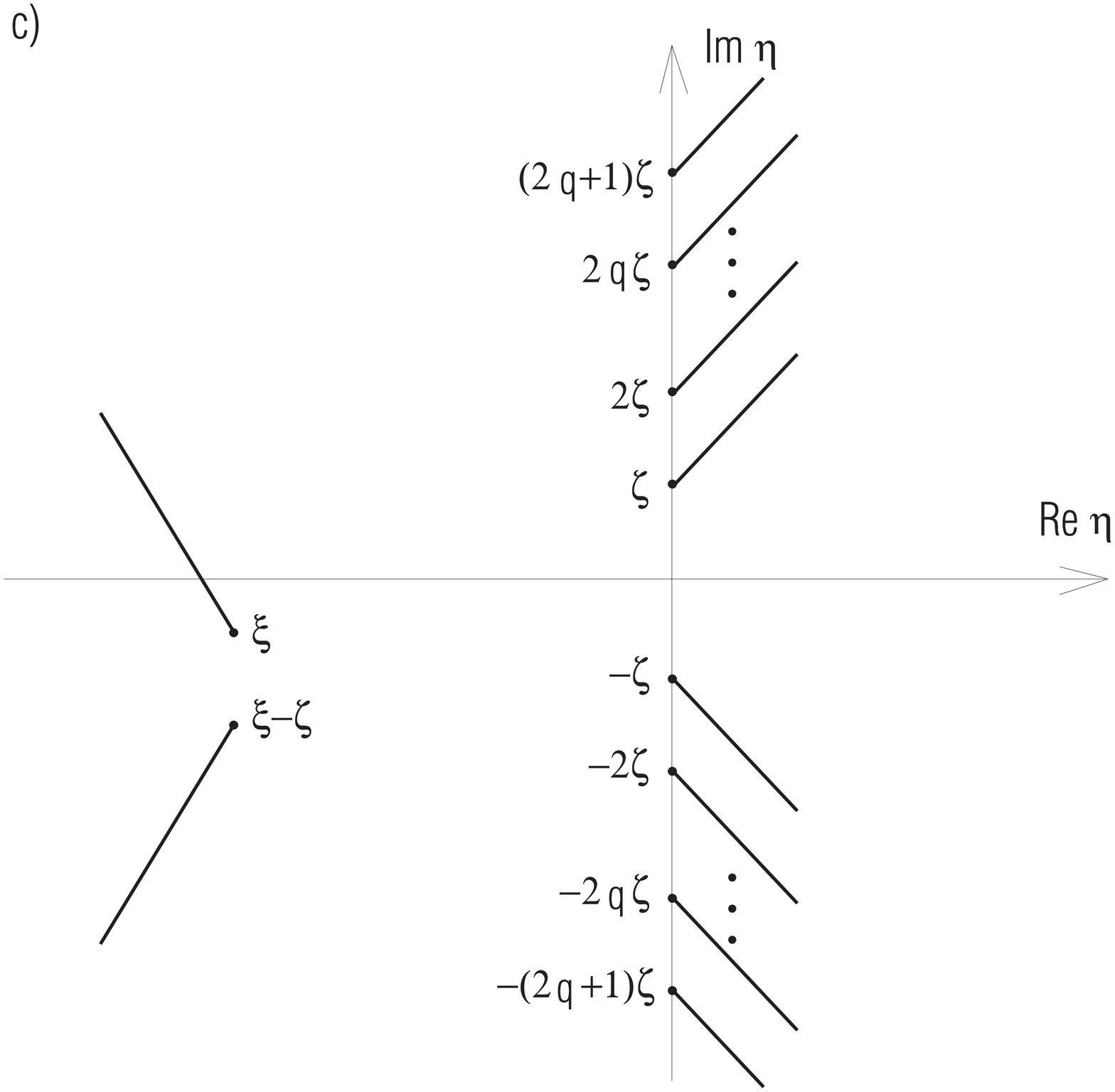,width=7cm} &
\parbox[b]{7cm}{Fig. 22. $\;\;\;$ The  subintegral singularities
in (\ref{A1.14}) defining  $\tilde{\Phi}_1^{(2q+2)}(\xi,s)$ 
in the $\xi_1$-plane (a), in the $\eta$-plane after the 
$\xi_1$-integration in \mref{A1.14} (b)
and in the $\eta$-plane (c) after the $\eta'$-integration}
\end{eqnarray*}
\vskip 18pt

The successive $\eta'$-integration leads
us to the '$\eta$-plane' pattern shown in Fig. 22c where the two
$\xi$-dependent singularities were produced by the EP-mechanism
whilst the fixed ones by the EP- and P-mechanisms simultaneously
with the exception of the highest two ones at $\eta=-(2q+1)\zeta$ and
$\eta=(2q+1)\zeta$ which are generated by the P-mechanism only.  

The final $\eta$-integration in (\ref{A1.14}) provide us with a pattern
analogous to the one of Fig. 14a by the EP- and P-mechanism and
with the pattern of Fig. 14b by the EP-mechanism when $q$ is
substituted by $q+1$.

\section*{ Appendix 4}
\renewcommand{\theequation}{A4.\arabic{equation}}
\zero

\hskip+2em  We describe here a procedure allowing us to construct
in a systematic way the optimum semiclassical representation for
the Borel summable quantity including both the main contribution
coming from the semiclassical series abbreviated at its least
term and the corresponding exponential contributions of an
arbitrary order. In its finite form the procedure provides us
with the exact formula for the quantity considered. However, if
continued infinitely the procedure give rise to the question of
convergence of the infinite functional series we get by it.  

To this goal we shall consider the basic quantity given by the
formula (\ref{2.8}). Integrating in it by parts we get 

\be\label{A4.1}
\chi_1 (\xi,\lambda)=
2\lambda \left(\sum_{k=0}^n 
\frac{(-1)^k}{(2\lambda)^{k+1}} \tilde{\chi}_1^{(k)}(\xi,0)+ 
\frac{(-1)^{n+1}}{(2\lambda)^{n+1}}
\int\limits_{\tilde{C}} e^{2\lambda s}
\tilde{\chi}_1^{(n+1)} (\xi,s) ds \right)
\ee

According to the well known prescribtion (which can be easily justified by
the analysis similar to the one performed below) we should put
$n=n_0=[\lambda|s_0|]$ in (\ref{A4.1}) where $[x]$ means the integer part 
of $x$ and $s_0$ is a singularity of
$\tilde{\chi}_1 (\xi,s)$ closest to the origin. Next we should
extract from the integral the exponentially small factor and
finally continue the procedure to the remaining Borel integral.
This could be done in the following way.  

First the $n_0 +1^{th}$ derivative of $\tilde{\chi}_1(\xi,s)$
can be given the form

\be\label{A4.2}
 \tilde{\chi}_1^{(n_0+1)}(\xi,s) =&
\frac{(-1)^{n_0+1}(n_0+1)!}{(2\pi i)}
\int_{K} \tilde{\chi}_1(\xi,s+t) t^{-n_0 -2} dt 
\ee
with the integration contour $K$ in (\ref{A4.2}) surrounding anticlockwise 
the negative half axis of the Borel plane (see Fig. 5).

As it follows from Fig. 2 $s_0=\xi-\zeta_1=\xi$ (since $\zeta_1=0$, 
see Fig. 4). Deforming the contour $K$ to surround the cuts 
generated by the points $\xi-\zeta_1$ and $\xi-\zeta_2$ of Fig. 5 
(for $\xi$ chosen as in the figure they are the unique cuts
visible in these positions) and shifting the integration
variable in the corresponding intagrals we get 

\be\label{A4.3}
 \tilde{\chi}_1^{(n+1)}(\xi,s) =
\frac{(-1)^{n_0+1}(n_0+1)!}{(2\pi i)}\sum_{j=1}^{2}
\int_{K_j} \tilde{\chi}_1(\xi-\zeta_j +t) (\xi-\zeta_j -s +t)^{-n_0 -2} dt 
\ee
where the contours $K_j$ surround (anticlockwise) the cuts whose origins are
at the point $t=0$.  

Further substituting (\ref{A4.3}) to (\ref{A4.1}) and
changing both the order of integrations in (\ref{A4.3}) and the
integration variables themselves we get as a result of these
calculations 

\be\label{A4.4}
\chi_1 (\xi,\lambda)=
\sum_{k=0}^{n_0} 
\frac{(-1)^k}{(2\lambda)^{k}} \tilde{\chi}_1^{(k)}(\xi,0)-
\sum_{j=1}^2 \frac{(n_0+1)!}{(2\lambda)^{n_0}(\xi-\zeta_j)^{n_0} }
\int\limits_{\tilde{C}} e^{2\lambda s}\kappa_j (\xi,s) ds 
\ee
where 

\be\label{A4.5}
{\kappa}_{j}(\xi,s) =
\frac{1}{2\pi i} \int\limits_{K_j} dt
\frac{ \tilde{\chi}_1(\xi,\xi-\zeta_j+t)}
{(1+\frac{t}{\xi - \zeta_j})^{n_0} } \nn \\
 \cdot \frac{1}
{\left(t+\xi -\zeta_j +\frac{n_0}{\lambda} \ln(1+\frac{t}{\xi-\zeta_j}) 
-s\right)
\left(t+\xi - \zeta_j +\frac{n_0}{\lambda} +
\frac{n_0}{\lambda}\ln(1+\frac{t}{\xi - \zeta_j}) -s\right)}  \\
 j=1,2 \nn
\ee
and where the contours $K_j$ run again around the cuts anchored at $t=0$.  

The form (\ref{A4.5}) for $\kappa$'s allows us to
continue the procedure of getting the asymptotic series
expansions for the integrals in (\ref{A4.4}) and next to abbreviate
the series at their least terms.  The latter are to be determind
by the singularities generated by the $t$-integrals in the $s$-plane
(as a result of the pinch mechanism) closest to the origin of
the plane. It is easy to see that among possible candidates for
the latter are the singularities at $s=\xi$, $\xi+n_0/\lambda$ for 
$\kappa_1(\xi,s)$ and the ones at $s=\xi-\zeta_2$, 
$\xi+n_0/\lambda-\zeta_2$ for $\kappa_2(\xi,s)$  (all the singularities
are generated by the P-mechanism at $t=0$). However, the
integrations in (\ref{A4.5}) along the corresponding cuts open
possibilities for new singularities to appear generated by the
t-singularities shared by the cuts. These possibillities still
enrich the variety of singularities which have to be taken into
account in choosing the one closest to the origin of the $s$-plane.

Therefore, to construct the representation (\ref{A4.1}) for
each of the two integrals in (\ref{A4.4}) we have to choose from the
singularities corresponding to each $\kappa$  the ones which are
closest to the origin. When these choices are done the procedure
described above can be repeated.  

Let us call $\kappa_j$, $j=1,2$, defined
by (\ref{A4.5}) the first generation family considering 
$\kappa_0(\xi,s)\equiv\tilde{\chi}_1(\xi,s)$ as
the zeroth generation one. It is clear that the general form of
the optimum semiclassical representation for $\chi_1(\xi,\lambda)$
is the following
\be\label{A4.6}
\chi_1(\xi,\lambda) = 2\lambda \sum_{m=0}^{n_0} 
\frac{(-1)^m \tilde{\chi_1(\xi,0)}}{(2\lambda)^{m+1}} \nn \\
 2\lambda \sum_{k=1}^{p}(-1)^k \sum_{j_1,...,j_k} \prod_{l=1}^k
\frac{(n_{j_{l-1}}+1)!}{(2\lambda)^{n_{j_{l-1}}+1}(\xi-\zeta_j)^{n_{j_{l-1}}}}
\sum_{m=0}^{n_{j_k}} \frac{(-1)^m \kappa_{j_1,...,j_k}^{(m)}(\xi,0)}
{(2\lambda)^{m+1}} \\
 2\lambda (-1)^{p+1} \sum_{j_1,...,j_{p+1}} \prod_{l=1}^{p+1}
\frac{(n_{j_{l-1}}+1)!}{(2\lambda)^{n_{j_{l-1}}+1}(\xi-\zeta_j)^{n_{j_{l-1}}}}
\int_{\tilde{C}} e^{2\lambda s}\kappa_{j_1,...,j_{p+1}} (\xi,s) ds \nn
\ee
where $j_0\equiv 0$ and $\kappa_{j_1...j_{p+1}}$'s constitute the $p+1^{th}$ 
generation family. The latter is constructed from the $p^{th}$ one (with 
$\zeta_{jp}$ as its singular points and with $n_{jp}/\lambda=|\zeta_{jp}^0|$
being the singularity closest to 
the origin) according to the formulae (\ref{A4.2})-(\ref{A4.5}).

It is important to stress that (\ref{A4.6}) is exact and its RHS
becomes an approximation to the left one only when the last sum
of the RHS containing the integrals is rejected.

\section*{ Appendix 5}
\renewcommand{\theequation}{A5.\arabic{equation}}
\zero

 We shall show below that the last term on the right
hand side sum in (\ref{4.15}) has to vanish when $\lambda_0 \to 0$. To this 
end let us note that we can rewrite the integral present in this term in
the following way: 

\be\label{A5.1}
\int\limits_{C'(\lambda_0)} \exp(2\lambda s) \log\chi{1\to 3}(\lambda)
d\lambda = 
\int\limits_{C_{\fr}}[ \exp(2\lambda s) \log\chi{1\to 3}(\lambda) \nn \\
+\exp(-2\lambda s) \log\chi{1\to 3}(-\lambda)]d\lambda = 
\int\limits_{C_{\fr}}[ \exp(2\lambda s)-\exp(-2\lambda s)]
 \log\chi{1\to 3}(\lambda) d\lambda \;\;\;\;\;\; \;\;\; \\
+\int\limits_{C_{\fr}^d}\exp(-2\lambda s)[1+\exp(-2\pi i\lambda)] d\lambda 
+ 
\int\limits_{C_{\fr}^u}\exp(-2\lambda s)[1\exp(2\pi i\lambda)] d\lambda \nn
\ee
where $C_{\fr}$ is the half-circle of radius $\lambda_0$
lying in the right half of the $\lambda$-plane, and $C_{\fr}^u$ and 
$C_{\fr}^d$ are the
corresponding upper and lower halves of $C_{\fr}$. We have also made
use of the relations (\ref{4.11}) to obtain the final form of (\ref{A3.1}).
It follows now from (\ref{4.10}) that we have: 

\be\label{A5.2}
\lim_{\lambda\to 0}\chi{1\to 3}(\lambda)=\sqrt{2} \mbox{for} 
\;\;\;\;\;\;\;\;\;\; |arg \lambda|
< \pi \;\;\;\;\;\;\;\;\;\;\;\;\;
\ee

Therefore, we can conclude that both $log|\chi_{1\to 3}(\lambda)|$ and 
$arg \chi_{1\to 3}(\lambda)$ are bounded in
the halfplane $\Re \lambda \geq 0$. The vanishing of all the integrals in
(\ref{A5.1}) when $\lambda_0 \to 0$ follows now directly from the last 
conclusion.

\end{document}